\documentclass[twocolumn]{aastex63}
\usepackage{amsmath}
\usepackage{amssymb}

\usepackage{makecell}
\submitjournal{AJ}

\shorttitle{A Mock Catalog of Gravitationally Lensed Quasars}
\shortauthors{Yue et al.}
\graphicspath{{./}{figures/}}
\begin{document}

\title{A Mock Catalog of Gravitationally Lensed Quasars for the LSST Survey}

\correspondingauthor{Minghao Yue}
\email{yuemh@email.arizona.edu}

\author[0000-0002-5367-8021]{Minghao Yue}
\affiliation{Steward Observatory, University of Arizona, 933 North Cherry Avenue, Tucson, AZ 85721, USA}

\author[0000-0003-3310-0131]{Xiaohui Fan}
\affiliation{Steward Observatory, University of Arizona, 933 North Cherry Avenue, Tucson, AZ 85721, USA}

\author[0000-0001-5287-4242]{Jinyi Yang}
\altaffiliation{Strittmatter Fellow}
\affil{Steward Observatory, University of Arizona, 933 North Cherry Avenue, Tucson, AZ 85721, USA}

\author[0000-0002-7633-431X]{Feige Wang}
\altaffiliation{NHFP Hubble Fellow}
\affiliation{Steward Observatory, University of Arizona, 933 North Cherry Avenue, Tucson, AZ 85721, USA}



\begin{abstract}
We present a mock catalog of gravitationally lensed quasars at $z_\text{qso}<7.5$
with simulated images for the Rubin Observatory Legacy Survey of Space and Time (LSST).
We adopt recent measurements of quasar luminosity functions to model the quasar population, 
and use the CosmoDC2 mock galaxy catalog to model the deflector galaxies, which successfully reproduces
the observed galaxy velocity dispersion functions up to $z_d\sim1.5$.
The mock catalog is highly complete for lensed quasars with Einstein radius $\theta_E>0\farcs07$ 
and quasar absolute magnitude $M_{i}<-20$. 
We estimate that there are $\sim10^3$ lensed quasars discoverable in current imaging surveys, 
and LSST will increase this number to $\sim2.4\times10^3$.
Most of the lensed quasars have image separation $\Delta \theta>0\farcs5$,
which will at least be marginally resolved in LSST images with seeing of $\sim0\farcs7$.
There will be $\sim200$ quadruply-lensed quasars discoverable in the LSST.
The fraction of quad lenses among all discoverable lensed quasars is about $\sim10\%-15\%$,
and this fraction decreases with survey depth.
This mock catalog shows a large diversity in the observational features of lensed quasars,
in terms of lensing separation and quasar-to-deflector flux ratio.
We discuss possible strategies for a complete search of lensed quasars in the LSST era. 
\end{abstract}


\section{Introduction} \label{sec:intro}

Gravitationally lensed quasars are valuable objects which enable a variety of unique studies
in extragalactic astronomy. Examples include high-resolution observations of quasar host galaxies
\citep[e.g.,][]{paraficz18},
studies of the dark matter profile and the circumgalactic medium of the foreground deflector galaxy
\citep[e.g.,][]{gilman20,cashman21},
measuring the accretion disk size of background quasars \citep[e.g.,][]{Cornachione20}, and constraining key 
cosmological parameters such as the Hubble constant \citep[e.g.,][]{suyu17}.
High-redshift lensed quasars are especially interesting, as they also probe
the faint quasar population that are inaccessible without lensing magnification 
\citep[e.g.,][]{mcgreer10, yang19, yue21}.

Despite their usefulness, gravitationally lensed quasars are rare,
with about 120 of them discovered prior to 2015 \citep[e.g.,][]{inada12}. 
In the last few years, many searches for lensed quasar have been carried out utilizing
recent wide-area imaging surveys, and have nearly doubled the sample size of lensed quasars 
\citep[e.g.,][]{more16,lemon18,lemon19,lemon20, agnello18, treu18, spiniello18, jaelani21}.
These searches have utilized effective candidate selection strategies for lensed quasars,
based on their observed features like colors, morphologies and variabilities.
In the near future, the Rubin Observatory Legacy Survey of Space and Time \citep[LSST,][]{lsst}
will deliver deep, multi-band and multi-epoch imaging with sub-arcsec spatial resolution in a wide sky area,
making it promising to build a large sample of lensed quasars in the next decade.

As the commissioning of LSST is approaching, there is a need to investigate the expected outputs of future surveys
for lensed quasars, and to design a complete and efficient candidate selection strategy with LSST data.
Mock catalogs and simulated observations are powerful tools to accomplish these tasks. 
In \citet{om10} (hereafter OM10), the authors present a mock catalog of lensed quasars,
and discuss the number of lensed quasars that can be found in a number of surveys.
\citet{agnello15} further generate simulated observations for the mock lensing systems using the properties
of real galaxies and quasars from various of sky surveys. The simulated observations,
as well as the candidate selection techniques developed based on it, have been proven to be effective 
in many lensed quasar searches \citep[e.g.,][]{agnello18, spiniello18, lemon20}.

However, the mock catalog in OM10 and the simulated observations have a few limitations. First,
the OM10 mock catalog is based on early measurements of quasar luminosity functions (QLFs) and galaxy
velocity dispersion functions (VDFs), while recent observations have provided more accurate measurements
of the quasar and galaxy populations, especially at high redshifts.
Second, the OM10 catalog only includes quasars at $z_\text{qso}<5.5$, while the depth and wavelength coverage of LSST
will allow finding lensed quasars at  redshift $z\gtrsim 7$.
In addition, the simulated observations in \citet{agnello15} were largely based on wide-area sky surveys like 
the Sloan Digital Sky Survey \citep[SDSS, ][]{sdss} and the Wide-field Infrared Survey Explorer \citep[{\em WISE}, ][]{wise},
which are usually not deep enough to characterize faint galaxies at relatively high redshift.
As such, the observed features of lensed quasars with faint (and thus less massive) deflector galaxies are not well-represented.

In this work, we present a new mock catalog of lensed quasars at $z_\text{qso}<7.5$ with simulated 
spectral energy distributions (SEDs) and LSST images,
extending previous studies to higher quasar redshifts and smaller deflector galaxy masses.
We use updated QLFs and VDFs to ensure accurate modeling of the lensed quasar population.
This mock catalog predicts the number of discoverable lensed quasars in current and next-generation sky surveys,
and can serve as training and testing sets in future searches for lensed quasars.
This paper is organized as follows.
We describe the method of generating the mock catalog in Section \ref{sec:method}.
Section \ref{sec:statistics} describes the predicted statistics of lensed quasars. 
Section \ref{sec:discussion} discusses some systematic uncertainties of the mock catalog
and investigate possible strategies for future lensed quasar searches.
We summarize in Section \ref{sec:conclusion}.
Throughout this paper we use a flat Lambda cold dark matter ($\Lambda$CDM) cosmology with 
$H_0=70 \text{ km s}^{-1}\text{ kpc}^{-1}$ and $\Omega_M=0.3$.

\section{Mock Catalog Generation} \label{sec:method}

\begin{figure*}
    \centering
    \includegraphics[width=0.98\textwidth,trim={0 6cm 6cm 6cm}]{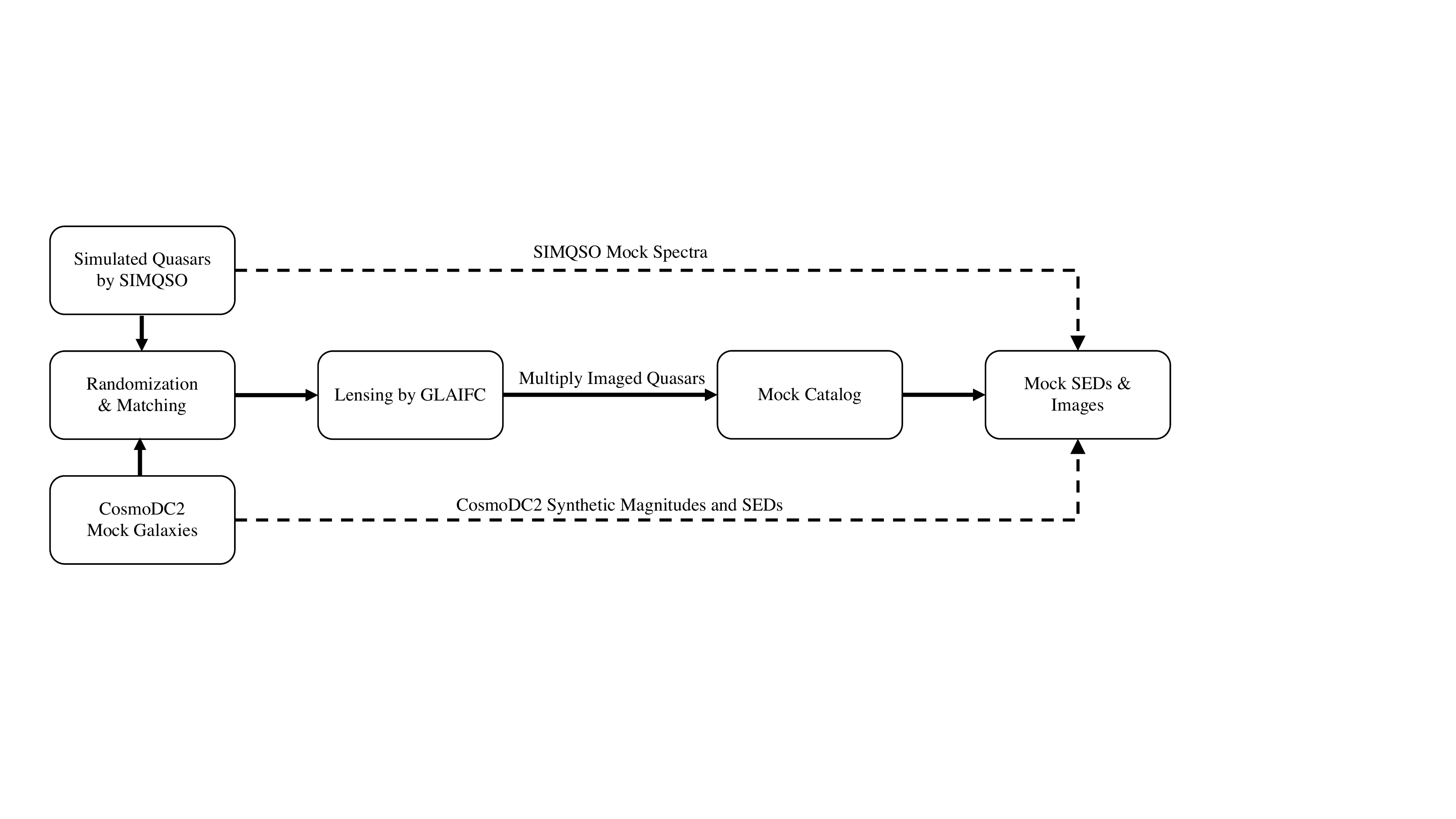}
    \caption{The flowchart of lensed quasar mock catalog generation process.}
    \label{fig:flowchart}
\end{figure*}

In this section, we describe the method we use to generate the mock catalog,
including simulations of the quasar population and the deflector galaxy population,
as well as the matching and lensing algorithm.
Figure \ref{fig:flowchart} shows the flow chart of these processes.
\footnote{The mock catalog and the code are available at https://github.com/yuemh/lensQSOsim.git}

\subsection{Deflector Population}\label{subsec:deflector}
We use the CosmoDC2 mock galaxy catalog \citep{dc2} to model the deflector galaxies.
CosmoDC2 is a synthetic galaxy catalog that covers an area of 440 deg$^2$ out to a redshift of $z=3$.
CosmoDC2 first identifies dark matter halos in the $N-$body simulation Outer Rim \citep{outerrim}, 
and assigns simulated galaxies from the UniverseMachine \citep{behroozi19} to the dark matter halos. 
The algorithm is tuned to match the observed stellar mass versus star formation rate $(M_*, \text{SFR})$ distributions of galaxies. 
CosmoDC2 then assigns various of properties to the mock galaxies, including stellar masses, 
optical-to-near infrared (NIR) SEDs, half-light radii and ellipticities, using empirical relations or 
simulated galaxy spectra library. 
\citet{dc2} shows that the mock catalog reproduces the observed colors and number counts of galaxies at a wide range of redshifts.
The large area of CosmoDC2 catalog ensures a small statistical error of the simulated lens sample,
and the galaxy properties provides all the information needed to generate mock observations
of  lensing systems.

We use singular isothermal ellipsoids \citep[SIE, e.g.,][]{SIE} to model the lensing potential 
of the deflector galaxies. An SIE lens is parameterized by its ellipticity, 
$e=1-q$ where $q$ is the axis ratio\footnote{The ellipticity defined here follows the convention in extragalactic astronomy; it is different from that of the eccentricity of an elliptical orbit, which is $\sqrt{1-q^2}$.}, and its Einstein radius,
\begin{equation} \label{eq:thetaE}
    \theta_E=4\pi\left(\frac{\sigma}{c}\right)^2\frac{D_\text{ds}}{D_\text{s}}
\end{equation}
where $\sigma$ is the line-of-sight velocity dispersion of the galaxy,
$D_\text{s}$ is the angular diameter distance from the observer to the source,
and $D_\text{ds}$ is the angular diameter distance from the deflector to the source.

We estimate the velocity dispersion of a galaxy using its stellar mass. 
Specifically, we convert stellar masses to dynamical masses using a stellar-to-dynamical mass ratio,
and calculate the velocity dispersion by applying the virial theorem \citep[e.g.,][]{cappellari06}:
\begin{equation} \label{eq:virial}
    \sigma = \sqrt{\frac{ GM_\text{dyn}}{K R_e}}
\end{equation}
We adopt $M_*/M_\text{dyn}=0.557$, which gives a good description for galaxies in
a wide mass $(100\text{ km s}^{-1}<\sigma<300\text{ km s}^{-1})$ 
and redshift $(0<z<1)$ range \citep[e.g.,][]{bezanson11,bezanson12}.
Since early-type galaxies dominates the deflector population
in galaxy-scale lenses \citep[e.g.,][]{oguri06,inada12},
we adopt the typical scaling factor $K=6$ for early-type galaxies \citep[e.g.,][]{beifiori14, nn16}.
We do not use the S\'ersic-index-dependent scaling factor as suggested in, e.g., \citet{bertin02},
as CosmoDC2 uses a bulge + disk composition to describe the galaxy morphology instead of a S\'ersic profile.

\begin{figure*}
    \centering
    \includegraphics[width=0.45\textwidth]{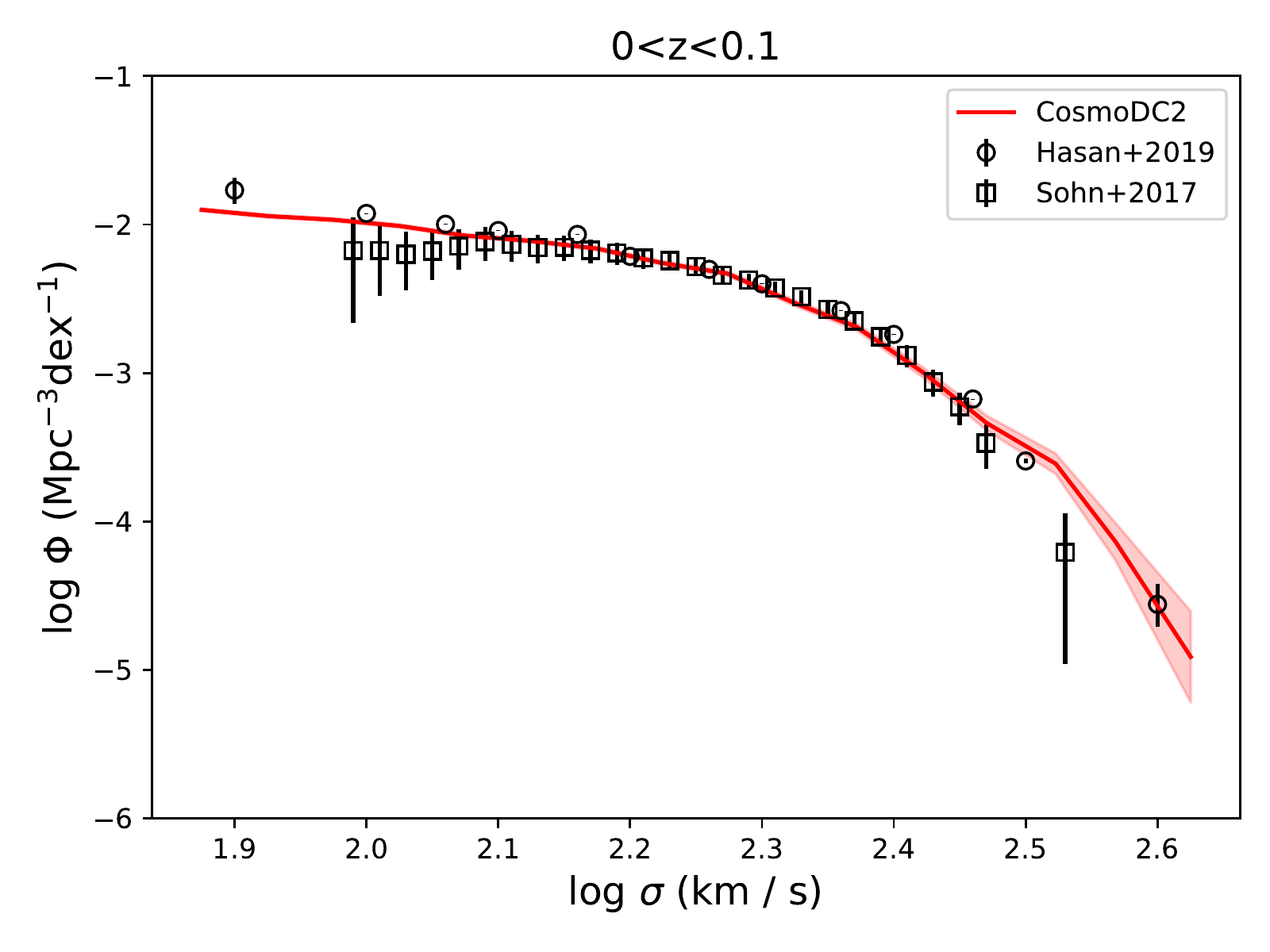}
    \includegraphics[width=0.45\textwidth]{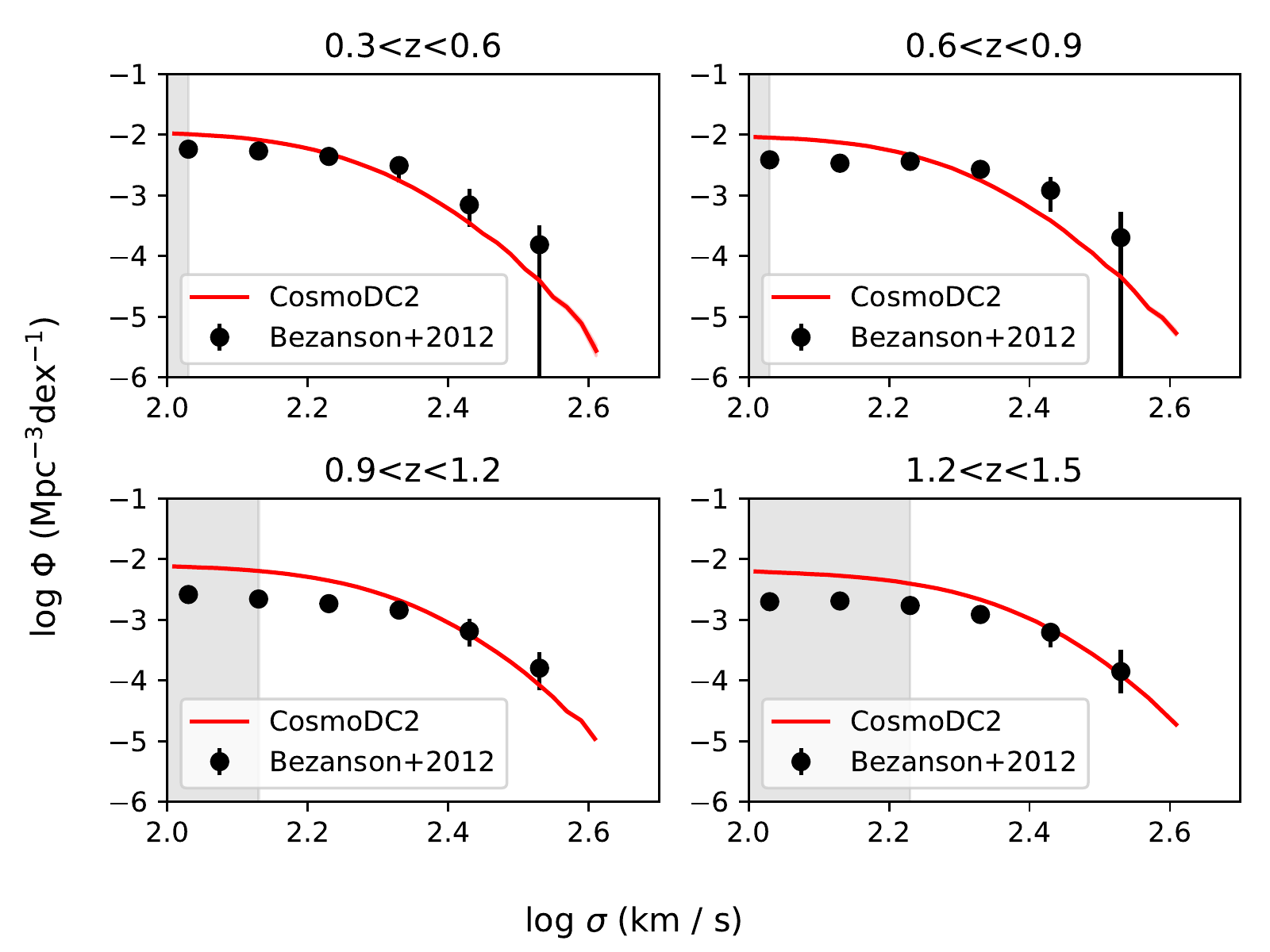}    
    \caption{The comparison between the derived VDFs of the CosmoDC2 catalog and the observed VDF at $z<1.5$.
    {\em Left}: The local $(z<0.1)$ VDF. The VDF of the CosmoDC2 catalog accurately matches the  
    spectroscopically measured VDF in \citet{sohn17} and \citet{hasan19}.
    {\em Right}: The VDF at $0.3<z<1.5$. The gray area marks the region where \citet{bezanson12} suggest that
    their galaxy sample has a low completeness. The CosmoDC2 VDF is in good agreement with the observed VDF up to
    $z=1.5$. At the less massive end, the CosmoDC2 catalog gives a higher number density than the observation,
    which can be explained by the incompleteness of the observed galaxy sample.
    }
    \label{fig:vdf}
\end{figure*}

Figure \ref{fig:vdf} compares the derived VDF of the CosmoDC2
mock galaxy catalog with the observed galaxy VDF. The local VDF has been accurately measured spectroscopically
\citep[e.g.,][]{choi07}. 
We find that the VDF of the CosmoDC2 catalog at $z<0.1$ is fully consistent with the recent spectroscopic measurements
\citep[][]{sohn17, hasan19}. 
At higher redshifts, it is challenging to measure the velocity dispersion of a large galaxy sample and thus the VDF
spectroscopically.
%and most of the previous studies use the SED and sizes to etimate the VDF beyond the local universe (cite).
The right panel of Figure \ref{fig:vdf} compares the CosmoDC2 catalog VDF 
with the observed VDF at $z<1.5$ from \citet{bezanson12}, who measure the dynamical mass of galaxies
in the COSMOS and UDF fields using photometric data.
The CosmoDC2 catalog is also in good agreement with the observed values except at the low-mass end at $z>0.6$, 
which can be due to incompleteness of the galaxy sample in the observations.
Figure \ref{fig:vdf} suggests that our method successfully reproduces the observed VDF
at least at $z<1.5$.

In this work, we only include galaxies with $\sigma>50\text{ km s}^{-1}$, which corresponds to
an Einstein radius limit of $\theta_{E, \text{lim}}=4\pi(\sigma_\text{lim}/c)^2=0\farcs072$.
Lensing systems with smaller $\theta_E$ are difficult to identify even with
space telescopes like the {\em Hubble Space Telescope (HST)}.

We notice that the observed $M_*-\sigma$ relation have significant scatters,
 which can be partly explained by the effect of galaxy sizes $(R_e)$.
 The observed scatter of Equation \ref{eq:virial} is $\sim0.02$dex for the velocity dispersion
 \citep[e.g.,][]{auger10}. This small scatter have little influence on the galaxy VDF,
 and is not considered in this work.

In addition to the deflector galaxy, we add an external shear and convergence to each lensing system.
%To obtain the probabilistic distributions of external shears and convergences,
We use the CosmoDC2 catalog to estimate the redshift evolution of external shears and convergences. 
Specifically, we divide the mock galaxy catalog into redshift bins with $\Delta z=0.01$, 
then fit the distribution of external convergences $(\kappa)$ 
and the two components of the external shear $(\gamma_1, \gamma_2)$
in each redshift bin as normal distributions centered at zero. 
For both $\kappa$ and $\gamma$, the best-fit standard deviation
can be well-described by a linear function of $\log(1+z)$:
\begin{equation}\label{eq:convshear}
\begin{split} 
    \sigma_\kappa &= 0.057\times\log(1+z) - 0.001\\
    \sigma_{(\gamma_1 \text{ or }\gamma_2)} &= 0.040\times\log(1+z) - 0.001
\end{split}
\end{equation}
For each lensing system, we use the best-fit normal distribution to generate a random set of
$(\kappa, \gamma_1, \gamma_2)$ values. 
\footnote{The actual distributions of external convergence and shears can be skewed.
 We use normal distributions in the sake of simplicity, since the convergence and shears
 have effectively no impact on the predicted number of lensed quasars.}

Note that the convergence and shears in the CosmoDC2 catalog have some
limitations. Specifically, the convergence and shears from ray-tracing simulations depend
on the resolution (i.e., the smoothing scale) of the density field, especially at small scales
\citep[e.g.,][]{hilbert09}. To perform the ray-tracing simulation,
CosmoDC2 divides the mass particles from the cosmological simulation
into discrete shells and estimates the surface mass density of each shell on a 
HEALPix grid with Nside=4096, which corresponds to a resolution of 0.5 arcminutes on the sky.
\citet{dc2} compares the cosmic shear power spectrum of the CosmoDC2 catalog and theoretical predictions,
showing that the difference is negligible at $\ell<1000$ and is $\lesssim10\%$ at $\ell\sim4000$.
As the values of $\kappa$ and $\gamma$ are small (Equation \ref{eq:convshear}),
this systematic uncertainty has effectively no impact on the number of lensed quasars.

\subsection{Quasar Population}
We simulate quasars at $z<7.5$ using python code SIMQSO\footnote{https://github.com/imcgreer/simqso} \citep{simqso}.
SIMQSO generates mock catalogs of quasars by randomly sampling
the absolute magnitude and the redshift of quasars according to the input QLF.
For each simulated quasar, SIMQSO generates a mock spectrum which includes a
broken power-law continuum and a number of emission line components, and sets 
a random sightline to model the intergalactic medium (IGM) absorption based on observed IGM models.
We use the default continuum, emission line and IGM parameters in SIMQSO 
which has been shown to successfully reproduce the observed quasar color distribution \citep[e.g.,][]{dr9}
and has been widely used to estimate the completeness of quasar surveys \citep[e.g.,][]{jiang16, yang16, McGreer18, wang19}.

We use a double power-law to describe the QLF:
\begin{equation} \label{eq:QLF}
    \Phi(M,z) = \frac{\Phi_*}{10^{0.4(M-M^*)(\alpha+1)}+10^{0.4(M-M^*)(\beta+1)}}
\end{equation}
where $\alpha$ and $\beta$ are faint and bright end slopes of the QLF,
$M^*$ is the break absolute magnitude and $\Phi_*$ is the normalization.
%Since we focus on high redshift in this work, we adopt the best-fit parametric QLFs from deep spectroscopic quasars surveys.
Specifically, the QLF we adopt at different redshifts are as follows:

\begin{enumerate}
    \item At $0<z<3.5$, we adopt the pure luminosity evolution (PLE, for $z<2.2$) and the luminosity evolution and density evolution (LEDE, for $2.2<z<3.5$) model from \citet{dr9}. We use the relations in the Appendix B of \citet{dr9}
    to convert the absolute magnitudes to $M_\text{1450}$, i.e., the absolute magnitude at rest-frame 1450\AA.
    \item At $3.5<z<7.5$, we compile a number of QLFs at different redshifts from the literature. These QLFs are summarized in Table \ref{tbl:QLF}. At $3.5<z<6.7$, we use linear interpolation to obtain the QLF parameters at a certain redshift. At $z>6.7$, we keep $\alpha, \beta$ and $M^*$ fixed to the $z=6.7$ value, and apply a density evolution of $\log \Phi_*(z) = \log \Phi_*(z=6.7) - 0.78 \times (z-6.7)$, following \citet{wang19}. 
\end{enumerate}

\begin{figure*}
    \centering
    \includegraphics[width=0.9\textwidth]{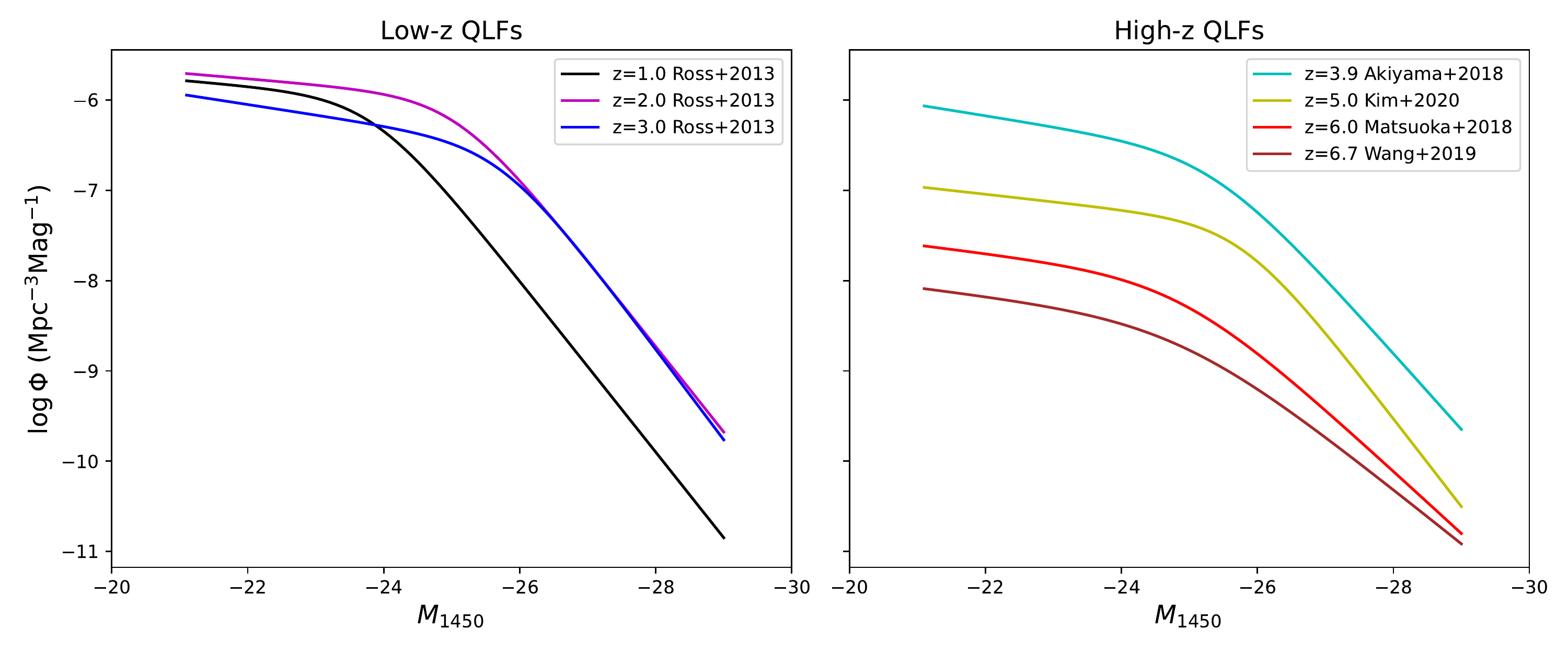}
    \caption{QLFs adopted by this work. We use the QLF from \citet{dr9} to 
    model quasars at $z<3.5$, and compile a number of literature to construct the QLF at $z>3.5$
    (see text for details). {\em Left}: The \citet{dr9} QLFs. 
    {\em Right}: The high-redshift QLFs we use from the literature. At $z\gtrsim4$,
    the values of $M^*, \alpha$ and $\beta$ do not show strong redshift evolution, 
    suggesting that a linear interpolation between the redshifts is adequate.}
    \label{fig:QLF}
\end{figure*}

Figure \ref{fig:QLF} shows the QLF used in this work at different redshifts. 
Both Table \ref{tbl:QLF} and the right panel of Figure \ref{fig:QLF} suggest a subtle redshift evolution
of $M^*, \alpha$ and $\beta$ at high redshift. In addition, a linear interpolation of
 $\log \Phi_*$ is equivalent to an exponential redshift evolution of $\Phi_*$,
which has been shown to well-describe the number density of high-redshift quasars
\citep[e.g.,][]{jiang16, McGreer18, wang19}.
Using linear interpolations between redshifts 
is thus sufficient to accurately capture the evolution of QLFs at high redshift.

Following the Active Galactic Nuclei (AGN) chapter in the LSST science book \citep{sciencebook}, 
we only include quasars with $M_{i}<-20$ in our simulated quasar sample, 
which is equivalent to $M_{1450}<-19.1$ according to the relation in \citet{dr9}.
For faint AGN below this flux limit, host galaxies start to dominate their emissions.
These objects require a survey strategy that is different from classical quasars and
 is out of the scope of this paper.

The area of the CosmoDC2 catalog is 440 deg$^2$, which is $1.067\%$ of the whole sky. 
To make the simulated lens catalog equivalent to an all-sky catalog, 
we scale the QLF by a factor of $2\%/1.067\%=1.874$, so that the number of expected lensed quasars
in the CosmoDC2 field equals to $1/50$ of the sky. By running 50 random realizations (see Section \ref{subsec:lensing}), 
we can obtain a mock catalog that is equivalent to the all-sky area.

\begin{deluxetable*}{ccccc}
\tablenum{1}
\label{tbl:QLF}
\tablecaption{QLF Parameters from the Literature}
\tablewidth{0pt}
\tablehead{
\colhead{Redshift} & \colhead{$\log{\Phi_*}$} & \colhead{$M^*$} & \colhead{$\alpha$}  & \colhead{$\beta$}\\
\colhead{} & \colhead{$(\text{Mpc}^{-3}\text{mag}^{-1})$} & \colhead{(mag)} & \colhead{}  & \colhead{}}
%\decimalcolnumbers
\startdata
\hline
% Low-Redshift QLFs \\\hline
$0.1<z<2.2$  \tablenotemark{1} & $-5.96^{+0.02}_{-0.06}$ & $-21.364^{+0.05}_{-0.11}-2.5\times(1.241^{+0.010}_{-0.028}z-0.249^{+0.006}_{-0.017}z^2)$ & $-1.16^{+0.02}_{-0.04}$ & $-3.37^{+0.03}_{-0.05}$\\
$2.2<z<3.5$ \tablenotemark{2} & $-5.83^{+0.15}_{-0.25}-0.689^{+0.021}_{-0.027}\times(z-2.2)$ & $-25.00^{+0.34}_{-0.24} - 0.809^{+0.033}_{-0.166}\times(z-2.2)$ & $-1.31^{+0.52}_{-0.19}$ & $-3.45^{+0.35}_{-0.21}$  \\
$z=3.9$ \tablenotemark{3} & $-6.58\pm0.01$ & $-25.36\pm0.13$ & $-1.30\pm0.05$ & $-3.11\pm0.07$ \\
$z=5.0$ \tablenotemark{4} & $-7.36^{+0.56}_{-0.81}$ & $-25.78^{+1.35}_{-1.10}$ & $-1.21^{+1.36}_{-0.64}$ & $-3.44^{+0.66}_{-0.84}$ \\
$z=6.0$ \tablenotemark{5} & $-7.96^{+0.28}_{-0.42}$ & $-24.90^{+0.75}_{-0.90}$ & $-1.23^{+0.44}_{-0.34}$ & $-2.73^{+0.23}_{-0.31}$ \\
$z=6.7$ \tablenotemark{6} & $-8.43\pm0.23$ & $[-24.90]$ & $[-1.23]$ & $-2.51\pm0.39$ \\\hline
%$z=5.0$ \tablenotemark{7} & $-8.97^{+0.15}_{-0.18}$ & $-27.47^{+0.22}_{-0.26}$ & $-1.97^{+0.09}_{-0.09}$ & $[-4.00]$ \\
%$z=6.0$ \tablenotemark{8} & $-9.14\pm0.27$ & $-27.16\pm0.44$ & $-1.79\pm0.11$ & $[-3.60]$ \\
%$z=6.7$ \tablenotemark{9} & $-9.56\pm0.38$ & $-27.20\pm0.74$ & $[-1.79]$ & $[-3.60]$ \\\hline
\enddata
\tablenotetext{1}{The pure luminosity evolution (PLE) model in \citet{dr9}}
\tablenotetext{2}{The luminosity evolution and density evolution (LEDE) model in \citet{dr9}}
\tablenotetext{3}{The parameterized QLF at $z=3.9$ from \citet{akiyama18}.}
\tablenotetext{4}{The parameterized QLF at $z=5$ from \citet{kim20}.}
\tablenotetext{5}{The parameterized QLF at $z=6$ from \citet{matsuoka18} (the standard model in their Table 5).}
\tablenotetext{6}{The QLF at $z\sim6.7$ by fitting the binned QLF from \citet{wang19}, with the faint-end slope and $M^*$ fixed to the values in \citet{matsuoka18}.}
%\tablenotetext{7}{The parameterized QLF at $z\sim3.9$ from \citet{mcgreer18}, where they fix the bright-end slope to $-4$.}
%\tablenotetext{8}{The QLF at $z\sim6$ by fitting the binned QLF from \citet{matsuoka18}, with the faint-end slope fixed to $-3.6$.}
%\tablenotetext{9}{The QLF at $z\sim6.7$ by fitting the binned QLF from \citet{wang19}, with the faint-end slope and the bright-end slope fixed.}
\tablecomments{The parameters are defined in Equation \ref{eq:QLF}.
Absolute magnitudes in this table are $M_{1450}$.
 The square brakets marks quantities that are fixed when fitting the QLFs. Specifically,
 \citet{wang19} only measures the bright end of $z\sim6.7$ QLF, and we fit the double power-law by fixing 
the $M^*$ and $\alpha$ values to that of \citet{matsuoka18}.
For $3.5<z<6.7$, we perform linear interpolation to obtain QLF model parameters.
For $z>6.7$, we assume a pure density evolution of 
$\log \Phi_*(z) = \log \Phi_*(z=6.7) - 0.78 \times (z-6.7)$, following \citet{wang19}.}
\end{deluxetable*}

\subsection{Lensing Pipeline} \label{subsec:lensing}

After producing the mock deflector galaxies and the simulated background quasars,
we generate one realization of the mock lens catalog as follows.
We first assign random positions in the CosmoDC2 field to the simulated quasars,
and assign random position angles to the mock galaxies.
Then, for each galaxy, we identify all quasars that have distances to the galaxy
smaller than $5\theta_{E, max}$, where $\theta_{E, max}=4\pi(\sigma/c)^2$ 
is the maximum possible Einstein radius for the deflector galaxy.
For each matched galaxy-quasar pair, we assign a random external convergence and a random external shear
according to the best-fit probabilistic distribution described in Section \ref{subsec:deflector},
and use \texttt{GLAFIC} \citep{glafic} to model the lensing configuration.
\texttt{GLAFIC} is a software that provides fast and flexible lensing analysis for
a variety of deflector mass and source emission models.
We then select multiply-imaged quasars into the mock lens catalog.

 We run 50 random realizations to generate a mock lens catalog that is equivalent to the all-sky area.
For quasars at $z>5$, due to their low number density, we run 1,000 realizations (i.e., $20\times$ all-sky area)
to reduce the Poisson noise. %The data product is further described in Appendix.

\subsection{Broad-Band Photometry and Mock Images}
We calculate the magnitdues of the simulated lensed quasars in the
LSST $ugrizy$, UKIDSS $JHK$  \citep{ukidss}, and {\it WISE} W1, W2 bands. 
We use the simulated spectra generated by SIMQSO to calculate the synthetic magnitudes of quasars.
For deflector galaxies, CosmoDC2 provides the synthetic LSST $ugrizy$ magnitudes
and 30 top-hat SED points in the wavelength range $1000\text{\AA}<\lambda<20000\text{\AA}$. 
We directly adopt the synthetic $ugrizy$ magnitudes 
from the CosmoDC2 catalog, and estimate the $JHK$ and W1, W2 magnitudes 
using the SED. Specifically, the SED covers $J$ and $H$ at all redshifts, 
and we linearly interpolate the SED points to build a spectra and calculate the magnitudes.
At  low redshifts, the SED does not cover the wavelengths of $K$, W1 and/or W2.
In this case, we fit the SED using the galaxy templates from
\citet{brown14}, and use the best-fit template to estimate the magnitude in that filter.

\begin{figure*}
    \centering
    \includegraphics[width=0.48\linewidth]{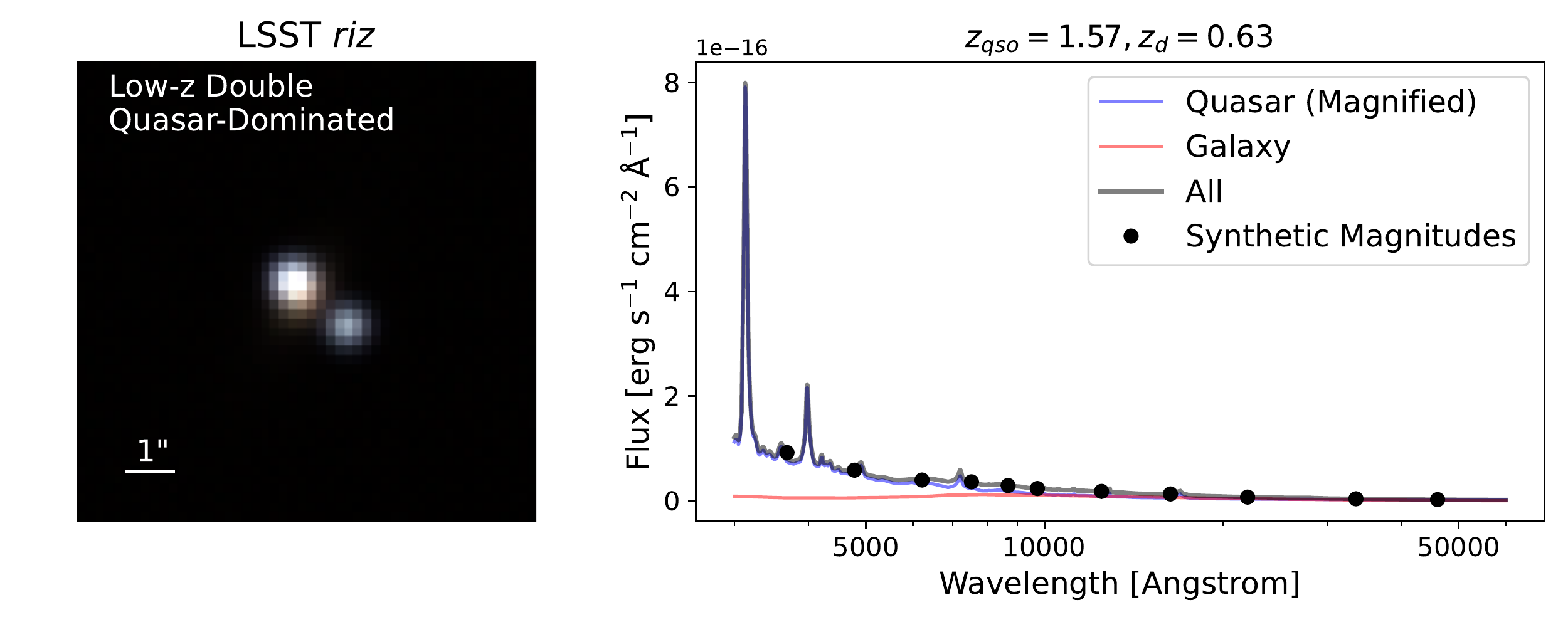}
    \includegraphics[width=0.48\linewidth]{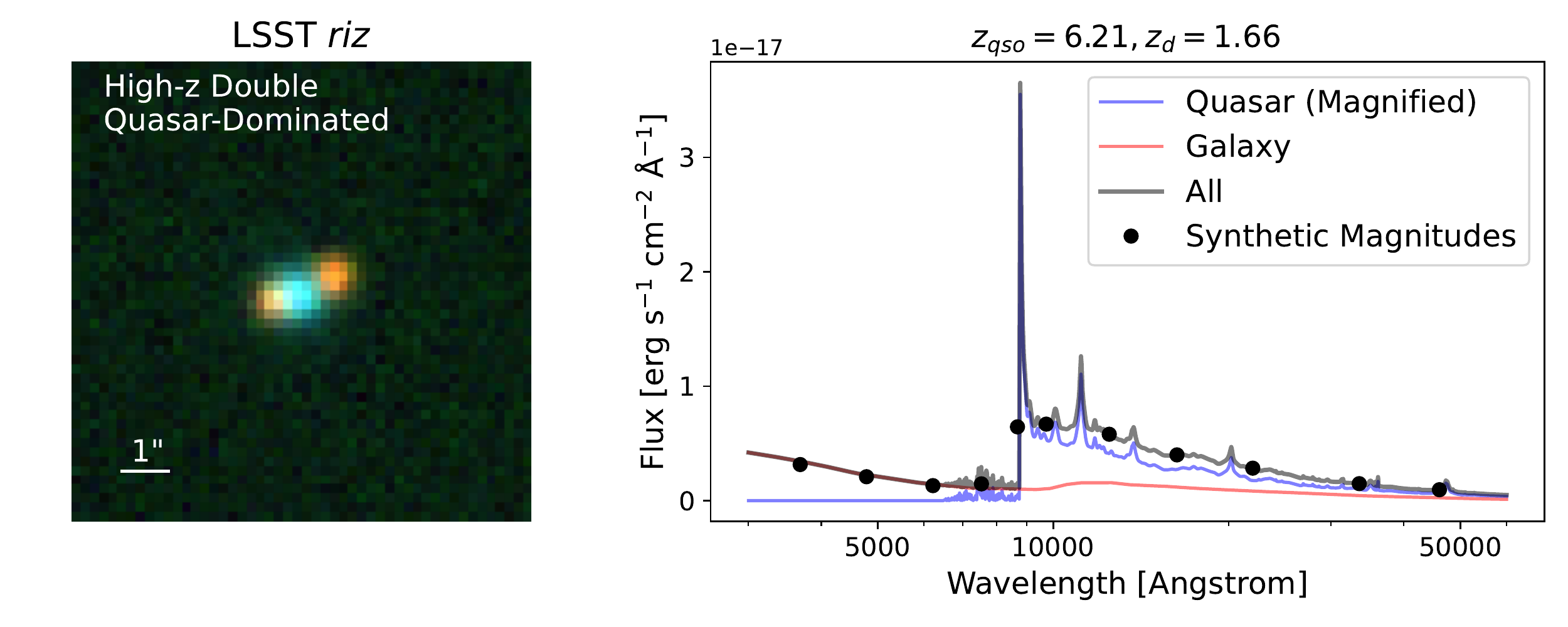}
    \includegraphics[width=0.48\linewidth]{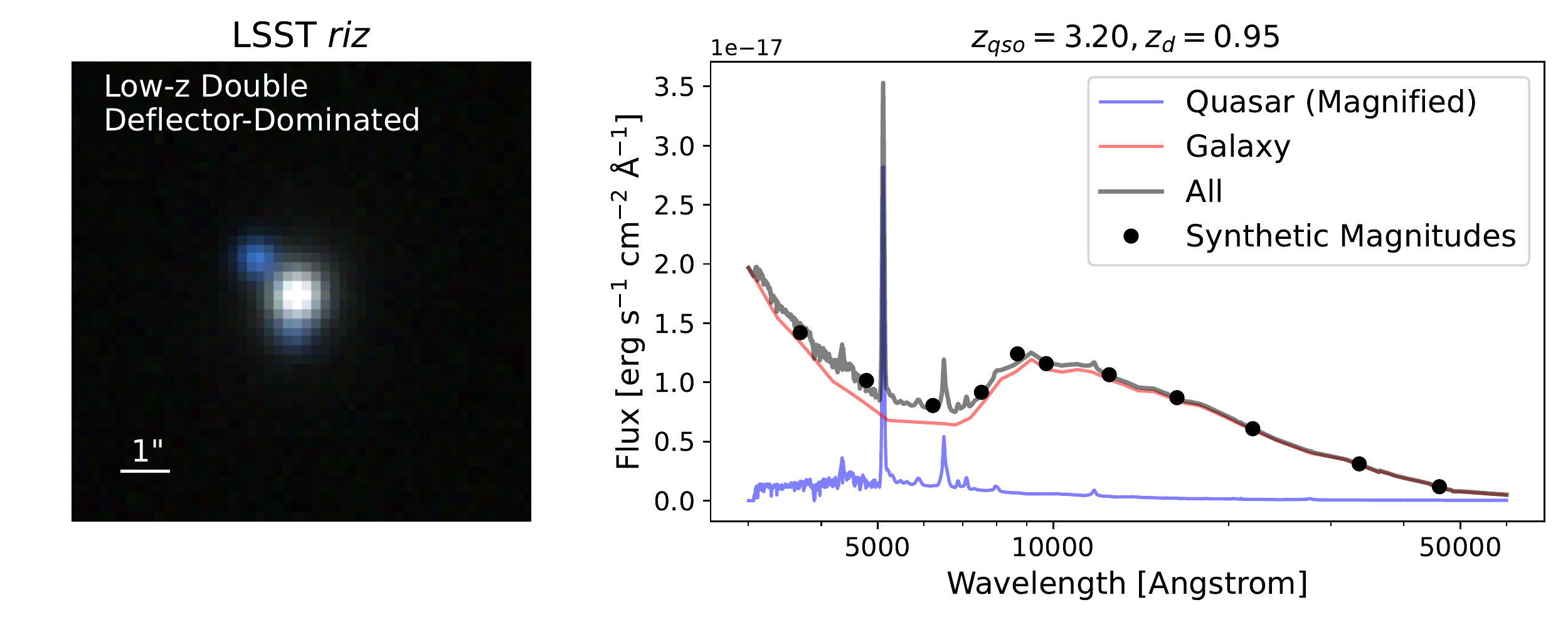}
    \includegraphics[width=0.48\linewidth]{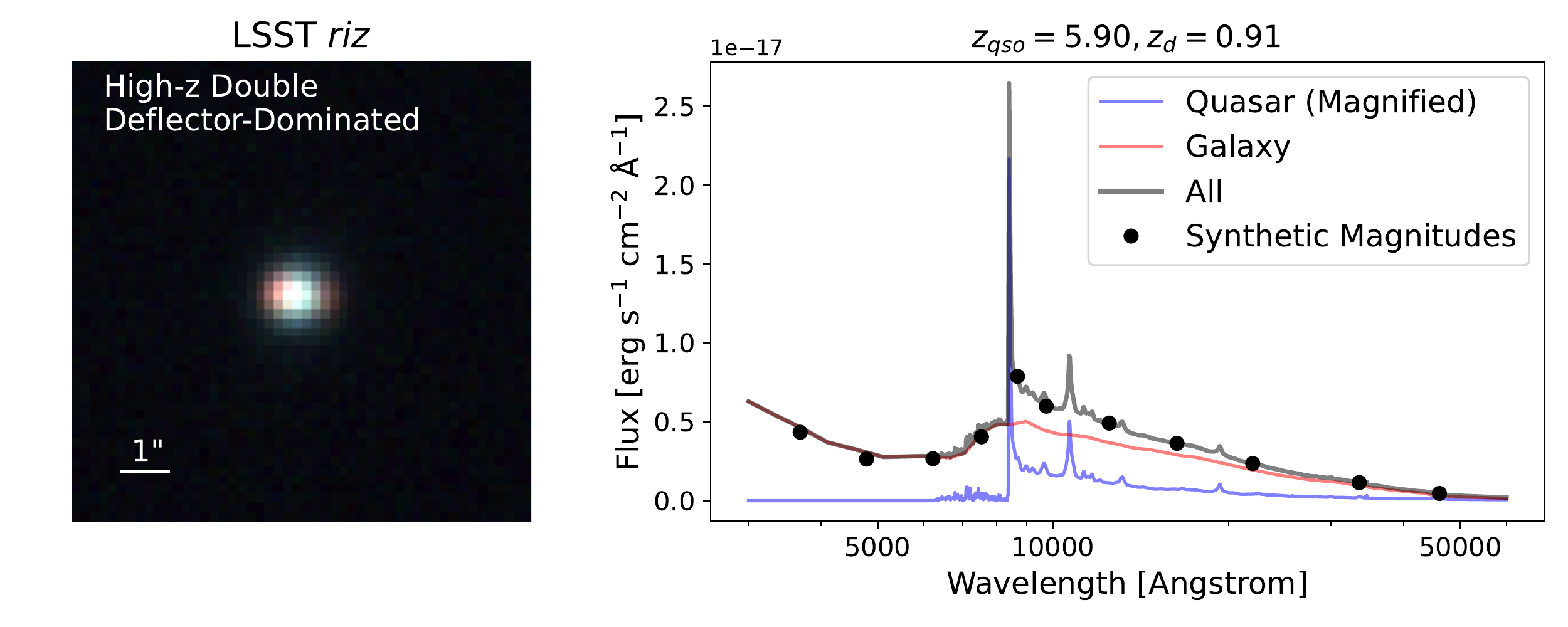}
    \includegraphics[width=0.48\linewidth]{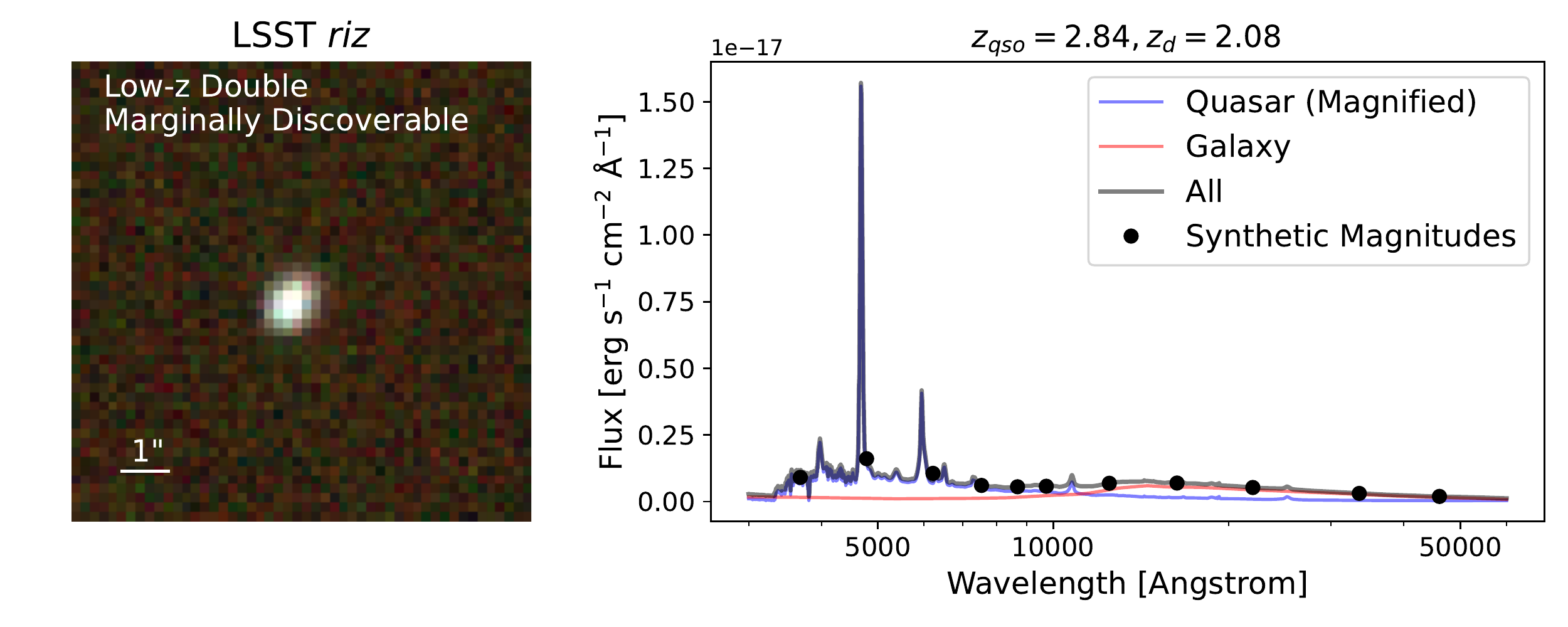}
    \includegraphics[width=0.48\linewidth]{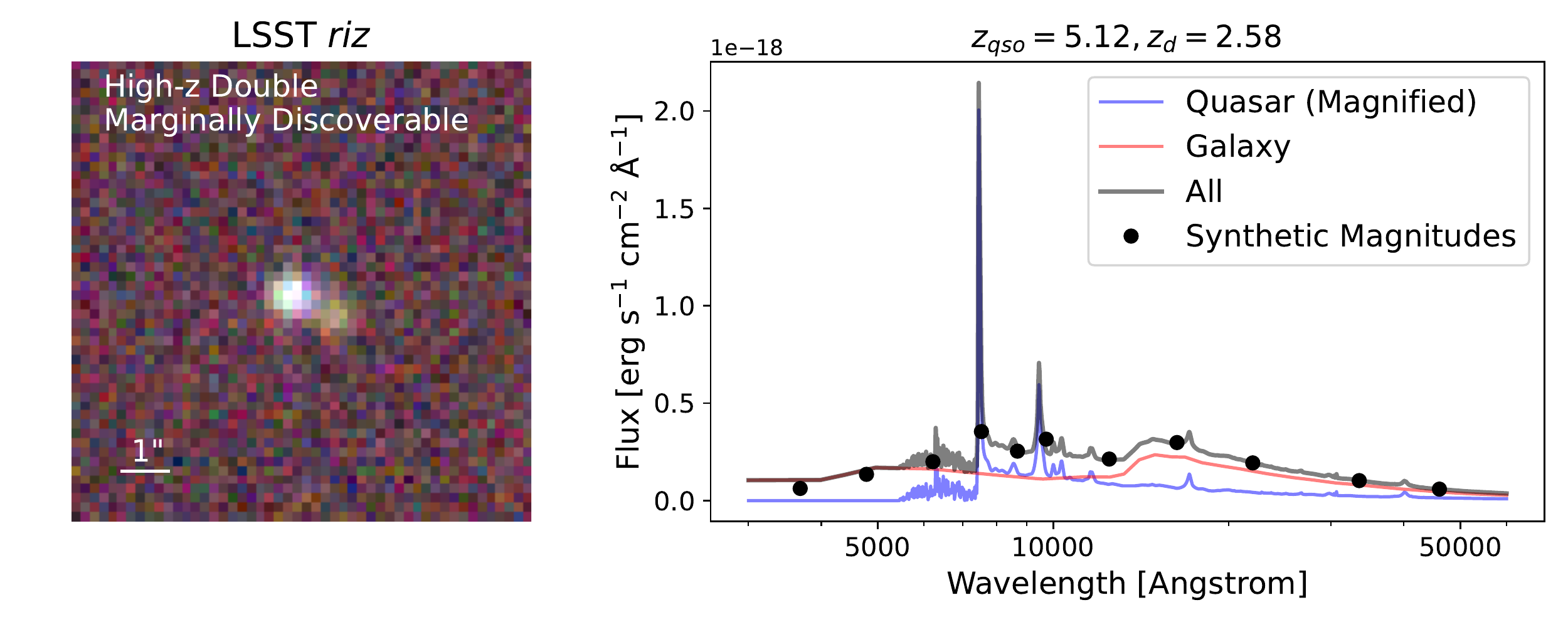}
    \includegraphics[width=0.48\linewidth]{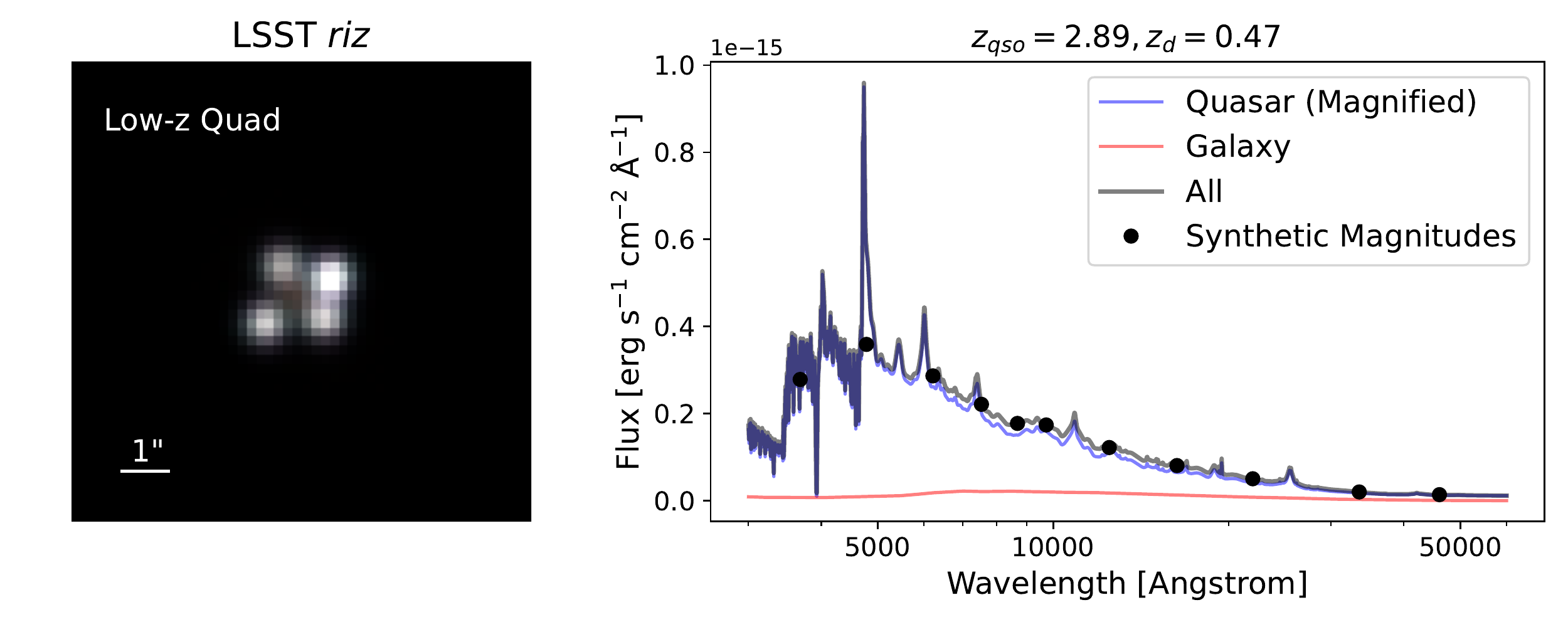}
    \includegraphics[width=0.48\linewidth]{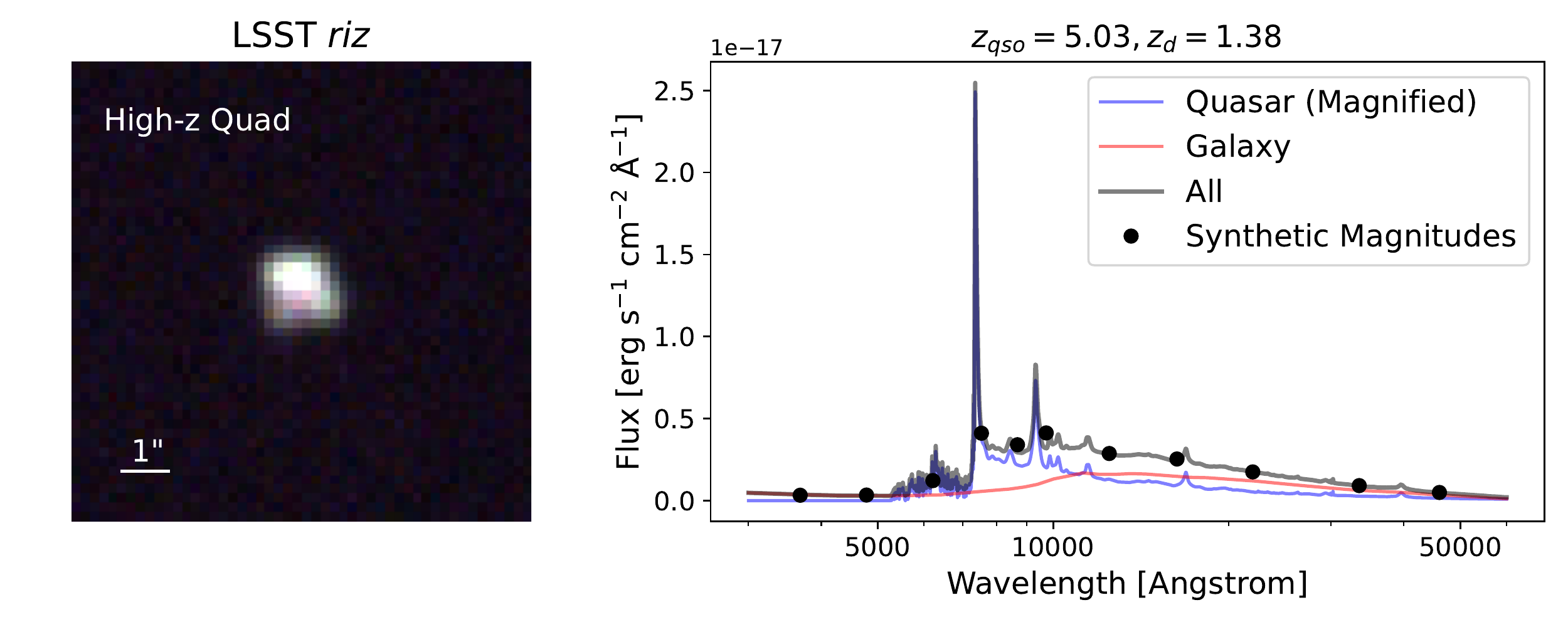}
    \caption{A gallery of mock lensed quasars. We show several typical cases
    for both the low-redshift and the high-redshift sample. 
    The synthetic magnitudes are in LSST $ugrizy$, UKIDSS $JHK$, and {\em WISE} W1, W2 bands.
    The mock lensed quasars show a large diversity in observational features in magnitudes, colors, lensing separations, etc.
    It is worth noticing that (1) the deflector galaxy
    is always at the center of the image, and is usually blended with the fainter lensed image in doubly-lensed systems;
    (2) the low-redshift example of a ``marginally discoverable" lens (third row, left) 
    has a small lensing separation of $0\farcs5$, and the high-redshift example (third row, right)
    has a marginally detectable fainter image (which is blended with the deflector galaxy).
    See section \ref{sec:statistics} for the definition of discoverable lenses.}
    \label{fig:example}
\end{figure*}

To further explore how survey strategies work for real data,
we generate simulated LSST images in $ugrizy$ bands for the mock lensed quasars.
We use point sources to describe the lensed images of the background quasar,
and use a bulge (de Vaucouleurs) + disk (exponential) two-component model to describe the foreground galaxy.
CosmoDC2 provides the radius and the SED of both components for each galaxy.
We then generate mock images using {\em galfit} \citep{galfit} for each lensing system.
Specifically, the images are created as follows:
\begin{enumerate}
    \item {\em The Point Spread Function (PSF)}: The PSF of each image is modeled as an 2-D Gaussian profile.
    The full-width half maximum (FWHM) is drawn from a normal distribution with a standard deviation
    $\sigma_\text{FWHM}=0.02\times mean$, where $mean$
    is the mean PSF size of the corresponding filter.
    The axis ratio is uniformly drawn in $[0.95, 1]$, and the position angle is
    uniformly drawn from $[0, 2\pi)$. This approach mimics the PSF variation
    in real observations.
    \item {\em Image Noises}: 
    We follow the methods in the LSST review paper \citep[][also see https://smtn-002.lsst.io/]{lsst}
    to calculate the errors of image pixels. 
    The random noise of an pixel consists of the background noise and the Poisson noise of the source flux.
    These noises are calculated based on Equation 5 in \citet{lsst}.
\end{enumerate}
Table \ref{tbl:survey} summarizes the mean PSF sizes and depths we used in the simulation. Figure \ref{fig:example}
shows some examples of the SED and simulated LSST images of a mock lensed quasar, which exhibit a large diversity in
observational features.

 \begin{deluxetable}{ccc}
\tablenum{2}
\label{tbl:survey}
\tablecaption{PSF Size and Depth of the Mock LSST Images}
\tablewidth{0pt}
\tablehead{\colhead{Filter} & \colhead{$5-\sigma$ Depth}\tablenotemark{1} & \colhead{PSF FWHM} \\
\colhead{} & \colhead{(mag)} & \colhead{$('')$}}
%\colhead{} & \multicolumn{5}{c}{Filter} & & \multicolumn{6}{c}{LSST-like}  \\
%\colhead{} & \colhead{$g$} & \colhead{$r$} & \colhead{$i$} & \colhead{$z$} & \colhead{$y$} &  & \colhead{$u$} & \colhead{$g$} & %\colhead{$r$} & \colhead{$i$} & \colhead{$z$} & \colhead{$y$}}
\startdata
\hline
$u$ & 26.1 & 0.81 \\
$g$ & 27.4 & 0.77 \\
$r$ & 27.5 & 0.73 \\
$i$ & 26.8 & 0.71 \\
$z$ & 26.1 & 0.69 \\
$y$ & 24.9 & 0.68\\\hline
\enddata
\tablenotetext{1}{For point sources.}
\end{deluxetable}

The final data products are a mock catalog of lensed quasars and mock LSST images for each simulated lensed quasar.
We provide a number of properties in the mock catalog, including the lensing configuration, 
the time delay of each lensed image, and the magnitudes of the quasar and the deflector. 
These properties are summarized in Table \ref{tbl:columns}.

\begin{deluxetable*}{ccl}
\tablenum{3}
\label{tbl:columns}
\tablecaption{Description of the columns in the mock catalog}
\tablewidth{0pt}
\tablehead{\colhead{Name} & \colhead{Format} & \colhead{Description}}
\startdata
\hline
\multicolumn{3}{c}{Quasar Parameters}\\\hline
qidx & int & The unique identifier of the mock quasar.\\
zq & float & The redshift of the mock quasar.\\
qra & float & The RA of the mock quasar, in degrees.\\
qdec & float & The Dec of the mock quasar, in degrees.\\
absMag & float & \makecell[tl]{The absolute magnitude at rest-frame 1450{\AA} $(M_{1450})$.} \\
QSOMags\tablenotemark{1} & list of floats & \makecell[tl]{The synthetic, unmagnified magnitudes of the mock quasar in LSST $ugrizy$,\\ UKIDSS $JHK$ and {\em WISE} W1, W2.}\\\hline
\multicolumn{3}{c}{Deflector Parameters}\\\hline
galaxy\_id & int & The ``galaxy\_id" field of the mock galaxy in the CosmoDC2 catalog.\\
zg &  float & The redshift of the mock galaxy.\\
gra & float & The RA of the mock galaxy, in degrees.\\
gdec & float & The Dec of the mock galaxy, in degrees.\\
rmaj & float & The semi-major axis of the galaxy, in arcsec.\\
rmin & float & The semi-minor axis of the galaxy, in arcsec.\\
pa\_lens & float & The position angle of the galaxy (following GLAFIC conversion).\\
stellar\_mass & float & The stellar mass in $M_\odot$.\\
sigma & float & The velocity dispersion, in km/s.\\
GalMags & list of floats & \makecell[tl]{The synthetic total magnitudes of the deflector galaxy in LSST $ugrizy$,\\ UKIDSS $JHK$ and {\em WISE} W1, W2.}\\
shear\_1, shear\_2 & float & The $x-$ and $y-$ components of the external shear.\\
convergence & float & The external convergence.\\\hline
\multicolumn{3}{c}{Lensing Configuration}\\\hline
Nimg & int & The number of lensed images.\\
image\_x\tablenotemark{2} & list of floats & the x-coordinates of the lensed images, in arcsec (following GLAFIC conversion).\\
image\_y & list of floats & the y-coordinates of the lensed images, in arcsec.\\
image\_mu & list of floats & the magnification of the lensed images.\\
image\_dt & list of floats & the time delay of the lensed images, in days.\\\hline
\enddata
\tablenotetext{1}{All magnitudes are noiseless.}
\tablenotetext{2}{In our conversion, the deflector galaxy locates at $x=0$ and $y=0$.}
\end{deluxetable*}

\section{Statistics of Lensed Quasars} \label{sec:statistics}

\begin{deluxetable*}{cccccccccc}
\tablenum{4}
\label{tbl:nlens}
\tablecaption{Predicted Number of Discoverable Lensed Quasars}
\tablewidth{0pt}
\tablehead{
\colhead{Survey} & \colhead{Area} & \colhead{Filter} & \colhead{$5\sigma$ Depth} & \colhead{FWHM} & \colhead{$N_\text{qso}$} & \colhead{$N_\text{lens}$} & \colhead{$N_\text{quad}$\tablenotemark{1}}\\
\colhead{} & \colhead{$(\text{deg}^2)$} & \colhead{} & \colhead{(mag)} & \colhead{$('')$} & \colhead{} & \colhead{} & \colhead{}}
\startdata
\hline
PS1\tablenotemark{2} & $3\times10^4$ & $i$ or $z$ & 23.1 or 22.3 & 1.11 or 1.07 & $4.8\times10^6$ & 928 & 132 \\
Legacy\tablenotemark{3} & $1.4\times10^4$ & $r$ or $z$ & 23.5 or 22.5 & 1.18 or 1.11 & $2.6\times10^6$ & 474 & 70\\
LSST\tablenotemark{4} & $2\times10^4$ & $i$ or $z$ & 26.8 or 26.1 & 0.71 or 0.69 & $7.4\times10^6$ & 2377 & 193\\
DES\tablenotemark{5} & $5\times10^3$ & $i$ or $z$ & 23.8 or 23.1 & 0.88 or 0.83 & $1.0\times10^6$ & 241 & 33\\
HSC\tablenotemark{6} & $1.4\times10^3$ & $i$ or $z$ & 26.4 or 25.5 & 0.56 or 0.63 & $5.1\times10^5$ & 165 & 14 \\\hline
\enddata
\tablenotetext{1}{The columns: $N_\text{qso}$ is the total number of quasars beyond the flux limit 
(estimated using the mock quasar sample), $N_\text{lens}$ is the predicted number of all discoverable lensed quasars,
and $N_\text{quad}$ is the predicted number of discoverable quad lenses. Magnitude limits are for point sources.}
\tablenotetext{2}{The Pan-STARRS1 (PS1) survey, adopted from \citet{ps1}}
\tablenotetext{3}{The DESI legacy imaging survey, adopted from \citet{legacy}}
\tablenotetext{4}{The LSST survey, adopted from \citet{lsst}}
\tablenotetext{5}{The Dark Energy Survey (DES), adopted from \citet{desdr2}}
\tablenotetext{6}{The Hyper Suprime-Cam Wide Survey (HSC), adopted from \citet{hscdr1} and \citet{hsc}}
\tablecomments{See Section \ref{sec:statistics} for the criteria of a discoverable lensed quasar.
We consider two filters for each survey, and a lensed quasar will be counted if it is discoverable in
either band.}
\end{deluxetable*}

\begin{figure*}
    \centering
    \includegraphics[width=0.9\textwidth]{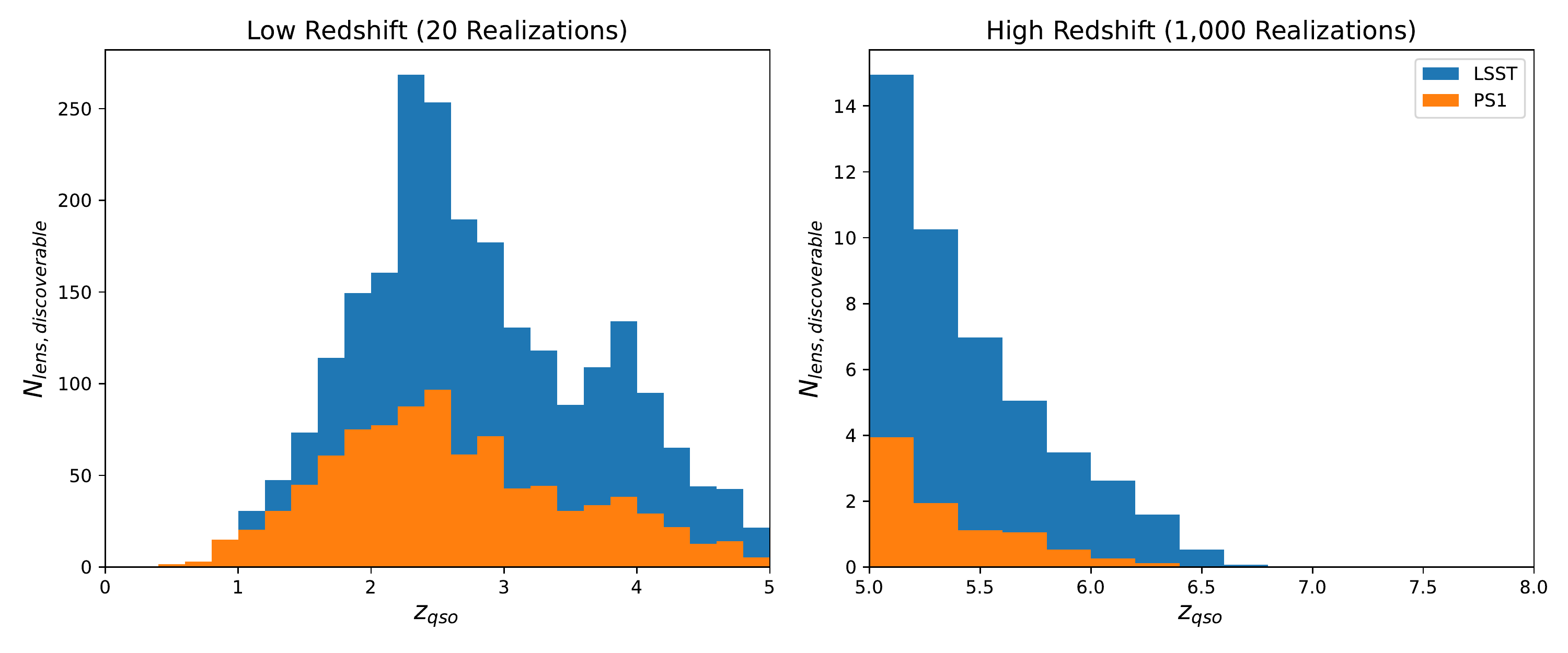}
    \caption{Predicted number of discoverable lensed quasars in LSST and PS1 survey, as a function of redshift.
    See Section \ref{sec:statistics} for the definition of a discoverable lensed quasar.
    {\em Left}: the low-redshift $(z_\text{qso}<5)$ sample. We run 50 realizations which is equivalent to 
    an all-sky area, and scale the predicted number of discoverable lenses according to the survey area 
    (listed in Table \ref{tbl:nlens}). 
    There are $\sim900$ discoverable lensed quasars in the PS1 survey, 
    and LSST will increase this number to about $2.4\times10^3$.
    Note that there is a bump at $z_\text{qso}\sim4$ in lensed quasar numbers, where we switch from the
    low-redshift QLF to the high-redshift one. 
    The faint-end values of the two QLFs are not fully consistent at this redshift.
    {\em Right:} the high-redshift $(z_\text{qso}>5)$ sample.
    We run 1,000 realizations to reduce the statistical error since the number density of high-redshift quasars 
    is low. The mock catalog suggests that only a handful of lensed quasars at $z_\text{qso}>5$ are discoverable
    in current imaging surveys like PS1, and there will be $\sim45$ discoverable in LSST.
    At $z\gtrsim6.5$, there are effectively no discoverable lensed quasars even for LSST.
    }
    \label{fig:nlens}
\end{figure*}

In this section we present the numbers of lensed quasars that are discoverable in various of
imaging surveys. 
Following  \citet{om10}, we define a lensed quasar to be {\em discoverable}
if: (1) it has a separation of $\Delta \theta>\frac{2}{3}\times\text{FWHM}$
and (2) has the second (for doubly-imaged lenses) or third (for quadruply-imaged lenses)
brightest lensed image detectable in a $5\sigma$ level. 
Here, the separation of a lensed quasar is defined as the largest separation between any pairs of the lensed images.
These criteria ensure that the lensing structure is at least marginally resolved.

Note that while lensed quasars that meet the critera above are in principle discoverable, in practice, their recovery faces a number of technical challenges. 
In real observations, the flux contribution of the lensing galaxy, low signal to noise ratios, 
and the presence of large number of both astrophysical and instrumental contaminants could result in
significant challenges and low success rate in lensed quasar selection, 
especially for the marginally discoverable cases. As a result, most of the lensed quasars
reported in recent surveys are bright ones with large separation \citep[e.g., ][]{lemon20,chan20}.
A main application of this catalog is to facilitate development of selection algorithm 
to enable effective recoveries of these discoverable systems.

In Table \ref{tbl:nlens}, we list the predictions for a number of current and next-generation imaging surveys. 
As specific examples, Figure \ref{fig:nlens} shows the redshift distribution
 of discoverable lensed quasars in the PS1 survey and the LSST survey.
 Our mock catalog suggests that current sky surveys such as PS1 can in principle find $\sim900$ of lensed quasars,
 and LSST will increase this number to $\sim2.4\times10^3$.
 Most of the lensed quasars have redshift $1.5<z_\text{qso}<3.5$, which is consistent with the statistics of the current sample
 of discovered lensed quasars.
Due to the decline of the quasar number density, the numbers of discoverable lensed quasars 
drop quickly at $z_\text{qso}\gtrsim4$. Figure \ref{fig:nlens} suggests that only a handful
lensed quasars at $z_\text{qso}>5$ are discoverable in imaging surveys such as PS1, and
 there will be $\sim45$ discoverable in LSST.
There are effectively no discoverable lensed quasars at $z>7$ even in the LSST survey.
We will discuss in more details the lensed fraction of $z\gtrsim6$ quasars using both the mock catalog 
and analytical models in a subsequent paper \citep{yue22}.

It is worth noticing that, even though LSST reaches more than three magnitude deeper than PS1, 
the number of total quasars and discoverable lensed quasars are only 1.5 and 2.6 times higher. 
This is due to the fact that we only include quasars with $M_i<-20$ in the catalog 
so that the AGNs are not host-dominated.
LSST reaches significantly fainter flux level than this limit especially at low redshift.

\begin{figure}
    \centering
    \includegraphics[width=0.98\columnwidth]{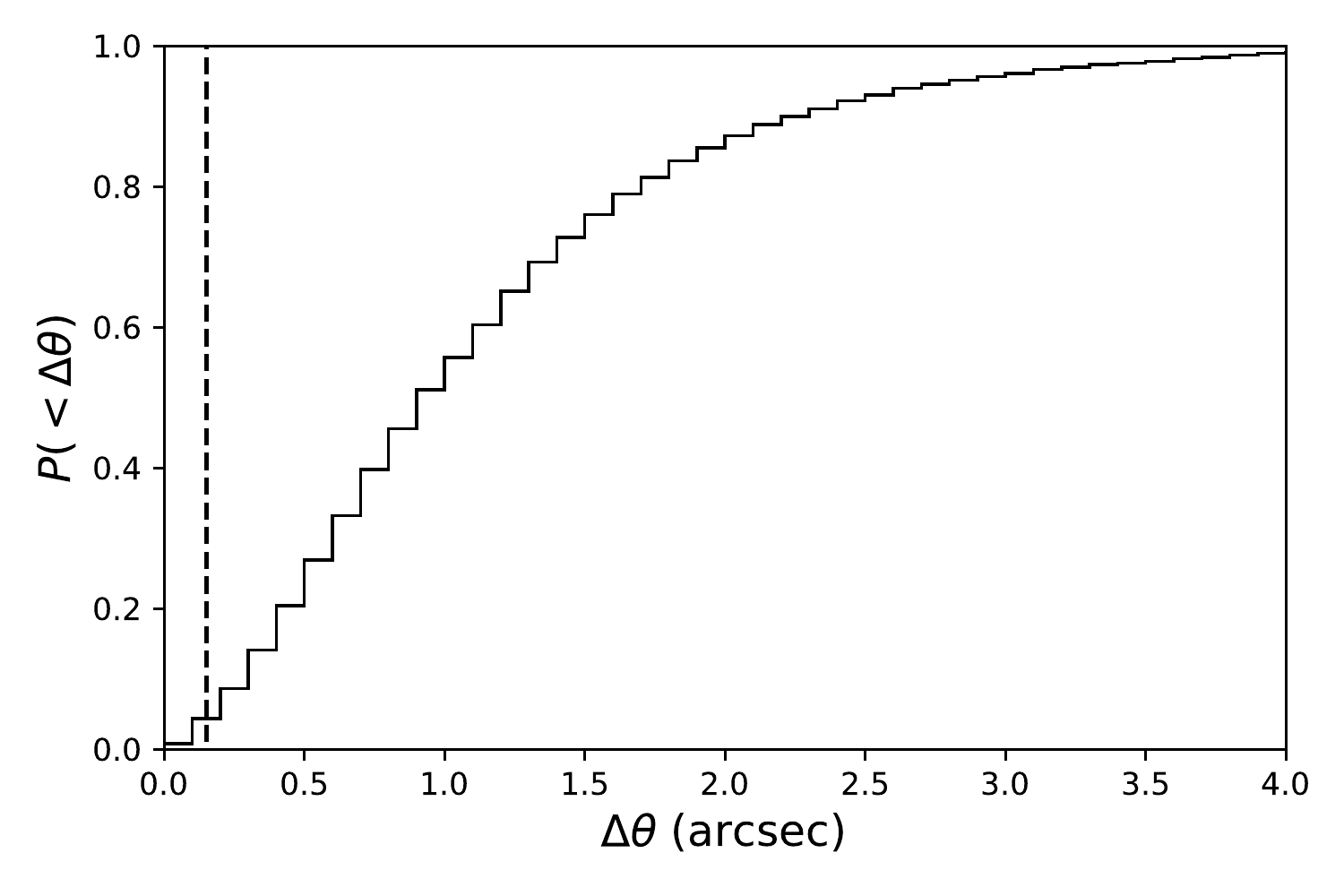}
    \caption{The distribution of the image separations of lensed quasars.
    The dashed line marks the limit above which the mock catalog is complete. About $\sim50\%$ of the lensed quasars
    have separation $\Delta \theta >1''$, and $\sim80\%$ of them have $\Delta \theta >0\farcs5$.}
    \label{fig:separation}
\end{figure}

Figure \ref{fig:separation} shows the distribution of lensing separations of the mock lensed quasars.
About $\sim80\% ~(50\%)$ of these lenses have separation larger than $0\farcs5 ~(1\farcs0)$. 
These numbers are consistent with the results in, e.g., \citet{hilbert08} and \citet{collett15}.
 A ground-based imaging survey with a PSF size of $\sim0\farcs7$ will be sufficient
to make most of the lenses marginally resolved, which suggests that
the bottleneck of lensed quasar surveys is the image depth rather than the PSF size.

We note that the predicted number of lensed quasars in this work is 
about two to three times lower 
than that of OM10, especially for deep surveys like HSC and LSST. 
The main reason is the difference in the simulated quasar sample; OM10 adopt
a steeper QLF faint-end slope, and more importantly, they do not apply a cut in the absolute magnitude,
while we only include quasars with $M_{i}<-20$. In addition, OM10 use a deflector VDF without redshift evolution,
in contrast to the CosmoDC2 VDF which declines with redshift. In Section \ref{subsec:om10} we discuss in more details the 
difference between our mock catalog and that of OM10, where we show that the number of lensed quasars are consistent
after taking the difference in QLFs and VDFs into account. 

\begin{figure*}
    \centering
    \includegraphics[width=0.32\textwidth]{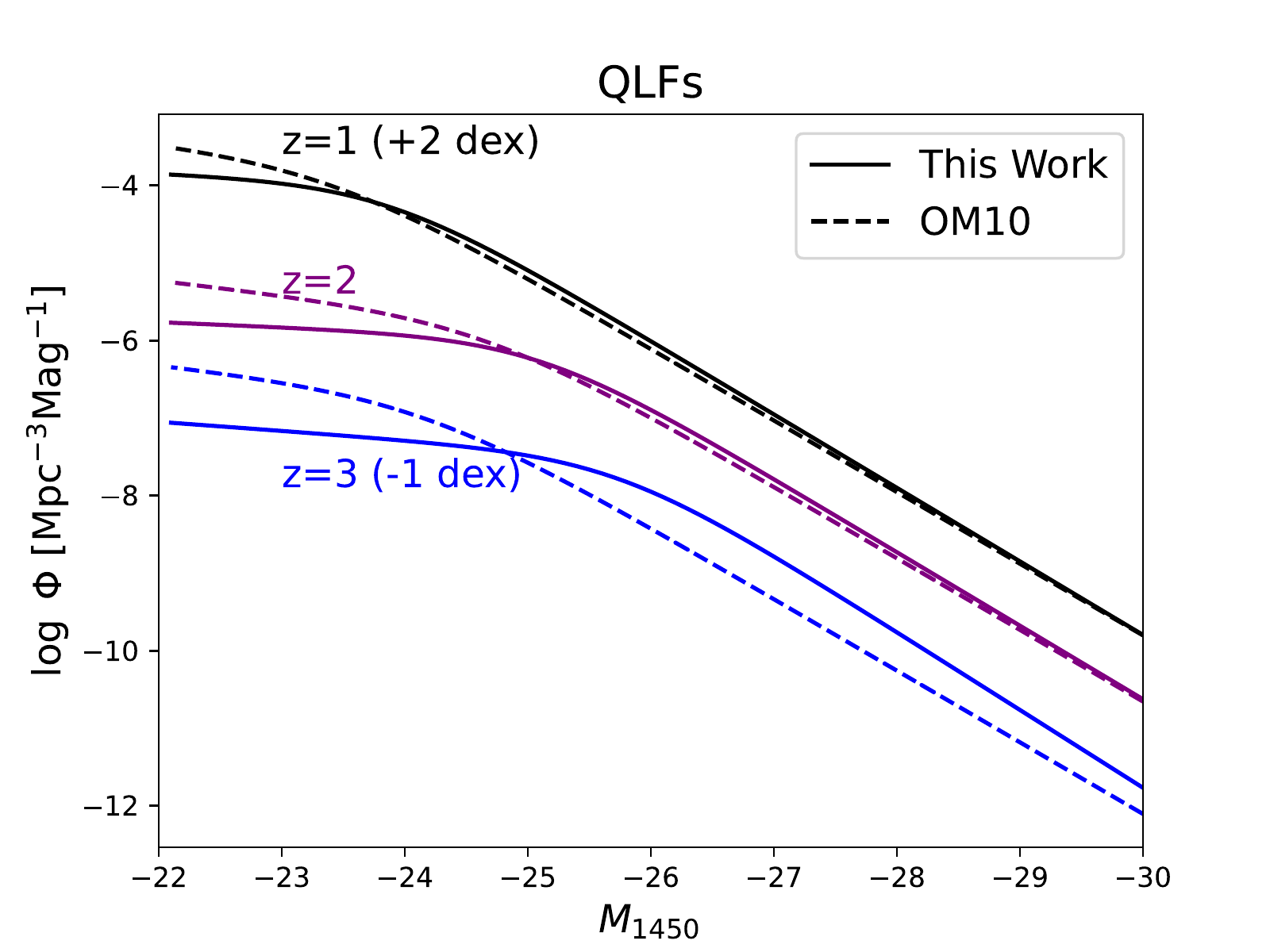}
    \includegraphics[width=0.32\textwidth]{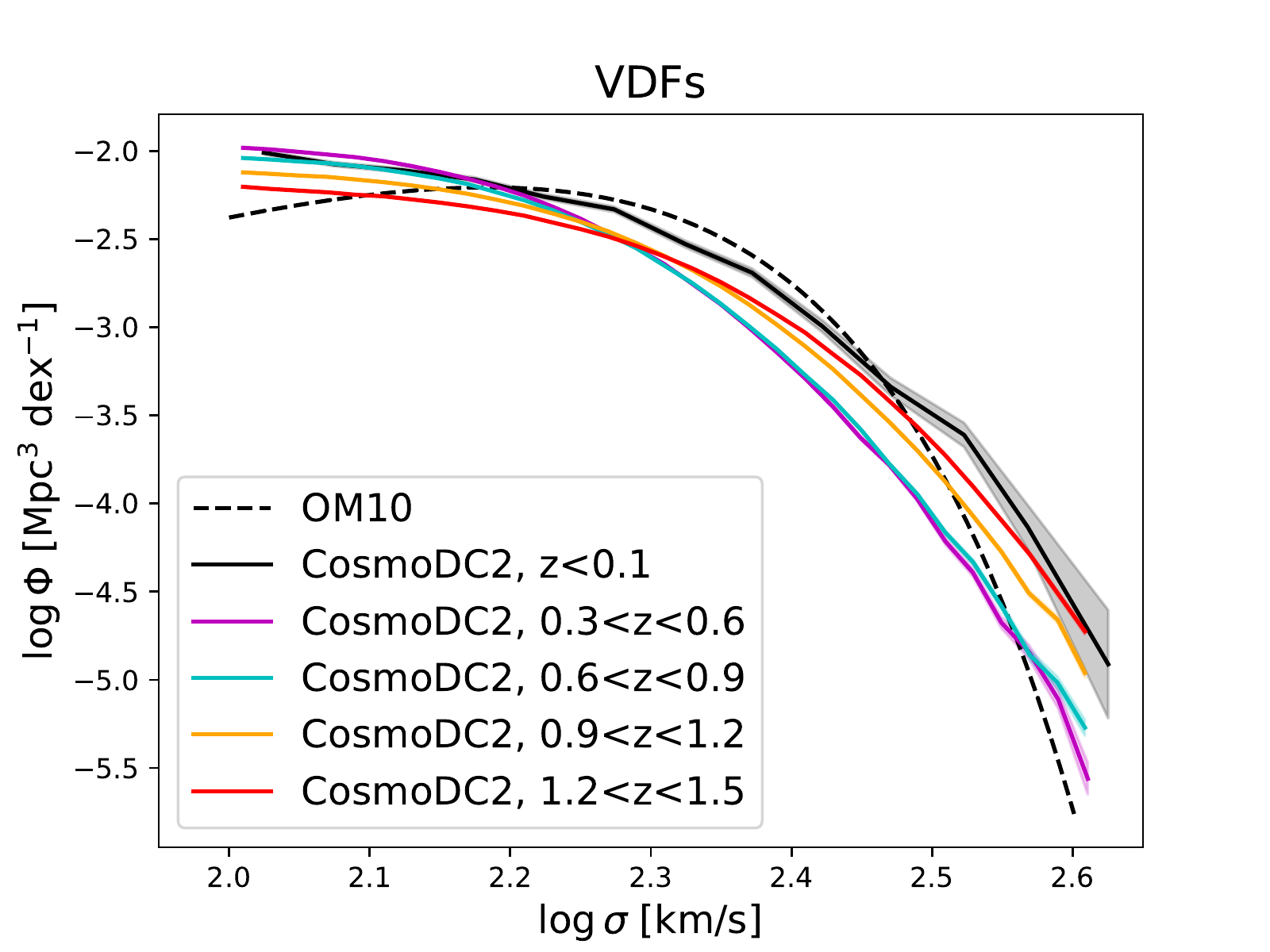}
    \includegraphics[width=0.32\textwidth]{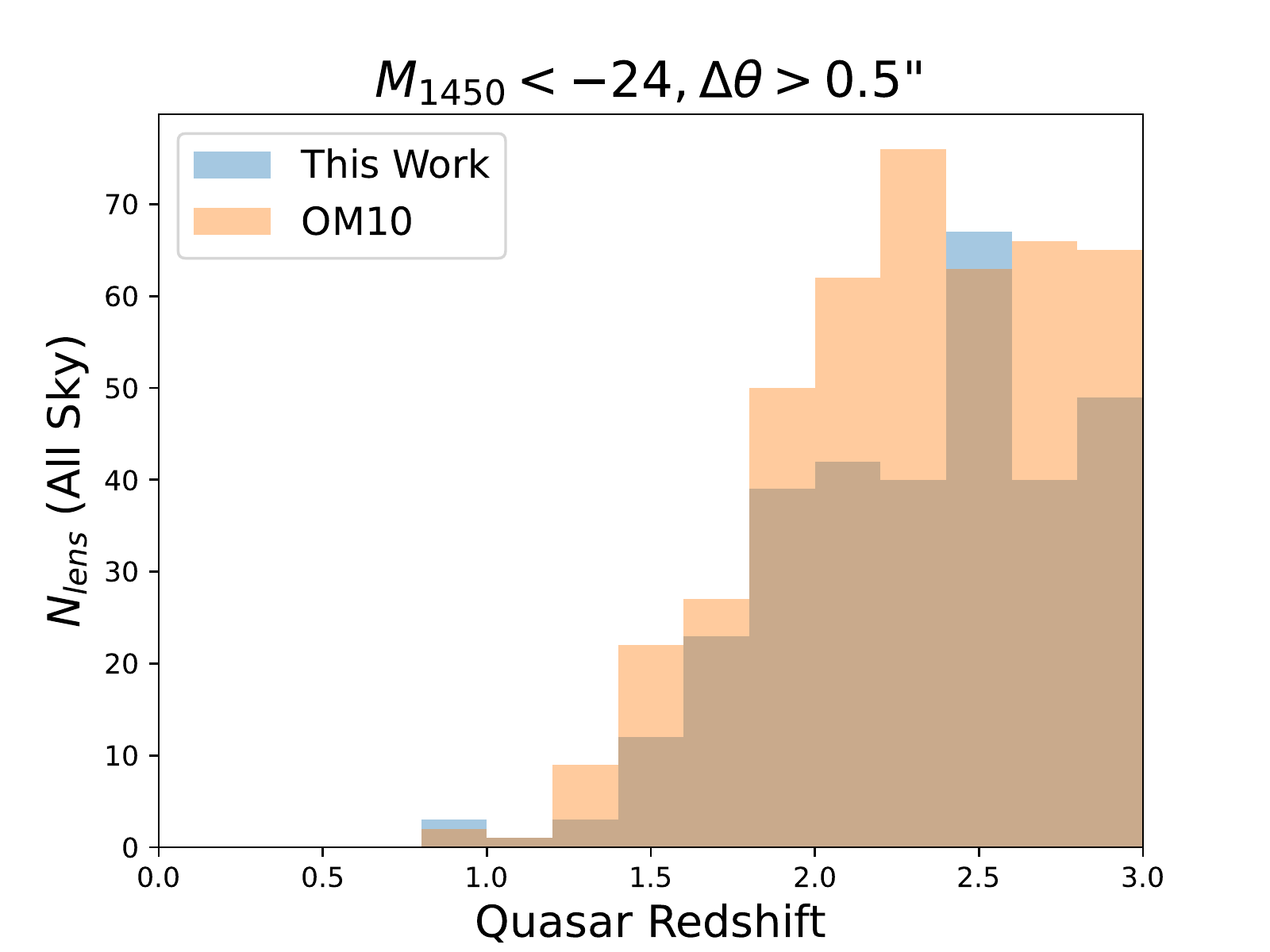}
    \caption{Comparison between this work and OM10. 
    {\em Left}: the QLF adopted in this work and OM10. The two QLFs are nearly identical for quasars 
    at $z\lesssim2.5$ with $M_{1450}<-24$. {\em Middle}: The VDF adopted in this work and OM10. 
    We show the CosmoDC2 VDF at $z<1.5$ which covers the majority of deflector galaxies for $z_\text{qso}\lesssim2.5$.
    OM10 uses the local VDF in \citet{choi07} and assumes no redshift evolution. 
    The local VDF of the CosmoDC2 catalog is close to the one used by OM10,
    and the OM10 VDF is slightly higher than the CosmoDC2 VDF at higher redshift.
    {\em Right}: the predicted number of lensed quasars at $z<3$.
     We only select lensed quasars with $\Delta \theta>0\farcs5$ and $M_{1450}<-24$,
     at which the QLFs used by OM10 and this work are close to each other. 
     Our mock catalog predicts slightly smaller numbers of lensed quasars than OM10 when only restricted to these parameter ranges,
     and the different can be fully explained by the adopted QLFs and VDFs (see text for details).
    }
    \label{fig:om10}
\end{figure*}

\subsection{Comparison with the OM10 Mock Catalog} \label{subsec:om10}

The mock catalog in OM10 and this work have a number of significant difference:
(1) The two mock catalogs adopt different QLFs and VDFs. (2) We only include quasars with $M_{i}<-20$,
while OM10 extend the quasar sample to as faint as $M_{i}\sim-18$.
And (3) OM10 only include lensing systems with separation $\Delta \theta>0\farcs5$, while our mock catalog have more compact lenses
down to $\Delta \theta\sim0\farcs1$.

To perform meaningful comparisons, we select subsamples of the two mock catalogs
that have $M_{1450}<-24$, $z_\text{qso}<3$ and $\Delta \theta>0\farcs5$. In this redshift and magnitude range,
the QLFs used in OM10 and this work are close to each other (Figure \ref{fig:om10}, left panel).
We also compare the VDFs adopted in OM10 and this work in the middle panel of Figure \ref{fig:om10}.
In short, OM10 adopt the local VDF from \citet{choi07} and assume no redshift evolution.
This VDF is close to the CosmoDC2 catalog VDF at $z<0.1$, but it predicts higher number density 
than the CosmoDC2 VDF beyond the local universe.
These differences can fully account for the predicted numbers of lensed quasars, 
shown in the right panel of Figure \ref{fig:om10}.
After applying the luminosity, redshift and lensing separation cuts mentioned above,
our mock catalog gives a slightly smaller number of lensed quasars compared to OM10. 
The main difference between the two mock catalogs appears at $z_\text{qso}\sim2.2$, 
where the OM10 catalog has $\sim30\%$ more lenses than this work.
This can be explained by the difference in the VDFs at $z_d\gtrsim0.3$, where the OM10 VDF is $\sim0.3$ dex higher than
the CosmoDC2 VDF.

The comparison suggests that our method of generating mock catalog is reliable,
and illustrates the importance of using accurate and updated QLFs and VDFs.
Another major difference between OM10 and this work is that we apply an absolute magnitude cut of $M_{i}<-20$ for quasars.
This cut is used in the LSST science book, and the main motivation is that fainter AGNs 
have distinct observational features compared to normal type-I quasars. For these objects,
host galaxy emissions start to dominate the SEDs and make their morphology extended.
Modeling the host galaxy emission is out of the scope of this work, and we expect that these faint AGNs require a
very different candidate selection strategy. 
%Nonetheless, modeling and searching for these faint strongly-lensed AGN
%is still important, as they enable many important studies that are similar to brighter lensed quasars.

AGN variability offers a possible way of 
 finding host-dominated lensed AGNs \citep[e.g.,][]{kochanek06}.
OM10 presented the number of lensed quasars that are brighter than 
the 10$\sigma$ limits of yearly-stacked images for each sky survey,
which can be identified via their variability.
Variability analysis will be especially useful in the LSST survey,
providing a critical supplement to image-based candidate selection methods.

%\textcolor{red}{It is also worth noticing that the survey depths adopted by
%OM10 are the 10$\sigma$ limits of yearly-stacked images for each sky survey. 
%OM10 suggests that lensed quasars beyond this flux limit can be identified via variability.
%Variability analysis can be especially useful in surveys for
%faint lensed AGNs where the host galaxy emission is prominent.}

\subsection{Quadruply-Lensed Quasars and Time Delays}

One important application of lensed quasars is to measure key cosmological parameters, especially the Hubble constant $H_0$.
This relies on measuring the time delays between the lensed images of the background quasar.
Compared to doubly-imaged ones, quadruply lensed quasars are preferred since they provide more time delays
and give stronger constraints.

Table \ref{tbl:nlens} gives the number of discoverable quadruply-lensed quasars in various sky surveys.
Our mock catalog suggests that there are $\sim200$ quadruply-lensed quasars discoverable in LSST.
The impact of survey depth on the number of discoverable lenses
is less significant for quads compared to doubly-lensed quasars,
which is a result of magnification bias and the criteria of discoverable lenses (Section \ref{sec:statistics}).
In general, the fainter image in doubly-lensed systems has a low magnification and is often de-magnified
$(\overline{\mu}=0.6)$, 
while the third brightest images in quad lenses have significantly higher magnifications $(\overline{\mu}=2.6)$.
Given the luminosity limit of our quasar sample $(M_{i}<-20)$,
most quadruply-lensed quasars are brighter than the LSST survey limit.
Nonetheless, the image quality of LSST is still critical to the search of these bright quads,
since the better depth and spatial resolution are essential for a high completeness and efficiency
in the candidate selection.

\begin{figure*}
    \centering
    \includegraphics[width=0.8\linewidth]{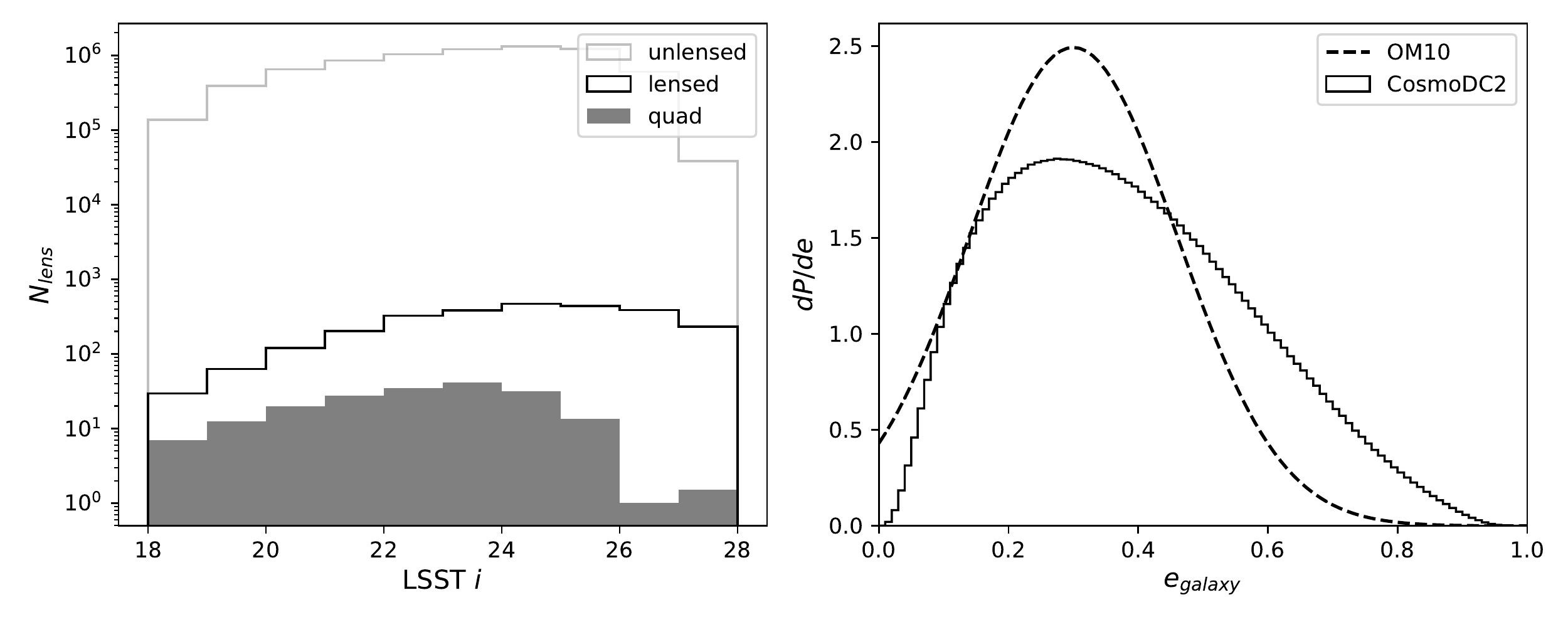}
    \caption{{\em Left:} The number of discoverable quadruply-lensed quasars as a function of survey depth. 
    We also include doubly-lensed quasars and unlensed quasars for comparison. 
    We use the LSST survey area and seeing condition when computing the numbers.
    The magnitudes correspond to the second brightest image for doubly lensed quasars,
    and the third brightest image for quad lenses.
    The fraction of quadruply lensed quasars among all lenses is $\sim10\%-15\%$ at $m_i\lesssim25$.
    This fraction drops quickly at $m_i\gtrsim26$, which is a result of the absolute magnitude cutoff
    $(M_{i}<-20)$.
    {\em Right}: the probabilistic distribution of the deflector ellipticity,
    $e_\text{galaxy}$, for the CosmoDC2 mock galaxies and the one adopted by OM10. 
    The two distributions are generally similar,  with some difference at the low- and high-ellipticity end.}
    \label{fig:quad}
\end{figure*}

Figure \ref{fig:quad} further illustrates the number of discoverable quaduply lensed quasars as a function of survey depth.
We assume an LSST-like PSF size of $0\farcs7$ when counting the number of discoverable lenses,
and normalize the numbers to the area of LSST.
We also include the number of doubly-lensed quasars and unlensed quasars for comparison.
%The number of lensed quasars is about $3.5$ dex fewer than unlensed ones.
Note that the magnitudes in Figure \ref{fig:quad} correspond to the second brightest image for doubly-lensed quasars
and the third brightest image for quadruply-lensed ones, in accordance with the definition of discoverable lensed quasars.
At $m_i\lesssim25$, about $\sim10\%-15\%$ of the discoverable lensed quasars are quads.
The magnification bias makes the fraction of quads decreases with survey depth, in agreement with the result in OM10.
The number of quad lenses drops quickly at $m_i\gtrsim26$, which is a result of the absolute magnitude cutoff.

The probability for a deflector to generate a quadruple lens increases with its ellipticity.
The right panel of Figure \ref{fig:quad} illustrates the distribution of the the deflector ellipticity,
$e_\text{galaxy}$, for the CosmoDC2 mock catalog and the one adopted by OM10.
OM10 assumes a Gaussian distribution of ellipticity for all the galaxies,
while CosmoDC2 use a more complicated approach, where the probabilistic distribution of $e_\text{galaxy}$
depends on the bulge-to-disk ratio and the galaxy luminosity. 
The overall ellipticity distribution of OM10 and CosmoDC2 are similar;
correspondingly, the predicted quad fractions at $m_i\lesssim26$ of this work and OM10
are consistent within statistical errors.

The external shears is another key factor in generating quadruply-lensed systems. 
\citet{luhtaru21} analyzed 39 quadruply-lensed quasars, 15 of which were found to be ``shear-dominated".
This result illustrates the significant contribution of external shears in making quad lenses.
OM10 assumes a log-normal distribution of the total external shear, $\gamma$, 
with a mean of 0.05 and a deviation of 0.2 dex.
In comparison, we adopt a redshift-evolving distribution for the external shear.
At most redshifts $(z\lesssim6)$, 
the shear distribution in this work has a lower mean value and a more prominent tail
at large shears. 
We thus expect that the quad lenses in our mock catalog and OM10 have different properties
(e.g., lensing configurations).
It might be useful to keep in mind that
the external convergence and shear distributions used in this work are derived from ray-tracing simulations,
which are subject to a number of systematic uncertainties (Section \ref{subsec:deflector}).

%\citet{luhtaru21} argues that the external shear in the OM10 mock catalog might be underestimated,
%emphasizing the importance of accurate models for external shears.
%It might be useful to keep in mind that
%the external convergence and shear distributions used in this work are from ray-tracing simulations,
%which are subject to a number of systematic uncertainties (Section \ref{subsec:deflector}).}

\begin{figure*}
    \centering
    \includegraphics[width=0.99\linewidth]{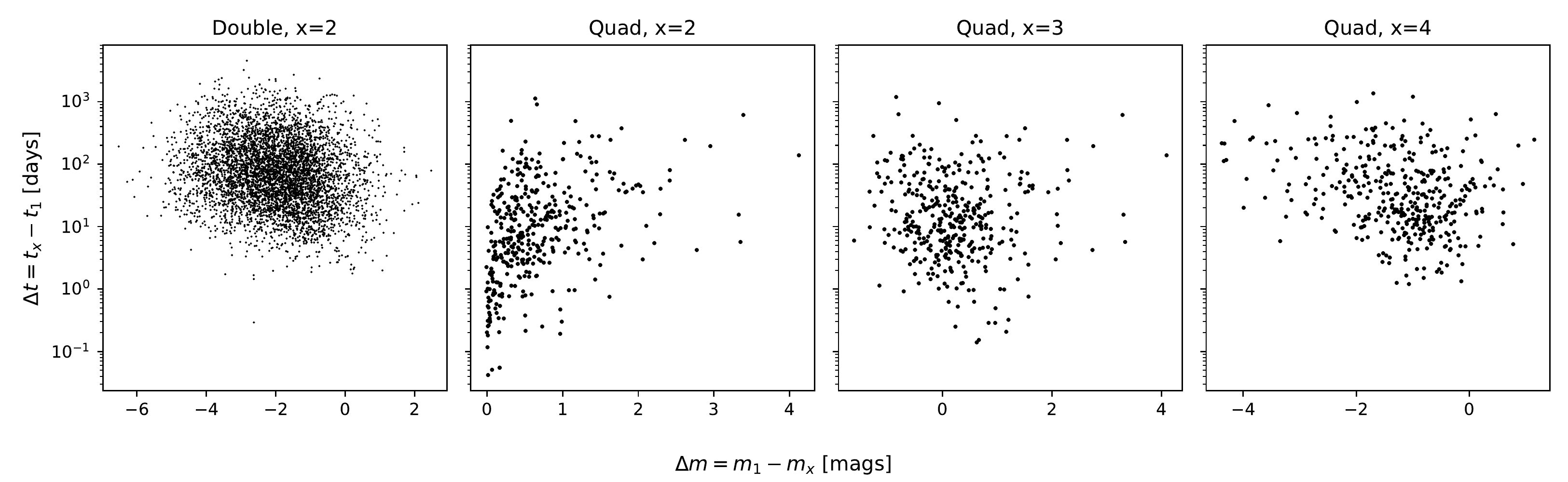}
    \caption{The magnitude difference and the time delays of the mock lensed quasars. 
    The subscripts represent the order of arrival, e.g., $m_1$ is the magnitude of the image that arrives first.
    Most lensed quasars have time lags $\sim10^1-10^3$ days. For doubly-imaged quasars, 
    the brighter image usually arrives earlier than the fainter one. For quad lenses, the first-arriving image
    is the second or third brightest in most cases.}
    \label{fig:dmdt}
\end{figure*}

Figure \ref{fig:dmdt} illustrates the distribution of time delays and the magnitude differences between lensed images.
The typical time delays are $10^1-10^3$ days. 
In most of the doubly-imaged lenses, the brighter image arrives earlier than the fainter one, while 
for quad lenses, the second or third brightest images arrives first in most cases.
These results are consistent with the findings in OM10.
For a typical time delay of $\sim10^2$ days, a cadence of $\sim5$ days
gives a good measurement of the light curves \citep{suyu17}. The cadence of LSST survey is not sufficient to provide 
accurate light curves for most of the lensed quasars. As such, LSST will mainly serve as a deep imaging survey for
discoveries, and follow-up monitoring is needed to measure the time delays of lensed quasars.

\subsection{Statistics of Deflector Galaxies}

\begin{figure}
    \centering
    \includegraphics[width=0.98\columnwidth]{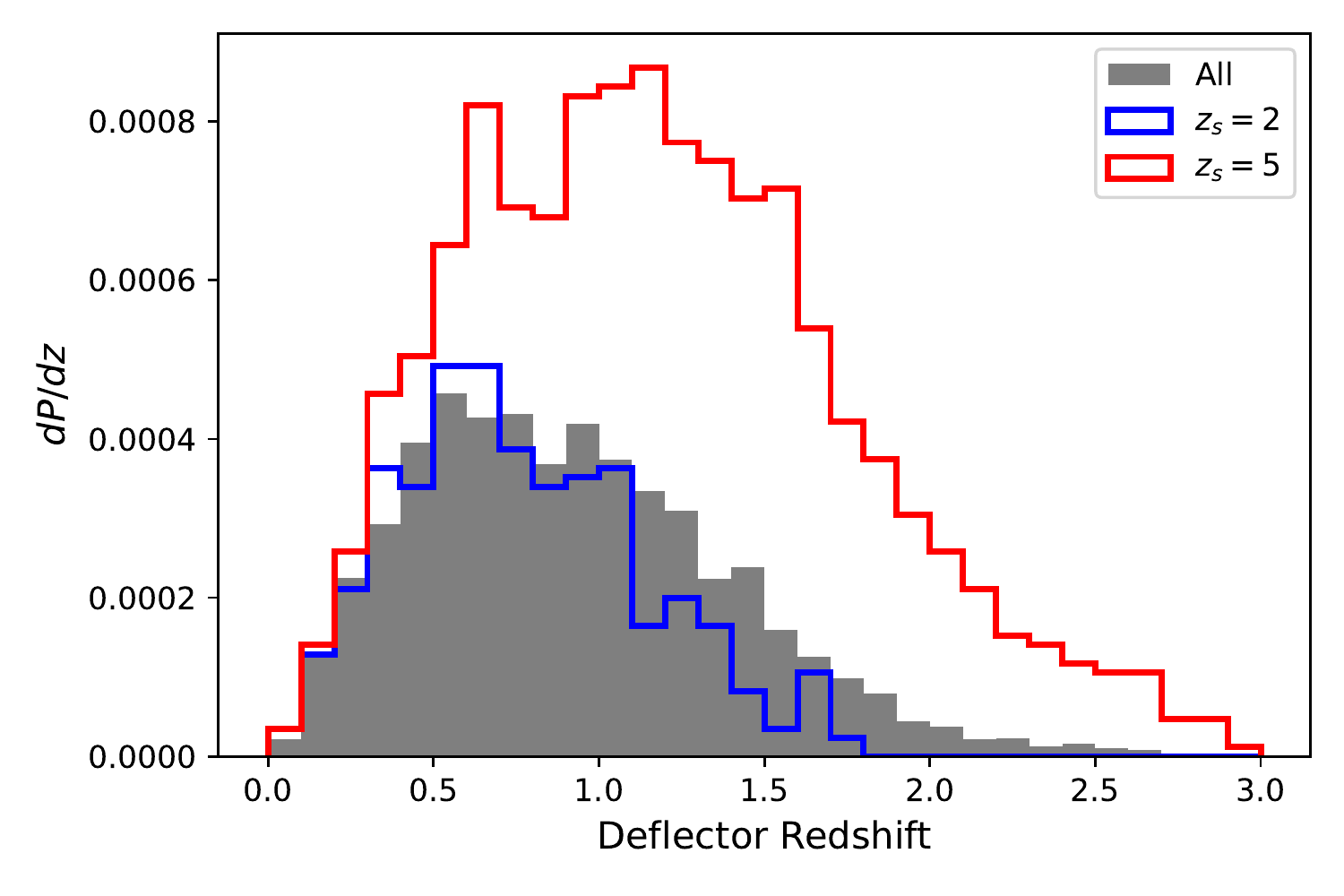}
    \caption{The distribution of deflector redshifts. The $y-$axis denotes the probability for a background source 
    to be lensed by a foreground deflector at a certain redshift. 
    The gray histogram represents all the simulated lensed quasars, which peaks at $\sim0.7$ and has a median of 0.87.
    The distribution drops to a negligible level at $z_d>2.5$. We also include two samples of background sources at $z_s=2$
    and $z_s=5$ for comparison.
    }
    \label{fig:zd}
\end{figure}

Figure \ref{fig:zd} shows the distribution of the deflector  redshift, $z_d$, of simulated lensed quasars.
The distribution peaks at $z_d\sim0.7$ and drops to a negligible level at $z_d\gtrsim2.5$. 
This picture is consistent with OM10 and previous modeling of deflector galaxy population \citep[e.g.,][]{hilbert08,wyithe11}.
For comparison, we also include two samples of sources at fixed redshift, $z_s=2$ and $z_s=5$.
Although CosmoDC2 only include galaxies at $z<3$, Figure \ref{fig:zd} indicates that only a small fraction
of lensed quasars have deflector  redshift $z_d>3$.

\begin{figure*}
    \centering
    \includegraphics[width=0.8\linewidth]{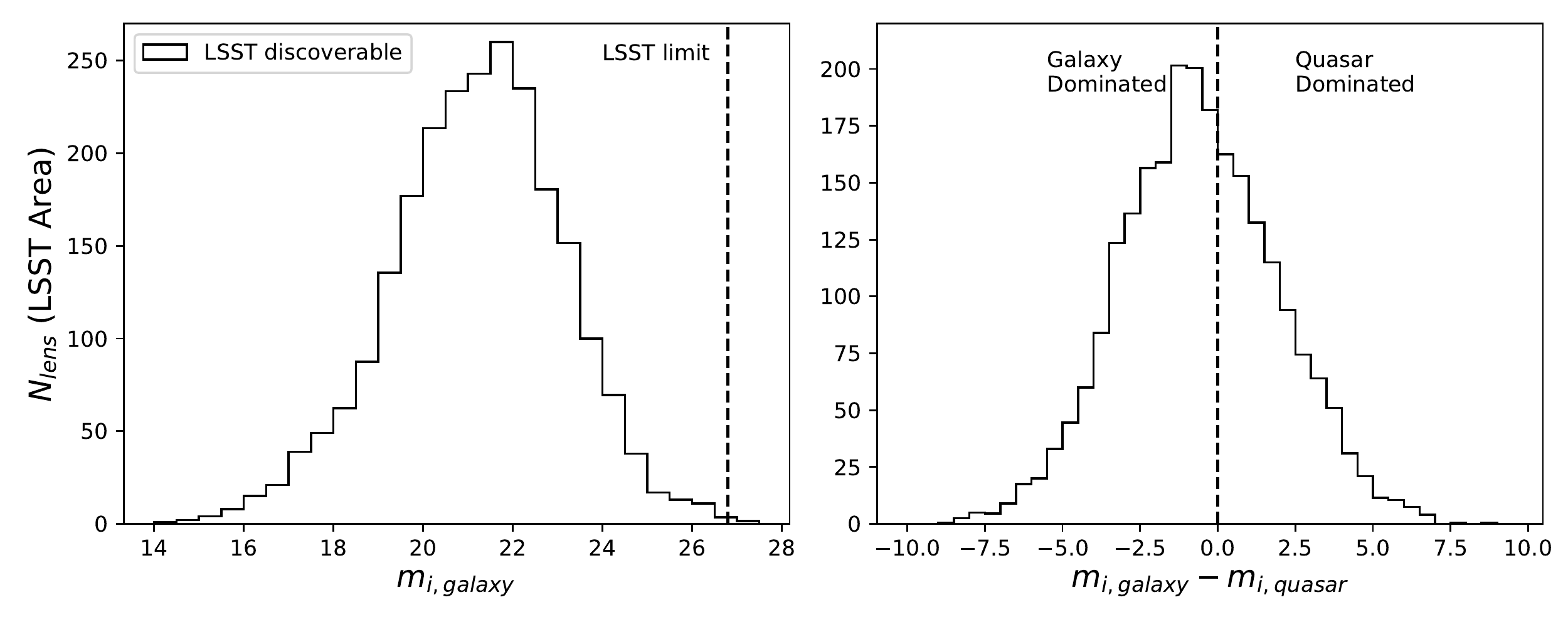}
    \caption{{\em Left:} The $i-$band magnitudes of the deflector galaxies of LSST-discoverable lensed quasars.
    Nearly all of them are beyond the LSST magnitude limit. The impact of deflector galaxy flux must be considered
    when selecting lensed quasar candidates.
    {\em Right:} the difference between the magnitude of the deflector galaxy and the magnified magnitude of the quasar. 
    The quasar magnitude corresponds to the total flux of all the 
    lensed images of the background quasar.
    The magnitude difference spreads over a wide range, which suggests a large diversity in observed features
    for lensed quasars.}
    \label{fig:deflectorflux}
\end{figure*}

In Figure \ref{fig:deflectorflux}, we present the $i-$band magnitudes of the deflectors 
and the deflector-to-quasar magnitude difference of the simulated lensed quasars.
In nearly all of the LSST-discoverable lensed quasars, the deflector galaxy is brighter than the LSST limit,
which suggests that the impact of the foreground galaxy must be considered when selecting lensed quasar candidates. The deflector-to-quasar magnitude difference spreads over a wide range, 
and the number of quasar-flux-dominated lenses are comparable to that of galaxy-flux-dominated ones.
Correspondingly, we expect a large diversity in observed features for lensed quasars.
The galaxy-dominated lenses are likely generated by a massive, luminous deflector and have large image separations,
and the quasar-dominated ones usually have compact lensing structures.
These objects may require very different candidate selection techniques.

\section{Discussions} \label{sec:discussion}

\subsection{Choices of QLFs and VDFs} \label{subsec:systematics}

Section \ref{subsec:om10} shows the impact of QLFs and VDFs on the modeling of the
lensed quasar population. In this section, we discuss the systematic uncertainties introduced by
our choices of QLFs and VDFs.

At $z\lesssim4$, the QLFs are well-measured down to at least $M_{1450}\sim-22$ \citep[e.g.,][]{richards06,dr9,pd16}, and
these measurements are in good agreement with each other.
Since most LSST-discoverable lensed quasars have $z_\text{qso}<4$,
the QLFs adopted by this study provide a good description of the quasar population.
Meanwhile, there are still debates about the QLFs at $z\gtrsim5$. 
A few studies give a faint-end slope of
$\alpha\sim-2$ \citep[e.g.,][]{jiang16,yang16,McGreer18}, while others suggest $\alpha\sim-1.3$
\citep{niida20,kim20}.
In this work, we use the HSC-based high-redshift QLFs which have $\alpha\sim-1.3$. If the faint-end slope
turns out to be steeper, there might be more faint lensed quasars at high redshift.

Unlike the QLFs, the deflector VDFs are not well constrained by spectroscopic surveys beyond the local universe.
Many previous studies assumes no evolution of the VDF \citep[e.g., OM10;][]{wyithe02}.
However, recent analysis of strong lensing systems indicate that the VDF drops towards high redshift \citep[e.g.,][]{geng21}, which is consistent 
with  photometric measurements \citep[e.g.,][]{bezanson11, bezanson12} and the CosmoDC2 VDFs used in this work.
It is thus critical to use redshift-evolving VDF models when modeling the lens population.

In addition, we consider the uncertainties introduced by the scaling relation used to calculate the velocity dispersion
(Equation \ref{eq:virial}). Many studies suggest that the $M_\text{dyn}/M_*$ ratio evolves with galaxy masses
\citep[e.g.,][]{auger10,nn16} and redshift \citep[e.g.,][]{beifiori14}. 
We thus consider the impact of an evolving $M_\text{dyn}/M_*$ ratio in the following form:
\begin{equation}
    \log\frac{M_\text{dyn}}{M_*} = a\log(M_*) + b\log(1+z) + c
\end{equation}
This is equivalent to an evolution of $\sigma \propto M_*^{a/2} (1+z)^{b/2}$.

\begin{figure*}
    \centering
    \includegraphics[width=0.45\textwidth]{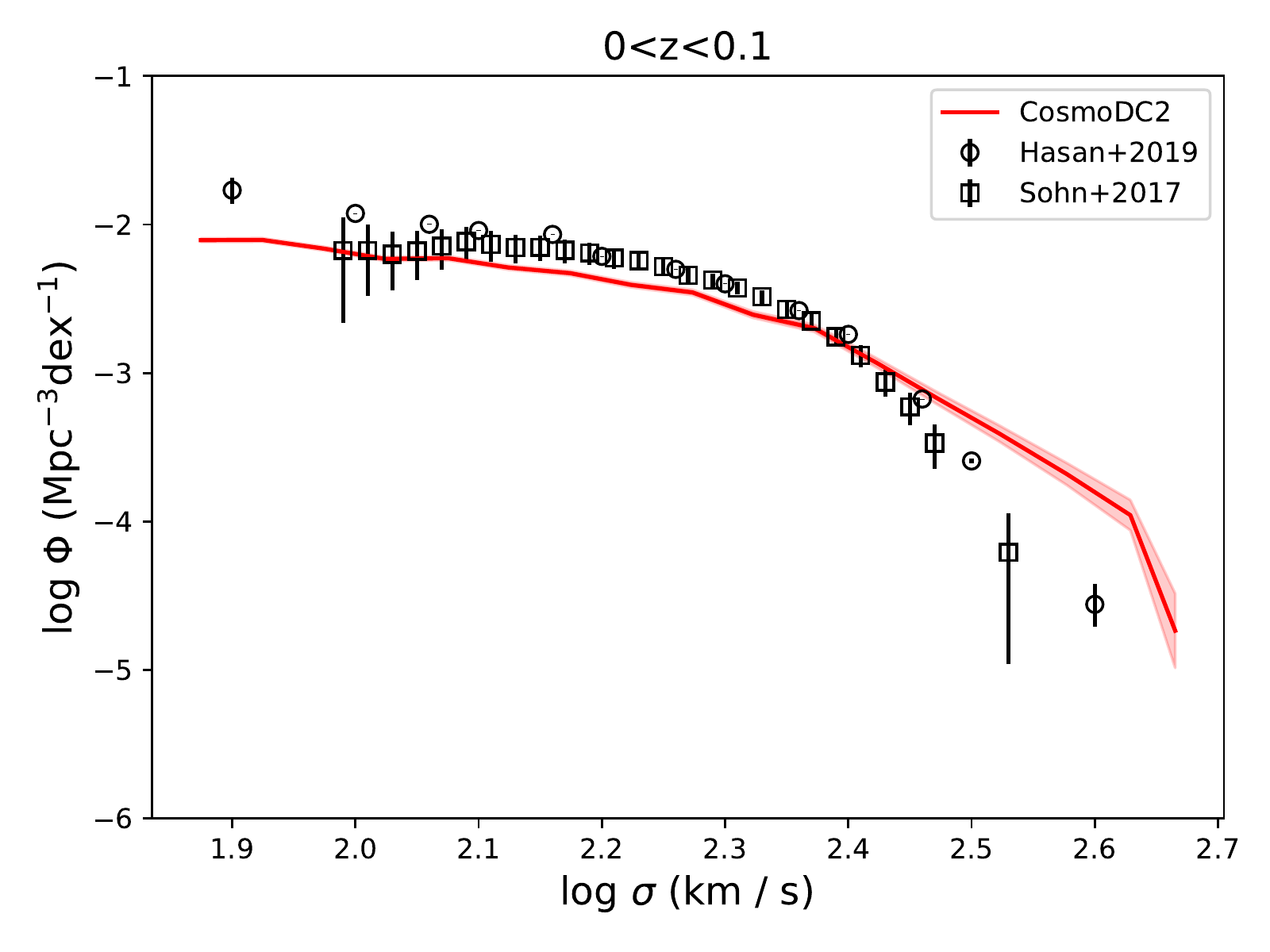}
    \includegraphics[width=0.45\textwidth]{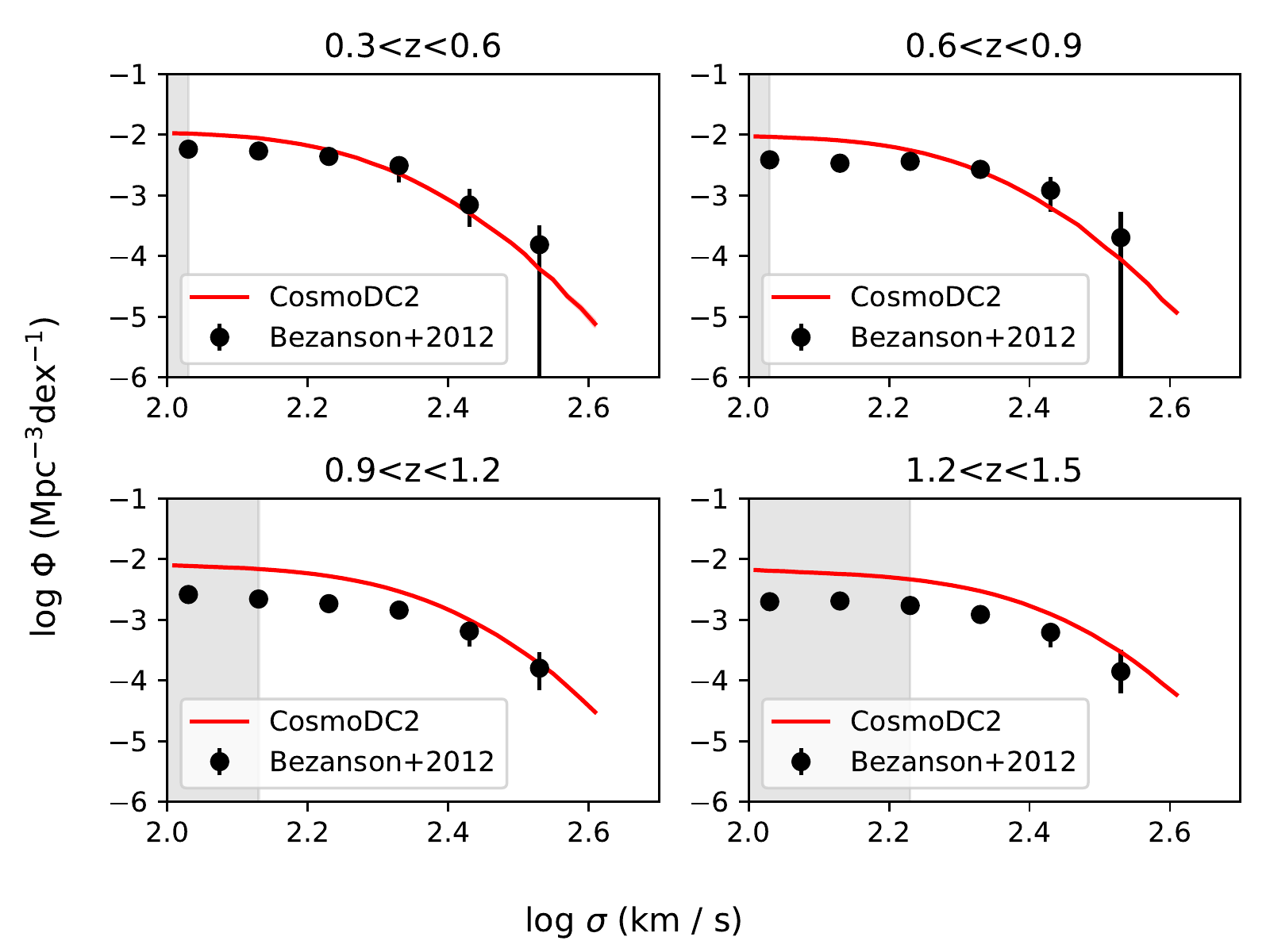}    
    \caption{{\em Left}: Testing the mass-dependence term in the velocity dispersion.
    We use the $M_*-M_\text{dyn}$ relation in \citet{auger10} to calculate the velocity dispersion of CosmoDC2 mock galaxies.
    The resulting local VDF does not match the observations.
    {\em Right:} Testing the redshift-evolution term in the velocity dispersion. We add the redshift-evolution
    from \citet{beifiori14} to the velocity dispersion inferred by Equation \ref{eq:virial} for mock galaxies.
    The redshift evolution is subtle, and the inferred VDF is consistent with observation at $z<1$. However, at $z>1$,
    the redshift evolution term starts to overpredict the number of galaxies at the massive end.
    }
    \label{fig:vdftest}
\end{figure*}

We first set $z=0$ to check the mass-dependence. We adopt the relation
in \citet{auger10}, which suggests $\log (M_* / M_\odot) = 0.885\log(M_\text{dyn}/M_\odot) + 0.905$.
The left panel of Figure \ref{fig:vdftest} shows the derived VDF at $z<0.1$, which does not reproduce the observed local VDF. 
We thus conclude that adding the mass-dependent term does not gives a good description
of the galaxy velocity dispersion for the CosmoDC2 catalog. 

We then keep the relation $M_*/M_\text{dyn}=0.557$, and adopt the redshift evolution from
\citet{beifiori14}, which gives $\sigma\propto(1+z)^{0.12}$.
The right panel of Figure \ref{fig:vdftest} compares the resulted CosmoDC2 VDF and the observed values in \citet{bezanson12}.
At $z<1$, the redshift evolution makes little difference compare to the original one,
and the resulting VDF is still consistent with observation. At $1.2<z<1.5$, the redshift-evolving estimation
starts to over-predict the number of galaxies at the massive end, where the observed galaxy sample should be complete.
This difference might be more significant at $z>1.5$ where the redshift evolution term gets larger.

In addition to the redshift evolution of the velocity dispersion, \citet{beifiori14} also
reports a redshift evolution of the $M_\text{dyn}/M_*$ ratio, which follows
$M_\text{dyn}/M_*\propto(1+z)^{-0.30\pm0.12}$. By including both trends
into Equation \ref{eq:virial}, we get a very subtle evolution of $\sigma \propto(1+z)^{-0.03}$.
This result might explain why a non-evolving $M_\text{dyn}/M_*$ ratio correctly reproduces the observed VDF.
In any case, the redshift dependence is subtle. 
Given the median redshift of deflectors ($z_d\sim0.7$, Figure \ref{fig:zd}),
the evolution will not strongly influence the resulting lensing statistics.

\subsection{Mock Catalog Completeness}

The CosmoDC2 catalog models dark matter halos down to $10^{9.8}M_\odot$. 
According to the $M_\text{galaxy}-M_\text{halo}$ relation \citep[e.g.,][]{behroozi13}, a halo of $\sim10^{10}M_\odot$
usually has $M_\text{galaxy}/M_\text{halo}\sim10^{-3}$. The CosmoDC2 catalog should thus
be complete down to $M_*\sim10^7M_\odot$. Our $\sigma-$selected deflector sample has a minimum
stellar mass of $M_*=10^{7.8}M_\odot$, which means that the mock catalog is complete 
for deflectors with $\sigma>50\text{ km s}^{-1}$ and $z_\text{d}<3$.

The main incompleteness comes from the fact that CosmoDC2 catalog only covers $0<z<3$.
Figure \ref{fig:zd} suggests that the majority of mock lenses have $z_d<3$.
\citet{wyithe11} studies the strong lensing statistics for $z\sim8$ galaxies, and their results suggest that 
$>80\%$ lensing systems have $z_d<3$. 
This fraction will be higher for lower source redshifts, and
we conclude that our mock catalog is highly complete.

\subsection{Finding More Lensed Quasars Beyond the ``Discoverable" Ones}

In the discussion above, we adopt the definition of a discoverable lensed quasar in OM10,
which requires an image separation $\Delta \theta>\frac{2}{3}\times\text{FWHM}$
and a $5\sigma$ detection of the second or third brightest lensed image.
The motivation of these criteria is that the lensed quasar images need to be at least marginally resolved and barely visible.
Under this definition,
about one third of the lensed quasars in the mock catalog are not discoverable in LSST.
In this subsection, we discuss possible strategies to find these undiscoverable lenses.

Identifying undiscoverable lensed quasars is important 
because the discoverable criteria can miss some unique and interesting objects.
As an example, \citet{fan19} report a compact lens with separation of $0\farcs2$ at $z=6.51$, J0439+1634,
which is the only known multiply-imaged quasar at $z>5$ to date.
J0439+1634 has a large magnification of $\mu=51$, making it the brightest quasar
at $z>6$ in nearly all wavelengths, and provides so far the best case study of a high-redshift quasar.
Although J0439+1634 is bright $(m_z=19.49\pm0.02)$,
the small separation makes it undiscoverable in ground-based imaging surveys.
Besides, undiscoverable lenses enable many studies that are impossible otherwise.
For instance, small-separation lenses can probe the mass distribution of less massive deflector galaxies.

We first consider the image separation criteria. Figure \ref{fig:separation} suggests that
LSST-like seeing $(\sim0\farcs7)$ will be sufficient to resolve the majority of lensed quasars.
Lensing systems with smaller separation have less massive and thus fainter deflector galaxies.
These objects are likely to have quasar-dominated flux with a marginally extended morphology.
Since the deflector galaxy and the lensed quasar images are blended, we expect that neither a PSF
nor a regular S\'ersic profile can provide a good description of their morphology,
according to which we can identify these compact lenses.
This method is used in \citet{fan19} to identify the $z=6.51$ compact lensed quasar
using ground-based imaging, which is later confirmed by {\em HST} imaging.

For lensed quasars that do not meet the second criteria, i.e., the fainter lensed image being brighter than the survey limit,
some of them can be discovered by identifying the brighter image as a quasar. If there is a bright galaxy next to it,
it is a hint that the quasar might be lensed. 
Figure \ref{fig:deflectorflux} suggests that the deflector galaxy is detectable in LSST survey in most lensing systems.
Follow-up deep imaging and/or spectroscopy can confirm 
if there is a fainter lensed image. This technique requires decent image de-blending
to disentangle the brighter lensed quasar image and the deflector galaxy.

As we show in Figure \ref{fig:deflectorflux}, the observational features of lensed quasars have a large variety.
A complete search of them requires a number of distinct candidate selection methods.
The mock catalog can serve as training and testing sets for future surveys of lensed quasars
with LSST data.

\section{Summary and Conclusion} \label{sec:conclusion}

We present a mock catalog of gravitationally lensed quasars at $z_\text{qso}<7.5$, which covers quasars with $M_{i}<-20$
and deflector galaxies with $\sigma>50\text{ km s}^{-1}$, using updated QLFs and deflector VDFs.
We generate synthetic magnitudes and simulated LSST images for the mock lensed quasars.
Using the mock catalog, we explore the expected outputs of lensed quasar surveys, 
and investigate possible strategies to find more lensed quasars.
Our main conclusions are:
\begin{enumerate}
    \item The number of discoverable lensed quasars in current sky surveys is $\sim10^3$,
        and LSST will increase this number to $\sim2.4\times10^3$.
        Most of the lensed quasars have image separation $\Delta \theta>0\farcs5$,
        which will at least get marginally resolved in the LSST survey.
    \item There will be $\sim200$ discoverable quadruply lensed quasars in the LSST survey.
    The fraction of quad lenses among all discoverable lensed quasars is $\sim10\%-15\%$ at $m_i\lesssim25$,
    and decreases with survey depth.
    The typical time delays between lensed images are $\sim10^1-10^3$ days. Follow-up high-cadence observations
    are necessary to accurately measure the time lags.
    \item The mock lensed quasars show a large diversity in their observational features,
    from deflector-dominated ones which are lensed by bright, massive galaxies with large lensing separations,
    to quasar-dominated ones which are likely more compact. 
    The variety in observational features requires complex candidate selection techniques.
\end{enumerate}

%The mock catalog and the simulated images will serve as training and testing sets
%in future surveys for these objects.

%The mock catalog can be used to explore the statistics of lensed quasars and to guide future surveys for these objects.
We estimate that the mock catalog has a high completeness for lensed quasars with Einstein radius $\theta_E>0\farcs07$.
This range covers nearly all the lensed quasars that can be identified in the near future,
including compact ones that can be discovered by the Euclid Telescope \citep{euclid} 
and the Roman Space Telescope \citep{wfirst}.
The mock catalog and the simulated images will be powerful tools in 
future lensed quasar surveys,
especially for high-redshift lensed quasars which are not well-explored by previous studies.
As an example, we have used the mock catalog to analyze the population of lensed quasars at $z>5$
and design a new survey for these objects (Yue et al. in prep).
Besides,
although we focus on traditional (point-like) type-I quasars, the code can be easily adapted to
study other background source population, including fainter AGNs and normal galaxies.


%\begin{acknowledgments}

\acknowledgments
We thank Peter Behroozi, Haowen Zhang, Fabio Pacucci and the anonymous referee for useful and inspiring comments. 
MY, XF and JY acknowledge supports by NSF grants AST 19-08284. 
FW thanks the support provided by NASA through the NASA Hubble Fellowship grant \#HST-HF2-51448.001-A awarded by the Space Telescope Science Institute, which is operated by the Association of Universities for Research in Astronomy, Incorporated, under NASA contract NAS5-26555.
\software{NumPy \citep{numpy1,numpy2},
        Astropy \citep{2013A&A...558A..33A,2018AJ....156..123A},  SIMQSO \citep{simqso},
          glafic \citep{glafic}, galfit \citep{galfit}, GCRcatalog \citep{gcrcatalog}}
          
%\end{acknowledgments}


%\vspace{5mm}
%\facilities{HST(STIS), Swift(XRT and UVOT), AAVSO, CTIO:1.3m,




%\appendix
%\section{Simulating Quasar Spectra Using SIMQSO}

\bibliography{sample631}{}

\begin{thebibliography}{}
\expandafter\ifx\csname natexlab\endcsname\relax\def\natexlab#1{#1}\fi
\providecommand{\url}[1]{\href{#1}{#1}}
\providecommand{\dodoi}[1]{doi:~\href{http://doi.org/#1}{\nolinkurl{#1}}}
\providecommand{\doeprint}[1]{\href{http://ascl.net/#1}{\nolinkurl{http://ascl.net/#1}}}
\providecommand{\doarXiv}[1]{\href{https://arxiv.org/abs/#1}{\nolinkurl{https://arxiv.org/abs/#1}}}

\bibitem[{{Agnello} {et~al.}(2015){Agnello}, {Kelly}, {Treu}, \&
  {Marshall}}]{agnello15}
{Agnello}, A., {Kelly}, B.~C., {Treu}, T., \& {Marshall}, P.~J. 2015, \mnras,
  448, 1446, \dodoi{10.1093/mnras/stv037}

\bibitem[{{Agnello} {et~al.}(2018){Agnello}, {Schechter}, {Morgan}, {Treu},
  {Grillo}, {Malesani}, {Anguita}, {Apostolovski}, {Rusu}, {Motta}, {Rojas},
  {Chehade}, \& {Shanks}}]{agnello18}
{Agnello}, A., {Schechter}, P.~L., {Morgan}, N.~D., {et~al.} 2018, \mnras, 475,
  2086, \dodoi{10.1093/mnras/stx3226}

\bibitem[{{Aihara} {et~al.}(2018{\natexlab{a}}){Aihara}, {Armstrong},
  {Bickerton}, {Bosch}, {Coupon}, {Furusawa}, {Hayashi}, {Ikeda}, {Kamata},
  {Karoji}, {Kawanomoto}, {Koike}, {Komiyama}, {Lang}, {Lupton}, {Mineo},
  {Miyatake}, {Miyazaki}, {Morokuma}, {Obuchi}, {Oishi}, {Okura}, {Price},
  {Takata}, {Tanaka}, {Tanaka}, {Tanaka}, {Uchida}, {Uraguchi}, {Utsumi},
  {Wang}, {Yamada}, {Yamanoi}, {Yasuda}, {Arimoto}, {Chiba}, {Finet},
  {Fujimori}, {Fujimoto}, {Furusawa}, {Goto}, {Goulding}, {Gunn}, {Harikane},
  {Hattori}, {Hayashi}, {He{\l}miniak}, {Higuchi}, {Hikage}, {Ho}, {Hsieh},
  {Huang}, {Huang}, {Imanishi}, {Iwata}, {Jaelani}, {Jian}, {Kashikawa},
  {Katayama}, {Kojima}, {Konno}, {Koshida}, {Kusakabe}, {Leauthaud}, {Lee},
  {Lin}, {Lin}, {Mandelbaum}, {Matsuoka}, {Medezinski}, {Miyama}, {Momose},
  {More}, {More}, {Mukae}, {Murata}, {Murayama}, {Nagao}, {Nakata}, {Niida},
  {Niikura}, {Nishizawa}, {Oguri}, {Okabe}, {Ono}, {Onodera}, {Onoue}, {Ouchi},
  {Pyo}, {Shibuya}, {Shimasaku}, {Simet}, {Speagle}, {Spergel}, {Strauss},
  {Sugahara}, {Sugiyama}, {Suto}, {Suzuki}, {Tait}, {Takada}, {Terai}, {Toba},
  {Turner}, {Uchiyama}, {Umetsu}, {Urata}, {Usuda}, {Yeh}, \& {Yuma}}]{hscdr1}
{Aihara}, H., {Armstrong}, R., {Bickerton}, S., {et~al.} 2018{\natexlab{a}},
  \pasj, 70, S8, \dodoi{10.1093/pasj/psx081}

\bibitem[{{Aihara} {et~al.}(2018{\natexlab{b}}){Aihara}, {Arimoto},
  {Armstrong}, {Arnouts}, {Bahcall}, {Bickerton}, {Bosch}, {Bundy}, {Capak},
  {Chan}, {Chiba}, {Coupon}, {Egami}, {Enoki}, {Finet}, {Fujimori}, {Fujimoto},
  {Furusawa}, {Furusawa}, {Goto}, {Goulding}, {Greco}, {Greene}, {Gunn},
  {Hamana}, {Harikane}, {Hashimoto}, {Hattori}, {Hayashi}, {Hayashi},
  {He{\l}miniak}, {Higuchi}, {Hikage}, {Ho}, {Hsieh}, {Huang}, {Huang},
  {Ikeda}, {Imanishi}, {Inoue}, {Iwasawa}, {Iwata}, {Jaelani}, {Jian},
  {Kamata}, {Karoji}, {Kashikawa}, {Katayama}, {Kawanomoto}, {Kayo}, {Koda},
  {Koike}, {Kojima}, {Komiyama}, {Konno}, {Koshida}, {Koyama}, {Kusakabe},
  {Leauthaud}, {Lee}, {Lin}, {Lin}, {Lupton}, {Mandelbaum}, {Matsuoka},
  {Medezinski}, {Mineo}, {Miyama}, {Miyatake}, {Miyazaki}, {Momose}, {More},
  {More}, {Moritani}, {Moriya}, {Morokuma}, {Mukae}, {Murata}, {Murayama},
  {Nagao}, {Nakata}, {Niida}, {Niikura}, {Nishizawa}, {Obuchi}, {Oguri},
  {Oishi}, {Okabe}, {Okamoto}, {Okura}, {Ono}, {Onodera}, {Onoue}, {Osato},
  {Ouchi}, {Price}, {Pyo}, {Sako}, {Sawicki}, {Shibuya}, {Shimasaku},
  {Shimono}, {Shirasaki}, {Silverman}, {Simet}, {Speagle}, {Spergel},
  {Strauss}, {Sugahara}, {Sugiyama}, {Suto}, {Suyu}, {Suzuki}, {Tait},
  {Takada}, {Takata}, {Tamura}, {Tanaka}, {Tanaka}, {Tanaka}, {Tanaka},
  {Terai}, {Terashima}, {Toba}, {Tominaga}, {Toshikawa}, {Turner}, {Uchida},
  {Uchiyama}, {Umetsu}, {Uraguchi}, {Urata}, {Usuda}, {Utsumi}, {Wang}, {Wang},
  {Wong}, {Yabe}, {Yamada}, {Yamanoi}, {Yasuda}, {Yeh}, {Yonehara}, \&
  {Yuma}}]{hsc}
{Aihara}, H., {Arimoto}, N., {Armstrong}, R., {et~al.} 2018{\natexlab{b}},
  \pasj, 70, S4, \dodoi{10.1093/pasj/psx066}

\bibitem[{{Akiyama} {et~al.}(2018){Akiyama}, {He}, {Ikeda}, {Niida}, {Nagao},
  {Bosch}, {Coupon}, {Enoki}, {Imanishi}, {Kashikawa}, {Kawaguchi}, {Komiyama},
  {Lee}, {Matsuoka}, {Miyazaki}, {Nishizawa}, {Oguri}, {Ono}, {Onoue}, {Ouchi},
  {Schulze}, {Silverman}, {Tanaka}, {Tanaka}, {Terashima}, {Toba}, \&
  {Ueda}}]{akiyama18}
{Akiyama}, M., {He}, W., {Ikeda}, H., {et~al.} 2018, \pasj, 70, S34,
  \dodoi{10.1093/pasj/psx091}

\bibitem[{{Astropy Collaboration} {et~al.}(2013){Astropy Collaboration},
  {Robitaille}, {Tollerud}, {Greenfield}, {Droettboom}, {Bray}, {Aldcroft},
  {Davis}, {Ginsburg}, {Price-Whelan}, {Kerzendorf}, {Conley}, {Crighton},
  {Barbary}, {Muna}, {Ferguson}, {Grollier}, {Parikh}, {Nair}, {Unther},
  {Deil}, {Woillez}, {Conseil}, {Kramer}, {Turner}, {Singer}, {Fox}, {Weaver},
  {Zabalza}, {Edwards}, {Azalee Bostroem}, {Burke}, {Casey}, {Crawford},
  {Dencheva}, {Ely}, {Jenness}, {Labrie}, {Lim}, {Pierfederici}, {Pontzen},
  {Ptak}, {Refsdal}, {Servillat}, \& {Streicher}}]{2013A&A...558A..33A}
{Astropy Collaboration}, {Robitaille}, T.~P., {Tollerud}, E.~J., {et~al.} 2013,
  \aap, 558, A33, \dodoi{10.1051/0004-6361/201322068}

\bibitem[{{Astropy Collaboration} {et~al.}(2018){Astropy Collaboration},
  {Price-Whelan}, {Sip{\H{o}}cz}, {G{\"u}nther}, {Lim}, {Crawford}, {Conseil},
  {Shupe}, {Craig}, {Dencheva}, {Ginsburg}, {VanderPlas}, {Bradley},
  {P{\'e}rez-Su{\'a}rez}, {de Val-Borro}, {Aldcroft}, {Cruz}, {Robitaille},
  {Tollerud}, {Ardelean}, {Babej}, {Bach}, {Bachetti}, {Bakanov}, {Bamford},
  {Barentsen}, {Barmby}, {Baumbach}, {Berry}, {Biscani}, {Boquien}, {Bostroem},
  {Bouma}, {Brammer}, {Bray}, {Breytenbach}, {Buddelmeijer}, {Burke},
  {Calderone}, {Cano Rodr{\'\i}guez}, {Cara}, {Cardoso}, {Cheedella}, {Copin},
  {Corrales}, {Crichton}, {D'Avella}, {Deil}, {Depagne}, {Dietrich}, {Donath},
  {Droettboom}, {Earl}, {Erben}, {Fabbro}, {Ferreira}, {Finethy}, {Fox},
  {Garrison}, {Gibbons}, {Goldstein}, {Gommers}, {Greco}, {Greenfield},
  {Groener}, {Grollier}, {Hagen}, {Hirst}, {Homeier}, {Horton}, {Hosseinzadeh},
  {Hu}, {Hunkeler}, {Ivezi{\'c}}, {Jain}, {Jenness}, {Kanarek}, {Kendrew},
  {Kern}, {Kerzendorf}, {Khvalko}, {King}, {Kirkby}, {Kulkarni}, {Kumar},
  {Lee}, {Lenz}, {Littlefair}, {Ma}, {Macleod}, {Mastropietro}, {McCully},
  {Montagnac}, {Morris}, {Mueller}, {Mumford}, {Muna}, {Murphy}, {Nelson},
  {Nguyen}, {Ninan}, {N{\"o}the}, {Ogaz}, {Oh}, {Parejko}, {Parley}, {Pascual},
  {Patil}, {Patil}, {Plunkett}, {Prochaska}, {Rastogi}, {Reddy Janga},
  {Sabater}, {Sakurikar}, {Seifert}, {Sherbert}, {Sherwood-Taylor}, {Shih},
  {Sick}, {Silbiger}, {Singanamalla}, {Singer}, {Sladen}, {Sooley},
  {Sornarajah}, {Streicher}, {Teuben}, {Thomas}, {Tremblay}, {Turner},
  {Terr{\'o}n}, {van Kerkwijk}, {de la Vega}, {Watkins}, {Weaver}, {Whitmore},
  {Woillez}, {Zabalza}, \& {Astropy Contributors}}]{2018AJ....156..123A}
{Astropy Collaboration}, {Price-Whelan}, A.~M., {Sip{\H{o}}cz}, B.~M., {et~al.}
  2018, \aj, 156, 123, \dodoi{10.3847/1538-3881/aabc4f}

\bibitem[{{Auger} {et~al.}(2010){Auger}, {Treu}, {Bolton}, {Gavazzi},
  {Koopmans}, {Marshall}, {Moustakas}, \& {Burles}}]{auger10}
{Auger}, M.~W., {Treu}, T., {Bolton}, A.~S., {et~al.} 2010, \apj, 724, 511,
  \dodoi{10.1088/0004-637X/724/1/511}

\bibitem[{{Behroozi} {et~al.}(2019){Behroozi}, {Wechsler}, {Hearin}, \&
  {Conroy}}]{behroozi19}
{Behroozi}, P., {Wechsler}, R.~H., {Hearin}, A.~P., \& {Conroy}, C. 2019,
  \mnras, 488, 3143, \dodoi{10.1093/mnras/stz1182}

\bibitem[{{Behroozi} {et~al.}(2013){Behroozi}, {Wechsler}, \&
  {Conroy}}]{behroozi13}
{Behroozi}, P.~S., {Wechsler}, R.~H., \& {Conroy}, C. 2013, \apj, 770, 57,
  \dodoi{10.1088/0004-637X/770/1/57}

\bibitem[{{Beifiori} {et~al.}(2014){Beifiori}, {Thomas}, {Maraston}, {Steele},
  {Masters}, {Pforr}, {Saglia}, {Bender}, {Tojeiro}, {Chen}, {Bolton},
  {Brownstein}, {Johansson}, {Leauthaud}, {Nichol}, {Schneider}, {Senger},
  {Skibba}, {Wake}, {Pan}, {Snedden}, {Bizyaev}, {Brewington}, {Malanushenko},
  {Malanushenko}, {Oravetz}, {Simmons}, {Shelden}, \& {Ebelke}}]{beifiori14}
{Beifiori}, A., {Thomas}, D., {Maraston}, C., {et~al.} 2014, \apj, 789, 92,
  \dodoi{10.1088/0004-637X/789/2/92}

\bibitem[{{Bertin} {et~al.}(2002){Bertin}, {Ciotti}, \& {Del
  Principe}}]{bertin02}
{Bertin}, G., {Ciotti}, L., \& {Del Principe}, M. 2002, \aap, 386, 149,
  \dodoi{10.1051/0004-6361:20020248}

\bibitem[{{Bezanson} {et~al.}(2012){Bezanson}, {van Dokkum}, \&
  {Franx}}]{bezanson12}
{Bezanson}, R., {van Dokkum}, P., \& {Franx}, M. 2012, \apj, 760, 62,
  \dodoi{10.1088/0004-637X/760/1/62}

\bibitem[{{Bezanson} {et~al.}(2011){Bezanson}, {van Dokkum}, {Franx},
  {Brammer}, {Brinchmann}, {Kriek}, {Labb{\'e}}, {Quadri}, {Rix}, {van de
  Sande}, {Whitaker}, \& {Williams}}]{bezanson11}
{Bezanson}, R., {van Dokkum}, P.~G., {Franx}, M., {et~al.} 2011, \apjl, 737,
  L31, \dodoi{10.1088/2041-8205/737/2/L31}

\bibitem[{{Brown} {et~al.}(2014){Brown}, {Moustakas}, {Smith}, {da Cunha},
  {Jarrett}, {Imanishi}, {Armus}, {Brandl}, \& {Peek}}]{brown14}
{Brown}, M. J.~I., {Moustakas}, J., {Smith}, J. D.~T., {et~al.} 2014, \apjs,
  212, 18, \dodoi{10.1088/0067-0049/212/2/18}

\bibitem[{{Cappellari} {et~al.}(2006){Cappellari}, {Bacon}, {Bureau}, {Damen},
  {Davies}, {de Zeeuw}, {Emsellem}, {Falc{\'o}n-Barroso}, {Krajnovi{\'c}},
  {Kuntschner}, {McDermid}, {Peletier}, {Sarzi}, {van den Bosch}, \& {van de
  Ven}}]{cappellari06}
{Cappellari}, M., {Bacon}, R., {Bureau}, M., {et~al.} 2006, \mnras, 366, 1126,
  \dodoi{10.1111/j.1365-2966.2005.09981.x}

\bibitem[{{Cashman} {et~al.}(2021){Cashman}, {Kulkarni}, \&
  {Lopez}}]{cashman21}
{Cashman}, F.~H., {Kulkarni}, V.~P., \& {Lopez}, S. 2021, \aj, 161, 90,
  \dodoi{10.3847/1538-3881/abd2b0}

\bibitem[{{Chambers} {et~al.}(2016){Chambers}, {Magnier}, {Metcalfe},
  {Flewelling}, {Huber}, {Waters}, {Denneau}, {Draper}, {Farrow}, {Finkbeiner},
  {Holmberg}, {Koppenhoefer}, {Price}, {Rest}, {Saglia}, {Schlafly}, {Smartt},
  {Sweeney}, {Wainscoat}, {Burgett}, {Chastel}, {Grav}, {Heasley}, {Hodapp},
  {Jedicke}, {Kaiser}, {Kudritzki}, {Luppino}, {Lupton}, {Monet}, {Morgan},
  {Onaka}, {Shiao}, {Stubbs}, {Tonry}, {White}, {Ba{\~n}ados}, {Bell},
  {Bender}, {Bernard}, {Boegner}, {Boffi}, {Botticella}, {Calamida},
  {Casertano}, {Chen}, {Chen}, {Cole}, {Deacon}, {Frenk}, {Fitzsimmons},
  {Gezari}, {Gibbs}, {Goessl}, {Goggia}, {Gourgue}, {Goldman}, {Grant},
  {Grebel}, {Hambly}, {Hasinger}, {Heavens}, {Heckman}, {Henderson}, {Henning},
  {Holman}, {Hopp}, {Ip}, {Isani}, {Jackson}, {Keyes}, {Koekemoer}, {Kotak},
  {Le}, {Liska}, {Long}, {Lucey}, {Liu}, {Martin}, {Masci}, {McLean}, {Mindel},
  {Misra}, {Morganson}, {Murphy}, {Obaika}, {Narayan}, {Nieto-Santisteban},
  {Norberg}, {Peacock}, {Pier}, {Postman}, {Primak}, {Rae}, {Rai}, {Riess},
  {Riffeser}, {Rix}, {R{\"o}ser}, {Russel}, {Rutz}, {Schilbach}, {Schultz},
  {Scolnic}, {Strolger}, {Szalay}, {Seitz}, {Small}, {Smith}, {Soderblom},
  {Taylor}, {Thomson}, {Taylor}, {Thakar}, {Thiel}, {Thilker}, {Unger},
  {Urata}, {Valenti}, {Wagner}, {Walder}, {Walter}, {Watters}, {Werner},
  {Wood-Vasey}, \& {Wyse}}]{ps1}
{Chambers}, K.~C., {Magnier}, E.~A., {Metcalfe}, N., {et~al.} 2016, arXiv
  e-prints, arXiv:1612.05560.
\newblock \doarXiv{1612.05560}

\bibitem[{{Chan} {et~al.}(2020){Chan}, {Suyu}, {Sonnenfeld}, {Jaelani}, {More},
  {Yonehara}, {Kubota}, {Coupon}, {Lee}, {Oguri}, {Rusu}, \& {Wong}}]{chan20}
{Chan}, J. H.~H., {Suyu}, S.~H., {Sonnenfeld}, A., {et~al.} 2020, \aap, 636,
  A87, \dodoi{10.1051/0004-6361/201937030}

\bibitem[{{Choi} {et~al.}(2007){Choi}, {Park}, \& {Vogeley}}]{choi07}
{Choi}, Y.-Y., {Park}, C., \& {Vogeley}, M.~S. 2007, \apj, 658, 884,
  \dodoi{10.1086/511060}

\bibitem[{{Collett}(2015)}]{collett15}
{Collett}, T.~E. 2015, \apj, 811, 20, \dodoi{10.1088/0004-637X/811/1/20}

\bibitem[{{Cornachione} \& {Morgan}(2020)}]{Cornachione20}
{Cornachione}, M.~A., \& {Morgan}, C.~W. 2020, \apj, 895, 93,
  \dodoi{10.3847/1538-4357/ab8aed}

\bibitem[{{DES Collaboration} {et~al.}(2021){DES Collaboration}, {Abbott},
  {Adamow}, {Aguena}, {Allam}, {Amon}, {Annis}, {Avila}, {Bacon}, {Banerji},
  {Bechtol}, {Becker}, {Bernstein}, {Bertin}, {Bhargava}, {Bridle}, {Brooks},
  {Burke}, {Carnero Rosell}, {Carrasco Kind}, {Carretero}, {Castander},
  {Cawthon}, {Chang}, {Choi}, {Conselice}, {Costanzi}, {Crocce}, {da Costa},
  {Davis}, {De Vicente}, {DeRose}, {Desai}, {Diehl}, {Dietrich},
  {Drlica-Wagner}, {Eckert}, {Elvin-Poole}, {Everett}, {Evrard}, {Ferrero},
  {Fert{\'e}}, {Flaugher}, {Fosalba}, {Friedel}, {Frieman},
  {Garc{\'\i}a-Bellido}, {Gelman}, {Gerdes}, {Giannantonio}, {Gill}, {Gruen},
  {Gruendl}, {Gschwend}, {Gutierrez}, {Hartley}, {Hinton}, {Hollowood},
  {Honscheid}, {Huterer}, {James}, {Jeltema}, {Johnson}, {Kent}, {Kron},
  {Kuehn}, {Kuropatkin}, {Lahav}, {Li}, {Lidman}, {Lin}, {MacCrann}, {Maia},
  {Manning}, {March}, {Marshall}, {Martini}, {Melchior}, {Menanteau}, {Miquel},
  {Morgan}, {Myles}, {Neilsen}, {Ogando}, {Palmese}, {Paz-Chinch{\'o}n},
  {Petravick}, {Pieres}, {Plazas}, {Pond}, {Rodriguez-Monroy}, {Romer},
  {Roodman}, {Rykoff}, {Sako}, {Sanchez}, {Santiago}, {Serrano},
  {Sevilla-Noarbe}, {Allyn. Smith}, {Smith}, {Soares-Santos}, {Suchyta},
  {Swanson}, {Tarle}, {Thomas}, {To}, {Tremblay}, {Troxel}, {Tucker}, {Turner},
  {Varga}, {Walker}, {Wechsler}, {Weller}, {Wester}, {Wilkinson}, {Yanny},
  {Zhang}, {Nikutta}, {Fitzpatrick}, {Jacques}, {Scott}, {Olsen}, {Huang},
  {Herrera}, {Juneau}, {Nidever}, {Weaver}, {Adean}, {Correia}, {de Freitas},
  {Freitas}, {Singulani}, \& {Vila-Verde}}]{desdr2}
{DES Collaboration}, {Abbott}, T.~M.~C., {Adamow}, M., {et~al.} 2021, arXiv
  e-prints, arXiv:2101.05765.
\newblock \doarXiv{2101.05765}

\bibitem[{{Dey} {et~al.}(2019){Dey}, {Schlegel}, {Lang}, {Blum}, {Burleigh},
  {Fan}, {Findlay}, {Finkbeiner}, {Herrera}, {Juneau}, {Landriau}, {Levi},
  {McGreer}, {Meisner}, {Myers}, {Moustakas}, {Nugent}, {Patej}, {Schlafly},
  {Walker}, {Valdes}, {Weaver}, {Y{\`e}che}, {Zou}, {Zhou}, {Abareshi},
  {Abbott}, {Abolfathi}, {Aguilera}, {Alam}, {Allen}, {Alvarez}, {Annis},
  {Ansarinejad}, {Aubert}, {Beechert}, {Bell}, {BenZvi}, {Beutler}, {Bielby},
  {Bolton}, {Brice{\~n}o}, {Buckley-Geer}, {Butler}, {Calamida}, {Carlberg},
  {Carter}, {Casas}, {Castander}, {Choi}, {Comparat}, {Cukanovaite}, {Delubac},
  {DeVries}, {Dey}, {Dhungana}, {Dickinson}, {Ding}, {Donaldson}, {Duan},
  {Duckworth}, {Eftekharzadeh}, {Eisenstein}, {Etourneau}, {Fagrelius},
  {Farihi}, {Fitzpatrick}, {Font-Ribera}, {Fulmer}, {G{\"a}nsicke},
  {Gaztanaga}, {George}, {Gerdes}, {Gontcho}, {Gorgoni}, {Green}, {Guy},
  {Harmer}, {Hernandez}, {Honscheid}, {Huang}, {James}, {Jannuzi}, {Jiang},
  {Joyce}, {Karcher}, {Karkar}, {Kehoe}, {Kneib}, {Kueter-Young}, {Lan},
  {Lauer}, {Le Guillou}, {Le Van Suu}, {Lee}, {Lesser}, {Perreault Levasseur},
  {Li}, {Mann}, {Marshall}, {Mart{\'\i}nez-V{\'a}zquez}, {Martini}, {du Mas des
  Bourboux}, {McManus}, {Meier}, {M{\'e}nard}, {Metcalfe},
  {Mu{\~n}oz-Guti{\'e}rrez}, {Najita}, {Napier}, {Narayan}, {Newman}, {Nie},
  {Nord}, {Norman}, {Olsen}, {Paat}, {Palanque-Delabrouille}, {Peng},
  {Poppett}, {Poremba}, {Prakash}, {Rabinowitz}, {Raichoor}, {Rezaie},
  {Robertson}, {Roe}, {Ross}, {Ross}, {Rudnick}, {Safonova}, {Saha},
  {S{\'a}nchez}, {Savary}, {Schweiker}, {Scott}, {Seo}, {Shan}, {Silva},
  {Slepian}, {Soto}, {Sprayberry}, {Staten}, {Stillman}, {Stupak}, {Summers},
  {Sien Tie}, {Tirado}, {Vargas-Maga{\~n}a}, {Vivas}, {Wechsler}, {Williams},
  {Yang}, {Yang}, {Yapici}, {Zaritsky}, {Zenteno}, {Zhang}, {Zhang}, {Zhou}, \&
  {Zhou}}]{legacy}
{Dey}, A., {Schlegel}, D.~J., {Lang}, D., {et~al.} 2019, \aj, 157, 168,
  \dodoi{10.3847/1538-3881/ab089d}

\bibitem[{{Fan} {et~al.}(2019){Fan}, {Wang}, {Yang}, {Keeton}, {Yue},
  {Zabludoff}, {Bian}, {Bonaglia}, {Georgiev}, {Hennawi}, {Li}, {McGreer},
  {Naidu}, {Pacucci}, {Rabien}, {Thompson}, {Venemans}, {Walter}, {Wang}, \&
  {Wu}}]{fan19}
{Fan}, X., {Wang}, F., {Yang}, J., {et~al.} 2019, \apjl, 870, L11,
  \dodoi{10.3847/2041-8213/aaeffe}

\bibitem[{{Geng} {et~al.}(2021){Geng}, {Cao}, {Liu}, {Liu}, {Biesiada}, \&
  {Lian}}]{geng21}
{Geng}, S., {Cao}, S., {Liu}, Y., {et~al.} 2021, \mnras, 503, 1319,
  \dodoi{10.1093/mnras/stab519}

\bibitem[{{Gilman} {et~al.}(2020){Gilman}, {Birrer}, {Nierenberg}, {Treu},
  {Du}, \& {Benson}}]{gilman20}
{Gilman}, D., {Birrer}, S., {Nierenberg}, A., {et~al.} 2020, \mnras, 491, 6077,
  \dodoi{10.1093/mnras/stz3480}

\bibitem[{{Harris} {et~al.}(2020){Harris}, {Millman}, {van der Walt},
  {Gommers}, {Virtanen}, {Cournapeau}, {Wieser}, {Taylor}, {Berg}, {Smith},
  {Kern}, {Picus}, {Hoyer}, {van Kerkwijk}, {Brett}, {Haldane}, {del R{\'\i}o},
  {Wiebe}, {Peterson}, {G{\'e}rard-Marchant}, {Sheppard}, {Reddy}, {Weckesser},
  {Abbasi}, {Gohlke}, \& {Oliphant}}]{numpy2}
{Harris}, C.~R., {Millman}, K.~J., {van der Walt}, S.~J., {et~al.} 2020, \nat,
  585, 357, \dodoi{10.1038/s41586-020-2649-2}

\bibitem[{{Hasan} \& {Crocker}(2019)}]{hasan19}
{Hasan}, F., \& {Crocker}, A. 2019, arXiv e-prints, arXiv:1904.00486.
\newblock \doarXiv{1904.00486}

\bibitem[{{Heitmann} {et~al.}(2019){Heitmann}, {Finkel}, {Pope}, {Morozov},
  {Frontiere}, {Habib}, {Rangel}, {Uram}, {Korytov}, {Child}, {Flender},
  {Insley}, \& {Rizzi}}]{outerrim}
{Heitmann}, K., {Finkel}, H., {Pope}, A., {et~al.} 2019, \apjs, 245, 16,
  \dodoi{10.3847/1538-4365/ab4da1}

\bibitem[{{Hilbert} {et~al.}(2009){Hilbert}, {Hartlap}, {White}, \&
  {Schneider}}]{hilbert09}
{Hilbert}, S., {Hartlap}, J., {White}, S.~D.~M., \& {Schneider}, P. 2009, \aap,
  499, 31, \dodoi{10.1051/0004-6361/200811054}

\bibitem[{{Hilbert} {et~al.}(2008){Hilbert}, {White}, {Hartlap}, \&
  {Schneider}}]{hilbert08}
{Hilbert}, S., {White}, S. D.~M., {Hartlap}, J., \& {Schneider}, P. 2008,
  \mnras, 386, 1845, \dodoi{10.1111/j.1365-2966.2008.13190.x}

\bibitem[{{Inada} {et~al.}(2012){Inada}, {Oguri}, {Shin}, {Kayo}, {Strauss},
  {Morokuma}, {Rusu}, {Fukugita}, {Kochanek}, {Richards}, {Schneider}, {York},
  {Bahcall}, {Frieman}, {Hall}, \& {White}}]{inada12}
{Inada}, N., {Oguri}, M., {Shin}, M.-S., {et~al.} 2012, \aj, 143, 119,
  \dodoi{10.1088/0004-6256/143/5/119}

\bibitem[{{Ivezi{\'c}} {et~al.}(2019){Ivezi{\'c}}, {Kahn}, {Tyson}, {Abel},
  {Acosta}, {Allsman}, {Alonso}, {AlSayyad}, {Anderson}, {Andrew}, {Angel},
  {Angeli}, {Ansari}, {Antilogus}, {Araujo}, {Armstrong}, {Arndt}, {Astier},
  {Aubourg}, {Auza}, {Axelrod}, {Bard}, {Barr}, {Barrau}, {Bartlett}, {Bauer},
  {Bauman}, {Baumont}, {Bechtol}, {Bechtol}, {Becker}, {Becla}, {Beldica},
  {Bellavia}, {Bianco}, {Biswas}, {Blanc}, {Blazek}, {Blandford}, {Bloom},
  {Bogart}, {Bond}, {Booth}, {Borgland}, {Borne}, {Bosch}, {Boutigny},
  {Brackett}, {Bradshaw}, {Brandt}, {Brown}, {Bullock}, {Burchat}, {Burke},
  {Cagnoli}, {Calabrese}, {Callahan}, {Callen}, {Carlin}, {Carlson},
  {Chandrasekharan}, {Charles-Emerson}, {Chesley}, {Cheu}, {Chiang}, {Chiang},
  {Chirino}, {Chow}, {Ciardi}, {Claver}, {Cohen-Tanugi}, {Cockrum}, {Coles},
  {Connolly}, {Cook}, {Cooray}, {Covey}, {Cribbs}, {Cui}, {Cutri}, {Daly},
  {Daniel}, {Daruich}, {Daubard}, {Daues}, {Dawson}, {Delgado}, {Dellapenna},
  {de Peyster}, {de Val-Borro}, {Digel}, {Doherty}, {Dubois},
  {Dubois-Felsmann}, {Durech}, {Economou}, {Eifler}, {Eracleous}, {Emmons},
  {Fausti Neto}, {Ferguson}, {Figueroa}, {Fisher-Levine}, {Focke}, {Foss},
  {Frank}, {Freemon}, {Gangler}, {Gawiser}, {Geary}, {Gee}, {Geha}, {Gessner},
  {Gibson}, {Gilmore}, {Glanzman}, {Glick}, {Goldina}, {Goldstein}, {Goodenow},
  {Graham}, {Gressler}, {Gris}, {Guy}, {Guyonnet}, {Haller}, {Harris},
  {Hascall}, {Haupt}, {Hernandez}, {Herrmann}, {Hileman}, {Hoblitt}, {Hodgson},
  {Hogan}, {Howard}, {Huang}, {Huffer}, {Ingraham}, {Innes}, {Jacoby}, {Jain},
  {Jammes}, {Jee}, {Jenness}, {Jernigan}, {Jevremovi{\'c}}, {Johns}, {Johnson},
  {Johnson}, {Jones}, {Juramy-Gilles}, {Juri{\'c}}, {Kalirai}, {Kallivayalil},
  {Kalmbach}, {Kantor}, {Karst}, {Kasliwal}, {Kelly}, {Kessler}, {Kinnison},
  {Kirkby}, {Knox}, {Kotov}, {Krabbendam}, {Krughoff}, {Kub{\'a}nek},
  {Kuczewski}, {Kulkarni}, {Ku}, {Kurita}, {Lage}, {Lambert}, {Lange},
  {Langton}, {Le Guillou}, {Levine}, {Liang}, {Lim}, {Lintott}, {Long},
  {Lopez}, {Lotz}, {Lupton}, {Lust}, {MacArthur}, {Mahabal}, {Mandelbaum},
  {Markiewicz}, {Marsh}, {Marshall}, {Marshall}, {May}, {McKercher}, {McQueen},
  {Meyers}, {Migliore}, {Miller}, {Mills}, {Miraval}, {Moeyens}, {Moolekamp},
  {Monet}, {Moniez}, {Monkewitz}, {Montgomery}, {Morrison}, {Mueller},
  {Muller}, {Mu{\~n}oz Arancibia}, {Neill}, {Newbry}, {Nief}, {Nomerotski},
  {Nordby}, {O'Connor}, {Oliver}, {Olivier}, {Olsen}, {O'Mullane}, {Ortiz},
  {Osier}, {Owen}, {Pain}, {Palecek}, {Parejko}, {Parsons}, {Pease},
  {Peterson}, {Peterson}, {Petravick}, {Libby Petrick}, {Petry},
  {Pierfederici}, {Pietrowicz}, {Pike}, {Pinto}, {Plante}, {Plate}, {Plutchak},
  {Price}, {Prouza}, {Radeka}, {Rajagopal}, {Rasmussen}, {Regnault}, {Reil},
  {Reiss}, {Reuter}, {Ridgway}, {Riot}, {Ritz}, {Robinson}, {Roby}, {Roodman},
  {Rosing}, {Roucelle}, {Rumore}, {Russo}, {Saha}, {Sassolas}, {Schalk},
  {Schellart}, {Schindler}, {Schmidt}, {Schneider}, {Schneider}, {Schoening},
  {Schumacher}, {Schwamb}, {Sebag}, {Selvy}, {Sembroski}, {Seppala}, {Serio},
  {Serrano}, {Shaw}, {Shipsey}, {Sick}, {Silvestri}, {Slater}, {Smith},
  {Smith}, {Sobhani}, {Soldahl}, {Storrie-Lombardi}, {Stover}, {Strauss},
  {Street}, {Stubbs}, {Sullivan}, {Sweeney}, {Swinbank}, {Szalay}, {Takacs},
  {Tether}, {Thaler}, {Thayer}, {Thomas}, {Thornton}, {Thukral}, {Tice},
  {Trilling}, {Turri}, {Van Berg}, {Vanden Berk}, {Vetter}, {Virieux},
  {Vucina}, {Wahl}, {Walkowicz}, {Walsh}, {Walter}, {Wang}, {Wang}, {Warner},
  {Wiecha}, {Willman}, {Winters}, {Wittman}, {Wolff}, {Wood-Vasey}, {Wu},
  {Xin}, {Yoachim}, \& {Zhan}}]{lsst}
{Ivezi{\'c}}, {\v{Z}}., {Kahn}, S.~M., {Tyson}, J.~A., {et~al.} 2019, \apj,
  873, 111, \dodoi{10.3847/1538-4357/ab042c}

\bibitem[{{Jaelani} {et~al.}(2021){Jaelani}, {Rusu}, {Kayo}, {More},
  {Sonnenfeld}, {Silverman}, {Schramm}, {Anguita}, {Inada}, {Kondo},
  {Schechter}, {Lee}, {Oguri}, {Chan}, {Wong}, \& {Inoue}}]{jaelani21}
{Jaelani}, A.~T., {Rusu}, C.~E., {Kayo}, I., {et~al.} 2021, \mnras, 502, 1487,
  \dodoi{10.1093/mnras/stab145}

\bibitem[{{Jiang} {et~al.}(2016){Jiang}, {McGreer}, {Fan}, {Strauss},
  {Ba{\~n}ados}, {Becker}, {Bian}, {Farnsworth}, {Shen}, {Wang}, {Wang},
  {Wang}, {White}, {Wu}, {Wu}, {Yang}, \& {Yang}}]{jiang16}
{Jiang}, L., {McGreer}, I.~D., {Fan}, X., {et~al.} 2016, \apj, 833, 222,
  \dodoi{10.3847/1538-4357/833/2/222}

\bibitem[{{Kim} {et~al.}(2020){Kim}, {Im}, {Jeon}, {Kim}, {Pak}, {Hyun},
  {Taak}, {Shin}, {Lim}, {Paek}, {Paek}, {Jiang}, {Choi}, {Hong}, {Ji}, {Jun},
  {Karouzos}, {Kim}, {Kim}, {Kim}, {Kim}, {Lee}, {Lee}, {Park}, {Yoon},
  {Byeon}, {Hwang}, {Kim}, {Kim}, \& {Park}}]{kim20}
{Kim}, Y., {Im}, M., {Jeon}, Y., {et~al.} 2020, \apj, 904, 111,
  \dodoi{10.3847/1538-4357/abc0ea}

\bibitem[{{Kochanek} {et~al.}(2006){Kochanek}, {Mochejska}, {Morgan}, \&
  {Stanek}}]{kochanek06}
{Kochanek}, C.~S., {Mochejska}, B., {Morgan}, N.~D., \& {Stanek}, K.~Z. 2006,
  \apjl, 637, L73, \dodoi{10.1086/500559}

\bibitem[{{Kormann} {et~al.}(1994){Kormann}, {Schneider}, \&
  {Bartelmann}}]{SIE}
{Kormann}, R., {Schneider}, P., \& {Bartelmann}, M. 1994, \aap, 284, 285

\bibitem[{{Korytov} {et~al.}(2019){Korytov}, {Hearin}, {Kovacs}, {Larsen},
  {Rangel}, {Hollowed}, {Benson}, {Heitmann}, {Mao}, {Bahmanyar}, {Chang},
  {Campbell}, {DeRose}, {Finkel}, {Frontiere}, {Gawiser}, {Habib}, {Joachimi},
  {Lanusse}, {Li}, {Mandelbaum}, {Morrison}, {Newman}, {Pope}, {Rykoff},
  {Simet}, {To}, {Vikraman}, {Wechsler}, {White}, \& {(The LSST Dark Energy
  Science Collaboration}}]{dc2}
{Korytov}, D., {Hearin}, A., {Kovacs}, E., {et~al.} 2019, \apjs, 245, 26,
  \dodoi{10.3847/1538-4365/ab510c}

\bibitem[{{Laureijs} {et~al.}(2011){Laureijs}, {Amiaux}, {Arduini},
  {Augu{\`e}res}, {Brinchmann}, {Cole}, {Cropper}, {Dabin}, {Duvet}, {Ealet},
  {Garilli}, {Gondoin}, {Guzzo}, {Hoar}, {Hoekstra}, {Holmes}, {Kitching},
  {Maciaszek}, {Mellier}, {Pasian}, {Percival}, {Rhodes}, {Saavedra Criado},
  {Sauvage}, {Scaramella}, {Valenziano}, {Warren}, {Bender}, {Castander},
  {Cimatti}, {Le F{\`e}vre}, {Kurki-Suonio}, {Levi}, {Lilje}, {Meylan},
  {Nichol}, {Pedersen}, {Popa}, {Rebolo Lopez}, {Rix}, {Rottgering},
  {Zeilinger}, {Grupp}, {Hudelot}, {Massey}, {Meneghetti}, {Miller}, {Paltani},
  {Paulin-Henriksson}, {Pires}, {Saxton}, {Schrabback}, {Seidel}, {Walsh},
  {Aghanim}, {Amendola}, {Bartlett}, {Baccigalupi}, {Beaulieu}, {Benabed},
  {Cuby}, {Elbaz}, {Fosalba}, {Gavazzi}, {Helmi}, {Hook}, {Irwin}, {Kneib},
  {Kunz}, {Mannucci}, {Moscardini}, {Tao}, {Teyssier}, {Weller}, {Zamorani},
  {Zapatero Osorio}, {Boulade}, {Foumond}, {Di Giorgio}, {Guttridge}, {James},
  {Kemp}, {Martignac}, {Spencer}, {Walton}, {Bl{\"u}mchen}, {Bonoli},
  {Bortoletto}, {Cerna}, {Corcione}, {Fabron}, {Jahnke}, {Ligori}, {Madrid},
  {Martin}, {Morgante}, {Pamplona}, {Prieto}, {Riva}, {Toledo}, {Trifoglio},
  {Zerbi}, {Abdalla}, {Douspis}, {Grenet}, {Borgani}, {Bouwens}, {Courbin},
  {Delouis}, {Dubath}, {Fontana}, {Frailis}, {Grazian}, {Koppenh{\"o}fer},
  {Mansutti}, {Melchior}, {Mignoli}, {Mohr}, {Neissner}, {Noddle}, {Poncet},
  {Scodeggio}, {Serrano}, {Shane}, {Starck}, {Surace}, {Taylor},
  {Verdoes-Kleijn}, {Vuerli}, {Williams}, {Zacchei}, {Altieri}, {Escudero
  Sanz}, {Kohley}, {Oosterbroek}, {Astier}, {Bacon}, {Bardelli}, {Baugh},
  {Bellagamba}, {Benoist}, {Bianchi}, {Biviano}, {Branchini}, {Carbone},
  {Cardone}, {Clements}, {Colombi}, {Conselice}, {Cresci}, {Deacon}, {Dunlop},
  {Fedeli}, {Fontanot}, {Franzetti}, {Giocoli}, {Garcia-Bellido}, {Gow},
  {Heavens}, {Hewett}, {Heymans}, {Holland}, {Huang}, {Ilbert}, {Joachimi},
  {Jennins}, {Kerins}, {Kiessling}, {Kirk}, {Kotak}, {Krause}, {Lahav}, {van
  Leeuwen}, {Lesgourgues}, {Lombardi}, {Magliocchetti}, {Maguire}, {Majerotto},
  {Maoli}, {Marulli}, {Maurogordato}, {McCracken}, {McLure}, {Melchiorri},
  {Merson}, {Moresco}, {Nonino}, {Norberg}, {Peacock}, {Pello}, {Penny},
  {Pettorino}, {Di Porto}, {Pozzetti}, {Quercellini}, {Radovich}, {Rassat},
  {Roche}, {Ronayette}, {Rossetti}, {Sartoris}, {Schneider}, {Semboloni},
  {Serjeant}, {Simpson}, {Skordis}, {Smadja}, {Smartt}, {Spano}, {Spiro},
  {Sullivan}, {Tilquin}, {Trotta}, {Verde}, {Wang}, {Williger}, {Zhao},
  {Zoubian}, \& {Zucca}}]{euclid}
{Laureijs}, R., {Amiaux}, J., {Arduini}, S., {et~al.} 2011, arXiv e-prints,
  arXiv:1110.3193.
\newblock \doarXiv{1110.3193}

\bibitem[{{Lawrence} {et~al.}(2007){Lawrence}, {Warren}, {Almaini}, {Edge},
  {Hambly}, {Jameson}, {Lucas}, {Casali}, {Adamson}, {Dye}, {Emerson},
  {Foucaud}, {Hewett}, {Hirst}, {Hodgkin}, {Irwin}, {Lodieu}, {McMahon},
  {Simpson}, {Smail}, {Mortlock}, \& {Folger}}]{ukidss}
{Lawrence}, A., {Warren}, S.~J., {Almaini}, O., {et~al.} 2007, \mnras, 379,
  1599, \dodoi{10.1111/j.1365-2966.2007.12040.x}

\bibitem[{{Lemon} {et~al.}(2020){Lemon}, {Auger}, {McMahon}, {Anguita},
  {Apostolovski}, {Chen}, {Fassnacht}, {Melo}, {Motta}, {Shajib}, {Treu},
  {Agnello}, {Buckley-Geer}, {Schechter}, {Birrer}, {Collett}, {Courbin},
  {Rusu}, {Abbott}, {Allam}, {Annis}, {Avila}, {Bertin}, {Brooks}, {Burke},
  {Carnero Rosell}, {Carrasco Kind}, {Carretero}, {Costanzi}, {da Costa}, {De
  Vicente}, {Desai}, {Eifler}, {Flaugher}, {Frieman}, {Garc{\'\i}a-Bellido},
  {Gaztanaga}, {Gerdes}, {Gruen}, {Gruendl}, {Gschwend}, {Gutierrez},
  {Honscheid}, {James}, {Kim}, {Krause}, {Kuehn}, {Kuropatkin}, {Lahav},
  {Lima}, {Lin}, {Maia}, {March}, {Marshall}, {Menanteau}, {Miquel}, {Palmese},
  {Paz-Chinch{\'o}n}, {Plazas}, {Roodman}, {Sanchez}, {Schubnell}, {Serrano},
  {Smith}, {Soares-Santos}, {Suchyta}, {Tarle}, \& {Walker}}]{lemon20}
{Lemon}, C., {Auger}, M.~W., {McMahon}, R., {et~al.} 2020, \mnras, 494, 3491,
  \dodoi{10.1093/mnras/staa652}

\bibitem[{{Lemon} {et~al.}(2019){Lemon}, {Auger}, \& {McMahon}}]{lemon19}
{Lemon}, C.~A., {Auger}, M.~W., \& {McMahon}, R.~G. 2019, \mnras, 483, 4242,
  \dodoi{10.1093/mnras/sty3366}

\bibitem[{{Lemon} {et~al.}(2018){Lemon}, {Auger}, {McMahon}, \&
  {Ostrovski}}]{lemon18}
{Lemon}, C.~A., {Auger}, M.~W., {McMahon}, R.~G., \& {Ostrovski}, F. 2018,
  \mnras, 479, 5060, \dodoi{10.1093/mnras/sty911}

\bibitem[{{LSST Science Collaboration} {et~al.}(2009){LSST Science
  Collaboration}, {Abell}, {Allison}, {Anderson}, {Andrew}, {Angel}, {Armus},
  {Arnett}, {Asztalos}, {Axelrod}, {Bailey}, {Ballantyne}, {Bankert},
  {Barkhouse}, {Barr}, {Barrientos}, {Barth}, {Bartlett}, {Becker}, {Becla},
  {Beers}, {Bernstein}, {Biswas}, {Blanton}, {Bloom}, {Bochanski}, {Boeshaar},
  {Borne}, {Bradac}, {Brandt}, {Bridge}, {Brown}, {Brunner}, {Bullock},
  {Burgasser}, {Burge}, {Burke}, {Cargile}, {Chandrasekharan}, {Chartas},
  {Chesley}, {Chu}, {Cinabro}, {Claire}, {Claver}, {Clowe}, {Connolly}, {Cook},
  {Cooke}, {Cooray}, {Covey}, {Culliton}, {de Jong}, {de Vries}, {Debattista},
  {Delgado}, {Dell'Antonio}, {Dhital}, {Di Stefano}, {Dickinson}, {Dilday},
  {Djorgovski}, {Dobler}, {Donalek}, {Dubois-Felsmann}, {Durech},
  {Eliasdottir}, {Eracleous}, {Eyer}, {Falco}, {Fan}, {Fassnacht}, {Ferguson},
  {Fernandez}, {Fields}, {Finkbeiner}, {Figueroa}, {Fox}, {Francke}, {Frank},
  {Frieman}, {Fromenteau}, {Furqan}, {Galaz}, {Gal-Yam}, {Garnavich},
  {Gawiser}, {Geary}, {Gee}, {Gibson}, {Gilmore}, {Grace}, {Green}, {Gressler},
  {Grillmair}, {Habib}, {Haggerty}, {Hamuy}, {Harris}, {Hawley}, {Heavens},
  {Hebb}, {Henry}, {Hileman}, {Hilton}, {Hoadley}, {Holberg}, {Holman},
  {Howell}, {Infante}, {Ivezic}, {Jacoby}, {Jain}, {R}, {Jedicke}, {Jee},
  {Garrett Jernigan}, {Jha}, {Johnston}, {Jones}, {Juric}, {Kaasalainen},
  {Styliani}, {Kafka}, {Kahn}, {Kaib}, {Kalirai}, {Kantor}, {Kasliwal},
  {Keeton}, {Kessler}, {Knezevic}, {Kowalski}, {Krabbendam}, {Krughoff},
  {Kulkarni}, {Kuhlman}, {Lacy}, {Lepine}, {Liang}, {Lien}, {Lira}, {Long},
  {Lorenz}, {Lotz}, {Lupton}, {Lutz}, {Macri}, {Mahabal}, {Mandelbaum},
  {Marshall}, {May}, {McGehee}, {Meadows}, {Meert}, {Milani}, {Miller},
  {Miller}, {Mills}, {Minniti}, {Monet}, {Mukadam}, {Nakar}, {Neill}, {Newman},
  {Nikolaev}, {Nordby}, {O'Connor}, {Oguri}, {Oliver}, {Olivier}, {Olsen},
  {Olsen}, {Olszewski}, {Oluseyi}, {Padilla}, {Parker}, {Pepper}, {Peterson},
  {Petry}, {Pinto}, {Pizagno}, {Popescu}, {Prsa}, {Radcka}, {Raddick},
  {Rasmussen}, {Rau}, {Rho}, {Rhoads}, {Richards}, {Ridgway}, {Robertson},
  {Roskar}, {Saha}, {Sarajedini}, {Scannapieco}, {Schalk}, {Schindler},
  {Schmidt}, {Schmidt}, {Schneider}, {Schumacher}, {Scranton}, {Sebag},
  {Seppala}, {Shemmer}, {Simon}, {Sivertz}, {Smith}, {Allyn Smith}, {Smith},
  {Spitz}, {Stanford}, {Stassun}, {Strader}, {Strauss}, {Stubbs}, {Sweeney},
  {Szalay}, {Szkody}, {Takada}, {Thorman}, {Trilling}, {Trimble}, {Tyson}, {Van
  Berg}, {Vanden Berk}, {VanderPlas}, {Verde}, {Vrsnak}, {Walkowicz},
  {Wandelt}, {Wang}, {Wang}, {Warner}, {Wechsler}, {West}, {Wiecha},
  {Williams}, {Willman}, {Wittman}, {Wolff}, {Wood-Vasey}, {Wozniak}, {Young},
  {Zentner}, \& {Zhan}}]{sciencebook}
{LSST Science Collaboration}, {Abell}, P.~A., {Allison}, J., {et~al.} 2009,
  arXiv e-prints, arXiv:0912.0201.
\newblock \doarXiv{0912.0201}

\bibitem[{{Luhtaru} {et~al.}(2021){Luhtaru}, {Schechter}, \& {de
  Soto}}]{luhtaru21}
{Luhtaru}, R., {Schechter}, P.~L., \& {de Soto}, K.~M. 2021, \apj, 915, 4,
  \dodoi{10.3847/1538-4357/abfda1}

\bibitem[{{Mao} {et~al.}(2018){Mao}, {Kovacs}, {Heitmann}, {Uram}, {Benson},
  {Campbell}, {Cora}, {DeRose}, {Di Matteo}, {Habib}, {Hearin}, {Bryce
  Kalmbach}, {Krughoff}, {Lanusse}, {Luki{\'c}}, {Mandelbaum}, {Newman},
  {Padilla}, {Paillas}, {Pope}, {Ricker}, {Ruiz}, {Tenneti},
  {Vega-Mart{\'\i}nez}, {Wechsler}, {Zhou}, {Zu}, \& {LSST Dark Energy Science
  Collaboration}}]{gcrcatalog}
{Mao}, Y.-Y., {Kovacs}, E., {Heitmann}, K., {et~al.} 2018, \apjs, 234, 36,
  \dodoi{10.3847/1538-4365/aaa6c3}

\bibitem[{{Matsuoka} {et~al.}(2018){Matsuoka}, {Strauss}, {Kashikawa}, {Onoue},
  {Iwasawa}, {Tang}, {Lee}, {Imanishi}, {Nagao}, {Akiyama}, {Asami}, {Bosch},
  {Furusawa}, {Goto}, {Gunn}, {Harikane}, {Ikeda}, {Izumi}, {Kawaguchi},
  {Kato}, {Kikuta}, {Kohno}, {Komiyama}, {Lupton}, {Minezaki}, {Miyazaki},
  {Murayama}, {Niida}, {Nishizawa}, {Noboriguchi}, {Oguri}, {Ono}, {Ouchi},
  {Price}, {Sameshima}, {Schulze}, {Shirakata}, {Silverman}, {Sugiyama},
  {Tait}, {Takada}, {Takata}, {Tanaka}, {Toba}, {Utsumi}, {Wang}, \&
  {Yamashita}}]{matsuoka18}
{Matsuoka}, Y., {Strauss}, M.~A., {Kashikawa}, N., {et~al.} 2018, \apj, 869,
  150, \dodoi{10.3847/1538-4357/aaee7a}

\bibitem[{{McGreer} {et~al.}(2018){McGreer}, {Fan}, {Jiang}, \&
  {Cai}}]{McGreer18}
{McGreer}, I.~D., {Fan}, X., {Jiang}, L., \& {Cai}, Z. 2018, \aj, 155, 131,
  \dodoi{10.3847/1538-3881/aaaab4}

\bibitem[{{McGreer} {et~al.}(2010){McGreer}, {Hall}, {Fan}, {Bian}, {Inada},
  {Oguri}, {Strauss}, {Schneider}, \& {Farnsworth}}]{mcgreer10}
{McGreer}, I.~D., {Hall}, P.~B., {Fan}, X., {et~al.} 2010, \aj, 140, 370,
  \dodoi{10.1088/0004-6256/140/2/370}

\bibitem[{{McGreer} {et~al.}(2013){McGreer}, {Jiang}, {Fan}, {Richards},
  {Strauss}, {Ross}, {White}, {Shen}, {Schneider}, {Myers}, {Brandt}, {DeGraf},
  {Glikman}, {Ge}, \& {Streblyanska}}]{simqso}
{McGreer}, I.~D., {Jiang}, L., {Fan}, X., {et~al.} 2013, \apj, 768, 105,
  \dodoi{10.1088/0004-637X/768/2/105}

\bibitem[{{More} {et~al.}(2016){More}, {Oguri}, {Kayo}, {Zinn}, {Strauss},
  {Santiago}, {Mosquera}, {Inada}, {Kochanek}, {Rusu}, {Brownstein}, {da
  Costa}, {Kneib}, {Maia}, {Quimby}, {Schneider}, {Streblyanska}, \&
  {York}}]{more16}
{More}, A., {Oguri}, M., {Kayo}, I., {et~al.} 2016, \mnras, 456, 1595,
  \dodoi{10.1093/mnras/stv2813}

\bibitem[{{Nigoche-Netro} {et~al.}(2016){Nigoche-Netro}, {Ramos-Larios},
  {Lagos}, {Ruelas-Mayorga}, {de la Fuente}, {Kemp}, {Navarro}, {Corral}, \&
  {Hidalgo-G{\'a}mez}}]{nn16}
{Nigoche-Netro}, A., {Ramos-Larios}, G., {Lagos}, P., {et~al.} 2016, \mnras,
  462, 951, \dodoi{10.1093/mnras/stw1661}

\bibitem[{{Niida} {et~al.}(2020){Niida}, {Nagao}, {Ikeda}, {Akiyama},
  {Matsuoka}, {He}, {Matsuoka}, {Toba}, {Onoue}, {Kobayashi}, {Taniguchi},
  {Furusawa}, {Harikane}, {Imanishi}, {Kashikawa}, {Kawaguchi}, {Komiyama},
  {Shirakata}, {Terashima}, \& {Ueda}}]{niida20}
{Niida}, M., {Nagao}, T., {Ikeda}, H., {et~al.} 2020, \apj, 904, 89,
  \dodoi{10.3847/1538-4357/abbe11}

\bibitem[{{Oguri}(2006)}]{oguri06}
{Oguri}, M. 2006, \mnras, 367, 1241, \dodoi{10.1111/j.1365-2966.2006.10043.x}

\bibitem[{{Oguri}(2010)}]{glafic}
---. 2010, \pasj, 62, 1017, \dodoi{10.1093/pasj/62.4.1017}

\bibitem[{{Oguri} \& {Marshall}(2010)}]{om10}
{Oguri}, M., \& {Marshall}, P.~J. 2010, \mnras, 405, 2579,
  \dodoi{10.1111/j.1365-2966.2010.16639.x}

\bibitem[{{Palanque-Delabrouille} {et~al.}(2016){Palanque-Delabrouille},
  {Magneville}, {Y{\`e}che}, {P{\^a}ris}, {Petitjean}, {Burtin}, {Dawson},
  {McGreer}, {Myers}, {Rossi}, {Schlegel}, {Schneider}, {Streblyanska}, \&
  {Tinker}}]{pd16}
{Palanque-Delabrouille}, N., {Magneville}, C., {Y{\`e}che}, C., {et~al.} 2016,
  \aap, 587, A41, \dodoi{10.1051/0004-6361/201527392}

\bibitem[{{Paraficz} {et~al.}(2018){Paraficz}, {Rybak}, {McKean}, {Vegetti},
  {Sluse}, {Courbin}, {Stacey}, {Suyu}, {Dessauges-Zavadsky}, {Fassnacht}, \&
  {Koopmans}}]{paraficz18}
{Paraficz}, D., {Rybak}, M., {McKean}, J.~P., {et~al.} 2018, \aap, 613, A34,
  \dodoi{10.1051/0004-6361/201731250}

\bibitem[{{Peng} {et~al.}(2002){Peng}, {Ho}, {Impey}, \& {Rix}}]{galfit}
{Peng}, C.~Y., {Ho}, L.~C., {Impey}, C.~D., \& {Rix}, H.-W. 2002, \aj, 124,
  266, \dodoi{10.1086/340952}

\bibitem[{{Richards} {et~al.}(2006){Richards}, {Strauss}, {Fan}, {Hall},
  {Jester}, {Schneider}, {Vanden Berk}, {Stoughton}, {Anderson}, {Brunner},
  {Gray}, {Gunn}, {Ivezi{\'c}}, {Kirkland}, {Knapp}, {Loveday}, {Meiksin},
  {Pope}, {Szalay}, {Thakar}, {Yanny}, {York}, {Barentine}, {Brewington},
  {Brinkmann}, {Fukugita}, {Harvanek}, {Kent}, {Kleinman}, {Krzesi{\'n}ski},
  {Long}, {Lupton}, {Nash}, {Neilsen}, {Nitta}, {Schlegel}, \&
  {Snedden}}]{richards06}
{Richards}, G.~T., {Strauss}, M.~A., {Fan}, X., {et~al.} 2006, \aj, 131, 2766,
  \dodoi{10.1086/503559}

\bibitem[{{Ross} {et~al.}(2013){Ross}, {McGreer}, {White}, {Richards}, {Myers},
  {Palanque-Delabrouille}, {Strauss}, {Anderson}, {Shen}, {Brandt},
  {Y{\`e}che}, {Swanson}, {Aubourg}, {Bailey}, {Bizyaev}, {Bovy}, {Brewington},
  {Brinkmann}, {DeGraf}, {Di Matteo}, {Ebelke}, {Fan}, {Ge}, {Malanushenko},
  {Malanushenko}, {Mandelbaum}, {Maraston}, {Muna}, {Oravetz}, {Pan},
  {P{\^a}ris}, {Petitjean}, {Schawinski}, {Schlegel}, {Schneider}, {Silverman},
  {Simmons}, {Snedden}, {Streblyanska}, {Suzuki}, {Weinberg}, \& {York}}]{dr9}
{Ross}, N.~P., {McGreer}, I.~D., {White}, M., {et~al.} 2013, \apj, 773, 14,
  \dodoi{10.1088/0004-637X/773/1/14}

\bibitem[{{Sohn} {et~al.}(2017){Sohn}, {Zahid}, \& {Geller}}]{sohn17}
{Sohn}, J., {Zahid}, H.~J., \& {Geller}, M.~J. 2017, \apj, 845, 73,
  \dodoi{10.3847/1538-4357/aa7de3}

\bibitem[{{Spergel} {et~al.}(2015){Spergel}, {Gehrels}, {Baltay}, {Bennett},
  {Breckinridge}, {Donahue}, {Dressler}, {Gaudi}, {Greene}, {Guyon}, {Hirata},
  {Kalirai}, {Kasdin}, {Macintosh}, {Moos}, {Perlmutter}, {Postman},
  {Rauscher}, {Rhodes}, {Wang}, {Weinberg}, {Benford}, {Hudson}, {Jeong},
  {Mellier}, {Traub}, {Yamada}, {Capak}, {Colbert}, {Masters}, {Penny},
  {Savransky}, {Stern}, {Zimmerman}, {Barry}, {Bartusek}, {Carpenter}, {Cheng},
  {Content}, {Dekens}, {Demers}, {Grady}, {Jackson}, {Kuan}, {Kruk}, {Melton},
  {Nemati}, {Parvin}, {Poberezhskiy}, {Peddie}, {Ruffa}, {Wallace}, {Whipple},
  {Wollack}, \& {Zhao}}]{wfirst}
{Spergel}, D., {Gehrels}, N., {Baltay}, C., {et~al.} 2015, arXiv e-prints,
  arXiv:1503.03757.
\newblock \doarXiv{1503.03757}

\bibitem[{{Spiniello} {et~al.}(2018){Spiniello}, {Agnello}, {Napolitano},
  {Sergeyev}, {Getman}, {Tortora}, {Spavone}, {Bilicki}, {Buddelmeijer},
  {Koopmans}, {Kuijken}, {Vernardos}, {Bannikova}, \&
  {Capaccioli}}]{spiniello18}
{Spiniello}, C., {Agnello}, A., {Napolitano}, N.~R., {et~al.} 2018, \mnras,
  480, 1163, \dodoi{10.1093/mnras/sty1923}

\bibitem[{{Suyu} {et~al.}(2017){Suyu}, {Bonvin}, {Courbin}, {Fassnacht},
  {Rusu}, {Sluse}, {Treu}, {Wong}, {Auger}, {Ding}, {Hilbert}, {Marshall},
  {Rumbaugh}, {Sonnenfeld}, {Tewes}, {Tihhonova}, {Agnello}, {Blandford},
  {Chen}, {Collett}, {Koopmans}, {Liao}, {Meylan}, \& {Spiniello}}]{suyu17}
{Suyu}, S.~H., {Bonvin}, V., {Courbin}, F., {et~al.} 2017, \mnras, 468, 2590,
  \dodoi{10.1093/mnras/stx483}

\bibitem[{{Treu} {et~al.}(2018){Treu}, {Agnello}, {Baumer}, {Birrer},
  {Buckley-Geer}, {Courbin}, {Kim}, {Lin}, {Marshall}, {Nord}, {Schechter},
  {Sivakumar}, {Abramson}, {Anguita}, {Apostolovski}, {Auger}, {Chan}, {Chen},
  {Collett}, {Fassnacht}, {Hsueh}, {Lemon}, {McMahon}, {Motta}, {Ostrovski},
  {Rojas}, {Rusu}, {Williams}, {Frieman}, {Meylan}, {Suyu}, {Abbott},
  {Abdalla}, {Allam}, {Annis}, {Avila}, {Banerji}, {Brooks}, {Carnero Rosell},
  {Carrasco Kind}, {Carretero}, {Castander}, {D'Andrea}, {da Costa}, {De
  Vicente}, {Doel}, {Eifler}, {Flaugher}, {Fosalba}, {Garc{\'\i}a-Bellido},
  {Goldstein}, {Gruen}, {Gruendl}, {Gutierrez}, {Hartley}, {Hollowood},
  {Honscheid}, {James}, {Kuehn}, {Kuropatkin}, {Lima}, {Maia}, {Martini},
  {Menanteau}, {Miquel}, {Plazas}, {Romer}, {Sanchez}, {Scarpine}, {Schindler},
  {Schubnell}, {Sevilla-Noarbe}, {Smith}, {Smith}, {Soares-Santos}, {Sobreira},
  {Suchyta}, {Swanson}, {Tarle}, {Thomas}, {Tucker}, \& {Walker}}]{treu18}
{Treu}, T., {Agnello}, A., {Baumer}, M.~A., {et~al.} 2018, \mnras, 481, 1041,
  \dodoi{10.1093/mnras/sty2329}

\bibitem[{{van der Walt} {et~al.}(2011){van der Walt}, {Colbert}, \&
  {Varoquaux}}]{numpy1}
{van der Walt}, S., {Colbert}, S.~C., \& {Varoquaux}, G. 2011, Computing in
  Science and Engineering, 13, 22, \dodoi{10.1109/MCSE.2011.37}

\bibitem[{{Wang} {et~al.}(2019){Wang}, {Yang}, {Fan}, {Wu}, {Yue}, {Li},
  {Bian}, {Jiang}, {Ba{\~n}ados}, {Schindler}, {Findlay}, {Davies}, {Decarli},
  {Farina}, {Green}, {Hennawi}, {Huang}, {Mazzuccheli}, {McGreer}, {Venemans},
  {Walter}, {Dye}, {Lyke}, {Myers}, \& {Haze Nunez}}]{wang19}
{Wang}, F., {Yang}, J., {Fan}, X., {et~al.} 2019, \apj, 884, 30,
  \dodoi{10.3847/1538-4357/ab2be5}

\bibitem[{{Wright} {et~al.}(2010){Wright}, {Eisenhardt}, {Mainzer}, {Ressler},
  {Cutri}, {Jarrett}, {Kirkpatrick}, {Padgett}, {McMillan}, {Skrutskie},
  {Stanford}, {Cohen}, {Walker}, {Mather}, {Leisawitz}, {Gautier}, {McLean},
  {Benford}, {Lonsdale}, {Blain}, {Mendez}, {Irace}, {Duval}, {Liu}, {Royer},
  {Heinrichsen}, {Howard}, {Shannon}, {Kendall}, {Walsh}, {Larsen}, {Cardon},
  {Schick}, {Schwalm}, {Abid}, {Fabinsky}, {Naes}, \& {Tsai}}]{wise}
{Wright}, E.~L., {Eisenhardt}, P. R.~M., {Mainzer}, A.~K., {et~al.} 2010, \aj,
  140, 1868, \dodoi{10.1088/0004-6256/140/6/1868}

\bibitem[{{Wyithe} \& {Loeb}(2002)}]{wyithe02}
{Wyithe}, J. S.~B., \& {Loeb}, A. 2002, \apj, 577, 57, \dodoi{10.1086/342181}

\bibitem[{{Wyithe} {et~al.}(2011){Wyithe}, {Yan}, {Windhorst}, \&
  {Mao}}]{wyithe11}
{Wyithe}, J. S.~B., {Yan}, H., {Windhorst}, R.~A., \& {Mao}, S. 2011, \nat,
  469, 181, \dodoi{10.1038/nature09619}

\bibitem[{{Yang} {et~al.}(2016){Yang}, {Wang}, {Wu}, {Fan}, {McGreer}, {Bian},
  {Yi}, {Yang}, {Ai}, {Dong}, {Zuo}, {Green}, {Jiang}, {Wang}, {Wang}, \&
  {Yue}}]{yang16}
{Yang}, J., {Wang}, F., {Wu}, X.-B., {et~al.} 2016, \apj, 829, 33,
  \dodoi{10.3847/0004-637X/829/1/33}

\bibitem[{{Yang} {et~al.}(2019){Yang}, {Venemans}, {Wang}, {Fan}, {Novak},
  {Decarli}, {Walter}, {Yue}, {Momjian}, {Keeton}, {Wang}, {Zabludoff}, {Wu},
  \& {Bian}}]{yang19}
{Yang}, J., {Venemans}, B., {Wang}, F., {et~al.} 2019, \apj, 880, 153,
  \dodoi{10.3847/1538-4357/ab2a02}

\bibitem[{{York} {et~al.}(2000){York}, {Adelman}, {Anderson}, {Anderson},
  {Annis}, {Bahcall}, {Bakken}, {Barkhouser}, {Bastian}, {Berman}, {Boroski},
  {Bracker}, {Briegel}, {Briggs}, {Brinkmann}, {Brunner}, {Burles}, {Carey},
  {Carr}, {Castander}, {Chen}, {Colestock}, {Connolly}, {Crocker}, {Csabai},
  {Czarapata}, {Davis}, {Doi}, {Dombeck}, {Eisenstein}, {Ellman}, {Elms},
  {Evans}, {Fan}, {Federwitz}, {Fiscelli}, {Friedman}, {Frieman}, {Fukugita},
  {Gillespie}, {Gunn}, {Gurbani}, {de Haas}, {Haldeman}, {Harris}, {Hayes},
  {Heckman}, {Hennessy}, {Hindsley}, {Holm}, {Holmgren}, {Huang}, {Hull},
  {Husby}, {Ichikawa}, {Ichikawa}, {Ivezi{\'c}}, {Kent}, {Kim}, {Kinney},
  {Klaene}, {Kleinman}, {Kleinman}, {Knapp}, {Korienek}, {Kron}, {Kunszt},
  {Lamb}, {Lee}, {Leger}, {Limmongkol}, {Lindenmeyer}, {Long}, {Loomis},
  {Loveday}, {Lucinio}, {Lupton}, {MacKinnon}, {Mannery}, {Mantsch}, {Margon},
  {McGehee}, {McKay}, {Meiksin}, {Merelli}, {Monet}, {Munn}, {Narayanan},
  {Nash}, {Neilsen}, {Neswold}, {Newberg}, {Nichol}, {Nicinski}, {Nonino},
  {Okada}, {Okamura}, {Ostriker}, {Owen}, {Pauls}, {Peoples}, {Peterson},
  {Petravick}, {Pier}, {Pope}, {Pordes}, {Prosapio}, {Rechenmacher}, {Quinn},
  {Richards}, {Richmond}, {Rivetta}, {Rockosi}, {Ruthmansdorfer}, {Sandford},
  {Schlegel}, {Schneider}, {Sekiguchi}, {Sergey}, {Shimasaku}, {Siegmund},
  {Smee}, {Smith}, {Snedden}, {Stone}, {Stoughton}, {Strauss}, {Stubbs},
  {SubbaRao}, {Szalay}, {Szapudi}, {Szokoly}, {Thakar}, {Tremonti}, {Tucker},
  {Uomoto}, {Vanden Berk}, {Vogeley}, {Waddell}, {Wang}, {Watanabe},
  {Weinberg}, {Yanny}, {Yasuda}, \& {SDSS Collaboration}}]{sdss}
{York}, D.~G., {Adelman}, J., {Anderson}, John~E., J., {et~al.} 2000, \aj, 120,
  1579, \dodoi{10.1086/301513}

\bibitem[{{Yue} {et~al.}(2021{\natexlab{a}}){Yue}, {Fan}, {Yang}, \&
  {Wang}}]{yue22}
{Yue}, M., {Fan}, X., {Yang}, J., \& {Wang}, F. 2021{\natexlab{a}}, arXiv
  e-prints, arXiv:2112.02821.
\newblock \doarXiv{2112.02821}

\bibitem[{{Yue} {et~al.}(2021{\natexlab{b}}){Yue}, {Yang}, {Fan}, {Wang},
  {Spilker}, {Georgiev}, {Keeton}, {Litke}, {Marrone}, {Walter}, {Wang}, {Wu},
  {Venemans}, \& {Zabludoff}}]{yue21}
{Yue}, M., {Yang}, J., {Fan}, X., {et~al.} 2021{\natexlab{b}}, \apj, 917, 99,
  \dodoi{10.3847/1538-4357/ac0af4}

\end{thebibliography}
\bibliographystyle{aasjournal}



\end{document}